\documentclass[11pt,a4paper]{article}
\usepackage{jheppub}

\usepackage{amsmath}
\usepackage{amsfonts}
\usepackage{slashed}
\usepackage{graphicx}
\usepackage{mathrsfs}
\usepackage{tikz}
\usepackage[numbers,sort&compress]{natbib}

\makeatletter
\g@addto@macro\bfseries{\boldmath}
\makeatother

\newcommand{\eps}{\epsilon}
\newcommand{\ord}{\begin{cal}O\end{cal}}

\def\beq{\begin{equation}}
\def\eeq{\end{equation}}
\def\bsp#1\esp{\begin{split}#1\end{split}}

\makeatletter
\newcommand{\IEIF}{%
  \def\@IEIFsep{(}%
  I_F\@IEIFi
}
\newcommand\@IEIFi{\@ifnextchar\stopIEIF{\@IEIFend}{\@IEIFii}}
\newcommand\@IEIFii[4]{%
  \big\@IEIFsep
  \begin{smallmatrix}
    #1 & #2 \\
    #3 & #4
  \end{smallmatrix}
  \def\@IEIFsep{|}
  \@IEIFi
}
\newcommand\@IEIFend[2]{% 
  ; #2 \bigr)
}
\makeatother
\newcommand{\cD}{\begin{cal}D\end{cal}}
\newcommand{\cE}{\begin{cal}E\end{cal}}

\newcommand{\cI}{\begin{cal}I\end{cal}}

\newcommand{\cM}{\begin{cal}M\end{cal}}
\newcommand{\cN}{\begin{cal}N\end{cal}}

\newcommand{\cP}{\begin{cal}P\end{cal}}

\newcommand{\cS}{\begin{cal}S\end{cal}}

\newcommand{\cW}{\begin{cal}W\end{cal}}

\def \sha{{\,\amalg\hskip -3.8pt\amalg\,}}

\newcommand{\gam}[3]{{\Gamma}\!\left(\begin{smallmatrix}#1\\#2\end{smallmatrix};#3\right)}
\newcommand{\gamtt}[4]{{\widetilde{\Gamma}}\!\left(\begin{smallmatrix}#1\\#2\end{smallmatrix};#3,#4\right)}

\newcommand{\IMF}[3]{{ I}\!\left( %\left[ 
\begin{smallmatrix}#1\\#2\end{smallmatrix}
;#3\right)}

\newcommand{\IMFF}[3]{{ I_F}\!\left( %\left[ 
\begin{smallmatrix}#1\\#2\end{smallmatrix}
;#3\right)}

\renewcommand{\ln}{\log}

\newcommand{\symb}{\mathscr{S}}

\title{Elliptic symbol calculus: from elliptic polylogarithms to iterated integrals of Eisenstein series}

\author[a]{Johannes Broedel} 
\author[b,c]{Claude Duhr}
\author[d]{Falko Dulat}
\author[b]{Brenda Penante}
\author[b]{Lorenzo Tancredi}

\affiliation[a]{Institut f\"{u}r Mathematik und Institut f\"{u}r Physik,
Humboldt-Universit\"{a}t zu Berlin,\\
IRIS Adlershof, Zum Grossen Windkanal 6, 12489 Berlin, Germany} 
\affiliation[b]{Theoretical Physics Department, CERN, Geneva, Switzerland} 
\affiliation[c]{Center for Cosmology, Particle Physics and Phenomenology (CP3),\\
Universit\'e Catholique de Louvain, 1348 Louvain-La-Neuve, Belgium}
\affiliation[d]{SLAC National Accelerator Laboratory, Stanford University, Stanford, CA 94309, USA}

\emailAdd{jbroedel@physik.hu-berlin.de}
\emailAdd{claude.duhr@cern.ch}
\emailAdd{dulatf@slac.stanford.edu}
\emailAdd{b.penante@cern.ch}
\emailAdd{lorenzo.tancredi@cern.ch}

\abstract{
We present a generalization of the symbol calculus from ordinary multiple polylogarithms to their elliptic counterparts. 
Our formalism is based on a special case of a coaction on large classes of periods that is applied in particular to elliptic polylogarithms and iterated integrals of modular forms. We illustrate how to use our formalism to derive relations among elliptic polylogarithms, in complete analogy with the non-elliptic case. We then analyze the symbol alphabet of elliptic polylogarithms evaluated at rational points, and we observe that it is given by Eisenstein series for a certain congruence subgroup. We apply our formalism to hypergeometric functions that can be expressed in terms of elliptic polylogarithms and show that they can equally be written in terms of iterated integrals of Eisenstein series. Finally, we present the symbol of the equal-mass sunrise integral in two space-time dimensions. The symbol alphabet involves Eisenstein series of level six and weight three, and we can easily integrate the symbol in terms of iterated integrals of Eisenstein series.}

\preprint{\begin{minipage}[t]{8cm}\begin{flushright}CP3-18-24, CERN-TH-2018-057\\
            HU-Mathematik-2018-03, HU-EP-18/09\\
            SLAC-PUB-17240\end{flushright}\end{minipage}}

\begin{document}

\maketitle

%%%%%%%%%%%%%%%%%%%%%%%%%%%%%%%%%%%%%%%%%%%%%%%%%%%%%%%%%%%%%%%%%%%%%%%%
% Shamelessly stolen from Thorsten's thohacks.sty
%%%%%%%%%%%%%%%%%%%%%%%%%%%%%%%%%%%%%%%%%%%%%%%%%%%%%%%%%%%%%%%%%%%%%%%%
\catcode`\@=11
\font\manfnt=manfnt
\def\Watchout{\@ifnextchar [{\W@tchout}{\W@tchout[1]}}
\def\W@tchout[#1]{{\manfnt\@tempcnta#1\relax%
  \@whilenum\@tempcnta>\z@\do{%
    \char"7F\hskip 0.3em\advance\@tempcnta\m@ne}}}
\let\foo\W@tchout
\def\dubious{\@ifnextchar[{\@dubious}{\@dubious[1]}}
\let\enddubious\endlist
\def\@dubious[#1]{%
  \setbox\@tempboxa\hbox{\@W@tchout#1}
  \@tempdima\wd\@tempboxa
  \list{}{\leftmargin\@tempdima}\item[\hbox to 0pt{\hss\@W@tchout#1}]}
\def\@W@tchout#1{\W@tchout[#1]}
\catcode`\@=12
%%%%%%%%%%%%%%%%%%%%%%%%%%%%%%%%%%%%%%%%%%%%%%%%%%%%%%%%%%%%%%%%%%%%%%%%

% !TEX root = elliptic_symbols.tex

\section{Introduction}
\label{sec:intro}

The discovery of the Higgs boson and the absence of new physics signals at the Large Hadron Collider (LHC) at CERN has established the Standard Model (SM) of particle physics as the prime candidate for the theory of the electroweak and strong interactions. In particular, the SM is now a complete and fully predictive theory that allows one, at least in principle, to make predictions at a high level of accuracy. However, new physics may still show up at the LHC via small deviations in key cross sections, distributions and decay rates. We are thus entering a new era of precision physics, where precise experimental measurements need to be compared to precise theoretical predictions at the percent level. 

The cornerstone of precision computations in quantum field theory is perturbation theory, where physical observables are expanded into a series in the coupling constants of the theory (assuming that they are not too large). The $n$-th order in perturbation theory then involves a sum of $n$-loop Feynman diagrams with a fixed set of external legs, and we need to integrate over the momentum flowing in each of the $n$ loops. The computation of higher orders in perturbation theory is therefore intimately related to the evaluation of multi-loop scattering amplitudes and Feynman integrals. While the evaluation of one-loop Feynman integrals in four space-time dimensions is well understood (see, e.g., ref.~\cite{Ellis:2007qk} and references therein), the computation of multi-loop integrals is still a bottleneck of precision computations. 

Over the last couple of years there has been a lot of progress in our ability to compute multi-loop scattering amplitudes. This progress has been fueled, among other things, by a better understanding of the class of special functions that show up in multi-loop Feynman integrals. In particular, it was realized that large classes of Feynman integrals evaluate to a class of special functions called \emph{multiple polylogarithms} (MPLs)~\cite{Goncharov:1995,Remiddi:1999ew,Goncharov:1998kja,Vollinga:2004sn}. These functions are not only of interest to physicists, but they have been an active area of research also in contemporary pure mathematics. In particular, MPLs are endowed with many additional algebraic structures that are exploited in novel techniques for Feynman integral calculations.

A particularly powerful tool for computing with MPLs is the so-called \emph{symbol}~\cite{Goncharov:2010jf,Goncharov:2009tja,ChenSymbol,Brown:2009qja,Duhr:2011zq}, a map which associates to an MPL a tensor whose entries are rational (or algebraic) functions. The main advantage of the symbol is that it trivializes complicated functional relations among MPLs. The symbol was introduced in physics for the first time in ref.~\cite{Goncharov:2010jf} in the context of the planar $\cN=4$ Super Yang--Mills theory to simplify the analytic result for the two-loop six-point remainder function of refs.~\cite{DelDuca:2009au,DelDuca:2010zg}. While the symbol of MPLs is easy to work with in practice, a lot of information is lost when passing from a function to its symbol, because the symbol maps  all constants to zero. In ref.~\cite{Duhr:2012fh} it was shown that some of this lost information can be recovered by lifting the symbol to the full coproduct (rather, the coaction) on MPLs~\cite{Goncharov:2005sla,GoncharovMixedTate,Brown:2011ik}. In this context, the symbol map can be seen as the maximal iteration of the coproduct.
The symbol and the coproduct of MPLs have led to the development of novel techniques to compute scattering amplitudes and Feynman integrals that evaluate to MPLs. For example, the coproduct on MPLs is one of the cornerstones of the bootstrap approach to compute scattering amplitudes in planar $\cN=4$ Super Yang--Mills, a program that has allowed for the determination of the six- and seven-particle amplitudes to a remarkably high loop order~\cite{Dixon:2011nj,Dixon:2011pw,Dixon:2013eka,Dixon:2014iba,Dixon:2014voa,Drummond:2014ffa,Dixon:2015iva,Caron-Huot:2016owq,Dixon:2016nkn}. It has also played an important role in more phenomenological applications, most notably in simplifying expressions for two-loop amplitudes for diboson production at the LHC~\cite{Gehrmann:2015ora} and in the computation of the N$^3$LO corrections to the inclusive Higgs production cross section~\cite{Anastasiou:2013mca,Anastasiou:2013srw,Duhr:2013msa,Anastasiou:2014lda,Anastasiou:2014vaa,Duhr:2014nda,Dulat:2014mda,Anastasiou:2015ema,Anastasiou:2015yha}.

While the mathematics of Feynman integrals that evaluate to MPLs is by now well understood, thanks also to the symbol and the coproduct, it is known that not every Feynman integral can be expressed in terms of MPLs, but more general classes of special functions are required. The first time a non-polylogarithmic function was shown to appear in a quantum field theory computation was in the context of the two-loop corrections to the electron self-energy in QED~\cite{Sabry}. Since then, non-polylogarithmic functions were observed also in other theories, and in particular in the SM and $\cN=4$ Super Yang--Mills~\cite{Aglietti:2007as,Brown:2013hda,CaronHuot:2012ab,Bonciani:2016qxi,vonManteuffel:2017hms,Primo:2017ipr,Ablinger:2017bjx,Chen:2017pyi,Bourjaily:2017bsb,Chen:2017soz,Broedel:2017kkb,Mistlberger:2018etf,Bloch:2016izu,Remiddi:2013joa,Laporta:2004rb,Bloch:2013tra,Adams:2013nia,Adams:2014vja,Adams:2015gva,Adams:2015ydq,Remiddi:2016gno,Adams:2016xah,Broedel:2017siw}. Over the last  couple of years a lot of effort has been put into trying to  
understand in a unified framework which functions can appear in the computation of a generic Feynman integral.
While the complete class of functions is very far from being understood, there is accumulating evidence that the next important case 
beyond MPLs should be of `elliptic type'. In this context, the so-called sunrise graph has received a lot of attention~\cite{Broadhurst:1987ei,Bauberger:1994by,Bauberger:1994hx,Laporta:2004rb,Kniehl:2005bc,Bloch:2013tra,Adams:2013nia,Remiddi:2013joa,Adams:2014vja,Adams:2015gva,Adams:2015ydq,Remiddi:2016gno,Adams:2016xah,Remiddi:2017har,Adams:2017ejb,Hidding:2017jkk,Broedel:2017siw}\footnote{A
  proposal for the numerical evaluation of the functions which appear in the
calculation of the two-loop massive sunrise graph has been recently put forward
in ref.~\cite{Passarino:2016zcd}.}, because it is the simplest graph which cannot be evaluated in terms of MPLs alone and requires the introduction of their 
elliptic generalizations.

From a mathematical point of view, (part of) the family of functions relevant to elliptic Feynman integrals seem to be the so-called \emph{multiple elliptic polylogarithms} (eMPLs)~\cite{BrownLevin}. These functions have recently been shown to describe also scattering amplitudes in superstring theory, where elliptic curves naturally occur as the worldsheet relevant to one-loop computations~\cite{Broedel:2014vla,Broedel:2015hia,Broedel:2017jdo,Broedel:2018izr}. Closely related to eMPLs are iterated integrals of modular forms~\cite{ManinModular,Brown:mmv}, and it was indeed observed that they show up in Feynman integral computations~
\cite{Ablinger:2017bjx,Adams:2017ejb}. Despite all this progress, we are still missing a complete mathematical picture of elliptic Feynman integrals and the class of special functions they evaluate to.

The purpose of this paper is to take some first steps in trying to improve our understanding of the mathematical properties of the class of eMPLs that show up in Feynman integral computations. In ref.~\cite{Broedel:2017kkb} some of the authors have shown how integrals over square roots that define elliptic curves naturally lead to (a variant of) the eMPLs defined in ref.~\cite{BrownLevin}, and some of the basic properties of these functions have been worked out. In this paper we take an additional step, and we define a variant of the symbol map and the coaction that extends to the elliptic case. Our starting point is a formula for the total differential of eMPLs that is very similar to the total differential of ordinary MPLs~\cite{Goncharov:1998kja}. The formula for the total differential shows that eMPLs satisfy a differential equation without a homogeneous term, which allows us to apply a very general construction due to Brown~\cite{Brown:coaction} to define a coaction on such functions. When applied to ordinary MPLs, this construction yields (a variant of) the well-known symbol and the coaction on these functions. We illustrate on several simple examples how one can use our coaction to derive relations among eMPLs, in the same spirit as in the non-elliptic case (cf. ref.~\cite{Duhr:2014woa}). Brown's construction is generic, and not restricted to MPLs and their elliptic generalizations, and so we can extend the coaction and the symbol to iterated integrals of modular forms.
We show how we can use our formalism to obtain a representation of certain hypergeometric functions that evaluate to eMPLs and of the sunrise integral in terms of iterated integrals of Eisenstein series. Our result clarifies in particular the observation of ref.~\cite{Adams:2017ejb} that in the case of the sunrise integral only Eisenstein series appear, and no cusp forms. 

This paper is organized as follows. In Section~\ref{sec:mpls} we give a short review of ordinary MPLs and their symbol calculus. In Section~\ref{sec:empls} we review the class of eMPLs relevant in this paper, and we present a simple and compact formula for their total differential. The knowledge of the differential can be turned into the definition of a symbol, similar to the case of ordinary MPLs. In Section~\ref{sec:de-Rham-symbols} we present Brown's general construction of symbols and a coaction on certain classes of periods. We apply this construction to the case of eMPLs in Section~\ref{sec:elliptic_symbol_calculus}, and we discuss some simple examples of how to use our formalism to derive functional relations among eMPLs in Section~\ref{sec:applications}. In Section~\ref{sec:modular_forms} we give a short introduction to iterated integrals of modular forms, and we show how one can apply Brown's construction to define a coaction and a symbol map on these integrals. We then use our results in Section~\ref{sec:eMPLs_to_modular} to show that eMPLs evaluated at rational points can always be expressed in terms of iterated integrals of Eisenstein series for a certain congruence subgroup. In Sections~\ref{sec:hyper} and~\ref{sec:sunrise} we present two applications of our formalism. In particular, we obtain representations for hypergeometric functions and for the sunrise integral in terms of iterated integrals of Eisenstein series. In Section~\ref{sec:conc} we draw our conclusions. We include various appendices where we present technical details that are omitted throughout the main text.

% !TEX root = elliptic_symbols.tex

\section{Multiple polylogarithms}
\label{sec:mpls}
Before we discuss how to extend (some of) the algebraic properties of polylogarithms beyond genus zero, we 
present in this section a concise review of ordinary multiple polylogarithms, as well as their symbols and coaction and how to use them to work out relations among MPLs. The material in this section is well known, see, e.g., ref.~\cite{Duhr:2014woa} for a pedagogical review.

\subsection{Multiple polylogarithms and their symbols}
Multiple polylogarithms (MPLs) are defined by the iterated integral~\cite{Lappo:1927,Goncharov:1998kja,GoncharovMixedTate}
 \beq\label{eq:Mult_PolyLog_def}
 G(a_1,\ldots,a_n;z)=\,\int_0^z\,\frac{d t}{t-a_1}\,G(a_2,\ldots,a_n;t)\,,
\eeq
 and the recursion starts with $G(;z)\equiv 1$. In the special case where all the $a_i$'s are zero, we define
\beq
G(\underbrace{0,\ldots,0}_{n\textrm{ times}};z) = \frac{1}{n!}\,\ln^n z\,.
\eeq
The number $n$ of integrations in eq.~\eqref{eq:Mult_PolyLog_def}, or equivalently the number of $a_i$'s, is called the \emph{weight} of the multiple polylogarithm.
A product of two MPLs with the same upper integration limit can be written as a linear combination
of MPLs. More precisely, MPLs form a \emph{shuffle algebra},
\beq\label{eq:G_shuffle}
G(a_1,\ldots,a_k;z)\,G(a_{k+1},\ldots,a_{k+l};z) = \sum_{\sigma\in \Sigma(k,l)}G(a_{\sigma(1)},\ldots,a_{\sigma(k+l)};z)\,,
\eeq
where $\Sigma(k,l)$ denotes the set of all shuffles of $(a_1,\ldots,a_k)$ and $(a_{k+1},\ldots,a_{k+l})$, i.e., the set of all permutations of their union that preserve the relative orderings within each set. Equation~\eqref{eq:G_shuffle} is in fact a special case of a general property of iterated integrals, see Appendix~\ref{app:iterated_integrals}. The shuffle product preserves the weight, i.e., the product of two MPLs of weight $k$ and $l$ is a linear combination of MPLs of weight $k+l$.

Multiple polylogarithms can be endowed with more algebraic structures (see, e.g., refs.~\cite{Goncharov:2005sla,GoncharovMixedTate}). If we work modulo $i\pi$, then the vector space of MPLs mod $i\pi$ (i.e., modulo their branch cuts) is conjectured to form a Hopf algebra with a coproduct $\Delta_{\textrm{MPL}}$ that respects the multiplication and the weight (see Appendix~\ref{app:algebras} for a review of the algebraic structures used in this paper). The coproduct is coassociative,
\beq\label{eq:MPL_coassoc}
(\textrm{id}\otimes \Delta_{\rm MPL})\Delta_{\rm MPL} = (\Delta_{\rm MPL}\otimes \textrm{id})\Delta_{\rm MPL}\,,
\eeq
and it is possible to lift $\Delta_{\textrm{MPL}}$ to a coaction on the full algebra of MPLs, with all factors of $i\pi$ kept. The coaction acts trivially on $i\pi$~\cite{Brown:2011ik,Duhr:2012fh},\footnote{By abuse of notation, we denote both the coproduct and the coaction by $\Delta_{\rm MPL}$, because, at least for our purposes, they are given by the same formula, and differ only by whether or not we work modulo $i\pi$ in the first factor.}
\beq\label{eq:Delta_ipi}
\Delta_{\rm MPL}(i\pi)=i\pi\otimes1\,.
\eeq 
In the generic case where all arguments are distinct, the coaction on MPLs can be cast in the following compact form
\beq\label{eq:Delta_MPL}
\Delta_{\rm MPL}(G(\vec a;z)) = \sum_{\vec b\subseteq \vec a} G(\vec b;z)\otimes G_{\vec b}(\vec a;z)\,,
\eeq
where the sum runs over all order-preserving subsets $\vec b$ of $\vec a$, including the empty set. The function $G_{\vec b}(\vec a;z)$ is defined as the iterated integral with the same integrand as $G(\vec a;z)$, but integrated over the contour $\gamma_{\vec b}$ that encircles the singularities at the points $z=a_i$, $a_i\in \vec b$, in the order in which the elements appear in $\vec b$. This is equivalent to taking the residues at these points, and we divide by $2\pi i$ per residue. Using the path composition formula for iterated integrals (see Appendix~\ref{app:iterated_integrals}), it is easy to check that $G_{\vec b}(\vec a;z)$ itself can be written as a product of MPLs. 

Let us conclude this basic introduction by reviewing how the operations of differentiation and taking discontinuities commute with the coaction~\cite{Duhr:2012fh,Brown:coaction}. One can show that discontinuities only act on the first entry of the coaction
\begin{align}
\label{eq:disc_coproduct}
\Delta_{\rm MPL}\,\textrm{Disc} & = (\textrm{Disc}\otimes\textrm{id})\,\Delta_{\rm MPL}\,,
\end{align}
while derivatives only act on the last entry,
\begin{align}
\label{eq:der_coproduct}
\Delta_{\rm MPL}\,{\partial_z} & = \left(\textrm{id}\otimes{\partial_z}\right)\,\Delta_{\rm MPL}\,.
\end{align}

\subsection{Symbols of MPLs}
\label{sec:MPL_symbol}
Besides the coaction $\Delta_{\textrm{MPL}}$, there is another very useful operation on MPLs, its so-called \emph{symbol}. There are various (essentially) equivalent definitions of the symbol of MPLs in the literature~\cite{ChenSymbol,Goncharov:2009tja,Brown:2009qja,Goncharov:2010jf,Duhr:2011zq}. Here, we follow the folklore in the physics literature and define the `symbol of a transcendental function $F_w$ of weight $w$' as follows: assume that the total differential of $F_w$ can be written in the form
\beq\label{eq:MPL_d_F}
dF_w = \sum_{i}F_{w-1,i}\,d\log R_i\,,
\eeq
where the $F_{w-1,i}$ are transcendental functions of weight $w-1$ and the $R_i$ are algebraic functions. We define the symbol of $F_w$ by the recursion
\beq\label{eq:SF_rec}
\cS(F_w) = \sum_{i}\cS(F_{w-1,i})\otimes R_i\,,
\eeq
and the recursion stops at $\cS(F_0) = F_0$. Let us make an obvious observation at this point: the recursive definition of the symbol only makes sense if the differential equation~\eqref{eq:MPL_d_F} does not have a homogeneous term, because otherwise the recursion does not close. Said in different words, not every function admits a symbol, and we can only define symbols for functions that satisfy a differential equation with trivial homogeneous part (such functions are often referred to as \emph{pure} functions in the physics literature).

It turns out that the total differential of MPLs indeed takes the form of eq.~\eqref{eq:MPL_d_F}. More precisely, we have~\cite{Goncharov:1998kja}
\beq\bsp\label{eq:MPL_tot_diff}
d G(a_1,\ldots,a_n;z)&\, = \sum_{i=1}^nG(a_1,\ldots,\hat a_i,\ldots,a_n;z)\,d\ln{a_{i-1}-a_i\over a_{i+1}-a_i}
%\\
%&\,= \sum_{i=1}^nG(a_1,\ldots,\hat a_i,\ldots,a_n;z)\,\left[dG(a_{i-1};a_i)-dG( a_{i+1};a_i)\right]\,,
\esp\eeq
where the hat indicates that the corresponding argument is absent, and we set $a_0=z$ and $a_{n+1}=0$. Combining eq.~\eqref{eq:SF_rec} and~\eqref{eq:MPL_tot_diff} we can easily see that the symbol of an MPL satisfies the recursion
\beq\bsp\label{eq:recursive_MPL_symbol}
\cS(G(a_1,\ldots,a_n;z))&\, = \sum_{i=1}^n\cS(G(a_1,\ldots,\hat a_i,\ldots,a_n;z))\otimes{a_{i-1}-a_i\over a_{i+1}-a_i}\,,
\esp\eeq
and the recursion stops at $\cS(G(;z)) = \cS(1)=1$. The symbol map has various well-known algebraic properties. In particular, it is linear and it maps any product of MPLs to the shuffle product of their symbols,
\beq\label{eq:symbol_shuffle}
\cS(a\cdot b) = \cS(a)\sha\cS(b)\,.
\eeq
Finally, the recursive definition shows that the symbol map sends constants to zero, because the differential of any constant vanishes.

The symbol is in fact closely connected to the coaction $\Delta_{\textrm{MPL}}$. Indeed, we can apply the coaction to either of the two factors in the tensor product in eq.~\eqref{eq:Delta_MPL}, and coassociativity implies that different ways of doing so lead to the same result. We can iterate this construction, and again coassociativity ensures that the result is unique. The iteration stops once we have decomposed an MPL of weight $w$ into a $w$-fold tensor product of logarithms. From eq.~\eqref{eq:der_coproduct} it is easy to see that this tensor satisfies the same recursion as the symbol given in eq. \eqref{eq:SF_rec}, and so we can identify the symbol of an MPL with the maximal iteration of its coaction. This in turn implies through eqs.~\eqref{eq:disc_coproduct} and~\eqref{eq:der_coproduct} that the symbol commutes nicely with the operations of taking discontinuities or derivatives. More precisely, if 
\beq
\cS(F) = \sum_{I=(i_1,\ldots,i_w)}c_I\,R_{i_1}\otimes\ldots\otimes R_{i_w}\,,
\eeq
then discontinuities correspond to clipping off entries from the left,
\beq
\cS(\textrm{Disc}\,F) = \sum_{I=(i_1,\ldots,i_w)}c_I\,(\textrm{Disc}\log R_{i_1})R_{i_2}\otimes\ldots\otimes R_{i_w}\,,
\eeq
while derivatives clip off entries from the right,
\beq
\cS(\partial_zF) = \sum_{I=(i_1,\ldots,i_w)}c_I\,R_{i_1}\otimes\ldots\otimes R_{i_{w-1}}\,(\partial_z\log R_{i_{w}})\,.
\eeq

The symbol map $\cS$ assigns to every MPL a tensor product of rational/algebraic functions. It is not injective, and not every tensor is the symbol of a transcendental function. Indeed, since $d^2=0$, we have
\beq
0=d^2F_w = \sum_{i,j}F_{w-2,ij}\,d\log R_i\wedge d\log R_j\,.
\eeq
This in turn imposes a constraint on the symbol through eq.~\eqref{eq:SF_rec}. More generally, one can show that a tensor
\beq
\sum_{I=(i_1,\ldots,i_w)}c_I\,R_{i_1}\otimes\ldots\otimes R_{i_w}
\eeq
is the symbol of a transcendental function if and only if it satisfies the following \emph{integrability condition} for each $1\le k<w$,
\beq\label{eq:MPL_integrability}
\sum_{I=(i_1,\ldots,i_w)}c_I\,(R_{i_1}\otimes\ldots\otimes R_{i_{k-1}}\otimes R_{i_{k+2}}\otimes\ldots\otimes R_{i_w})\,d\log R_{i_k}\wedge d\log R_{i_{k+1}} = 0\,.
\eeq

%%%%%%%%%%%%%%%%%%%%%%%%%%%%%%%%%%%%%%%%%%%%%%%%%%
\subsection{Relations among MPLs}
\label{sec:functional_eqs}
In this section, we review how to use the symbol and the coaction on MPLs to study, and derive, relations among MPLs. We only discuss an example which highlights the main points. For a pedagogical introduction and further examples we refer to refs.~\cite{Duhr:2012fh,Duhr:2014woa}.

Consider the function
\beq
f(z)=G(1,0,1,0;1-z)\,,
\eeq
our goal is to find a relation that allows us to write $f$ as a linear combination of MPLs of the form $G(\vec a;z)$. We start by computing the symbol of $f$,
\beq\label{eq:MPL_example_1}
\cS(f(z)) = (1-z)\otimes z\otimes(1-z) \otimes z = \cS(G(0,1,0,1;z))\,.
\eeq
While the function $G(0,1,0,1;z)$ has the same symbol as $f$, it would be premature to conclude that the two functions are identical, because they may differ by elements that lie in the kernel of $\cS$. Some of these elements can, however, be detected by the coaction $\Delta_{\textrm{MPL}}$. It will be useful to introduce the following notation for different components of the coaction on an MPL,
\beq
\Delta_{\textrm{MPL}}(G(\vec a;z)) = \sum_{p_1+p_2 = |\vec a|}\Delta_{p_1,p_2}(G(\vec a;z))\,,
\eeq
where $\Delta_{p_1,p_2}$ contains all terms in the coaction that have objects of weight $p_1$ and $p_2$ in the first and second factors respectively.
More generally, for any partition $p_1+\ldots+p_k=|\vec a|$, we denote by $\Delta_{p_1,\ldots,p_k}$ the terms in the $(k-1)$-th iteration of the coaction that have an object of weight $p_i$ in the $i$-th entry of the tensor. The maximal iteration $\Delta_{1,\ldots,1}$ of the coaction can be identified with the symbol. 

Using this notation, we compute $\Delta_{2,1,1}(f(z)-G(0,1,0,1;z))$. Since the argument of $\Delta_{2,1,1}$ has vanishing symbol by eq.~\eqref{eq:MPL_example_1}, the first entry must lie in the kernel of the symbol map. We find
\beq
\Delta_{2,1,1}(f(z)-G(0,1,0,1;z)) = \zeta_2\otimes G(0,1;z) = \Delta_{2,1,1}(\zeta_2\,G(0,1;z))\,.
\eeq
By the same argument, we see that $f(z)-G(0,1,0,1;z)-\zeta_2\,G(0,1;z)$ lies in the kernel of $\Delta_{2,1,1}$, but it does not vanish when acting with $\Delta_{3,1}$,
\beq
\Delta_{3,1}(f(z)-G(0,1,0,1;z)-\zeta_2\,G(0,1;z)) = 2\zeta_3\otimes G(0;z) = \Delta_{3,1}(2\zeta_3\,G(0;z))\,.
\eeq
Finally, the combination $f(z)-G(0,1,0,1;z)-\zeta_2\,G(0,1;z)-2\zeta_3\,G(0;z)$ lies in the kernel of $\Delta_{3,1}$,
and it must therefore be constant. Indeed, we have
\beq
f(z) = G(0,1,0,1;z)+\zeta_2\,G(0,1;z)+2\zeta_3\,G(0;z) + \frac{7}{4}\zeta_4\,,
\eeq
which is the desired relation.

Let us conclude this section with an observation. We see that in the previous example we did not use the full information contained in $\Delta_{\textrm{MPL}}$, but only the symbol $\cS=\Delta_{1,1,1,1}$ and the components $\Delta_{2,1,1}$ and $\Delta_{3,1}$ were actively used. More generally, in applications one rarely works with all the components $\Delta_{p_1,\ldots,p_k}$ for all possible partitions $(p_1,\ldots,p_k)$, but one usually only needs the subset of components $\Delta_{p,1,\ldots,1}$. The knowledge of these components is equivalent to replacing the second entries in eq.~\eqref{eq:Delta_MPL} by their symbols,
\beq\bsp\label{eq:Delta_MPL_decompose}
(\textrm{id}\otimes \cS)\Delta_{\textrm{MPL}}(G(\vec a;z)) &\,= \sum_{p=0}^{|\vec a|}\Delta_{p,1,\ldots,1}(G(\vec a;z)) = \sum_{\vec b\subseteq \vec a} G(\vec b;z)\otimes \cS(G_{\vec b}(\vec a;z))\,.
\esp\eeq
This particular variant of the coaction, with the second entries replaced by their symbols, will play an important role in the remainder of this paper.

% !TEX root = elliptic_symbols.tex

\section{Elliptic multiple polylogarithms and their symbols}
\label{sec:empls}

The goal of this paper is to extend some of the algebraic properties of ordinary MPLs to their elliptic analogues. 
In this section we give a lightning review of elliptic polylogarithms (eMPLs), and we present a formula for their total differential
that is very similar to the corresponding formula for ordinary MPLs in eq.~\eqref{eq:MPL_tot_diff}. We then use 
this formula to define symbols of eMPLs, through a recursive procedure very similar to eq.~\eqref{eq:recursive_MPL_symbol}.

\subsection{Elliptic multiple polylogarithms}

The aim of this section is to give a short review of how to define the incarnation of elliptic multiple polylogarithms considered in this paper. The functions that we define are closely related to the multiple elliptic polylogarithms studied in the mathematics~\cite{BrownLevin,MatthesThesis} and string theory literature~\cite{Broedel:2014vla,Broedel:2015hia,Broedel:2017jdo}. However, as we will see below, we prefer to work with functions that are manifestly holomorphic and do not depend on the complex conjugated variables. This choice is motivated (among other things) by the fact that Feynman integrals give rise to holomorphic quantities, and we prefer not to introduce any explicit dependence on antiholomorphic variables.

We start by defining elliptic curves. We only introduce the bare minimum of mathematical background on elliptic curves to understand the definition of elliptic polylogarithms, and we refer to the literature for a detailed discussion, e.g., ref.~\cite{silverman} (see also ref.~\cite{Broedel:2017kkb}). 
Loosely speaking, an elliptic curve can be defined as the zero set of a polynomial equation of the form $y^2=P(x)$, where $P$ is a polynomial of degree three or four (with distinct roots). Every elliptic curve defines a (compact) Riemann surface of genus one, and so it is equivalent to a complex one-dimensional torus $\mathbb{C}/\Lambda$, where $\Lambda = \mathbb{Z}\omega_1+ \mathbb{Z}\omega_2$ is a lattice and $\omega_i$ are complex numbers that are linearly independent over $\mathbb{R}$, called the \emph{periods} of the elliptic curve. We can perform a rescaling and assume without loss of generality that $\omega_1=1$. In other words, every elliptic curve is isomorphic to an elliptic curve of the form $\mathbb{C}/\Lambda_{\tau}$, with $\Lambda_{\tau} = \mathbb{Z}+ \mathbb{Z}\tau$, with $\textrm{Im}\,\tau>0$. Unless stated otherwise, we will always assume that the periods are $(\omega_1,\omega_2) = (1,\tau)$. Different values of $\tau$ may still correspond to the same elliptic curve. More precisely, $\tau$ and $\tau'$ define the same elliptic curve if and only if they are related by an $SL(2,\mathbb{Z})$ transformation, where $SL(2,\mathbb{Z})$  acts on the upper half-plane via M\"obius transformations, called \emph{modular transformations},
\beq\label{eq:modular_trafo}
\tau\mapsto \gamma\cdot \tau = \frac{a\tau+b}{c\tau+d}\,,\quad \gamma= \left(\begin{smallmatrix}
a& b\\ c&d \end{smallmatrix}\right)\in SL(2,\mathbb{Z})\,. 
\eeq

We now define a class of iterated integrals with at most logarithmic singularities. Inspired by refs.~\cite{MatthesThesis,Broedel:2014vla,BrownLevin}, we define \emph{elliptic multiple polylogarithms} (eMPLs) as
\beq\label{eq:gamt_def}
\gamtt{n_1 &\ldots& n_k}{z_1 & \ldots & z_k}{z}{\tau} = \int_0^zdz'\,g^{(n_1)}(z'-z_1,\tau)\,\gamtt{n_2 &\ldots& n_k}{z_2 & \ldots & z_k}{z'}{\tau}\,,
\eeq
where $z_i$ are complex numbers and $n_i\in \mathbb{N}$ are positive integers. 
The integers $k$ and $\sum_in_i$ are called the \emph{length} and the \emph{weight} of the eMPL. Just like ordinary MPLs, eMPLs form a shuffle algebra, 
\beq
\widetilde{\Gamma}(A_1 \cdots A_k;z,\tau)\,\widetilde{\Gamma}(A_{k+1}\cdots A_{k+l};z,\tau) = \sum_{\sigma\in \Sigma(k,l)}\widetilde{\Gamma}(A_{\sigma(1)} \cdots A_{\sigma(k+l)};z,\tau)\,,
\eeq
where we have introduced the notation $A_i=\left(\begin{smallmatrix}n_i\\ z_i\end{smallmatrix}\right)$. The shuffle product preserves both the weight and the length of eMPLs.

In the case where $(n_k,z_k) = (1,0)$, the integral in eq.~\eqref{eq:gamt_def} is divergent and requires regularization. Here we follow closely ref.~\cite{Broedel:2014vla} for the choice of the regularization scheme. We can always use the shuffle algebra to write every eMPL as a linear combination of products where the only divergent quantities are
\beq
\widetilde{\Gamma}\big(\!\underbrace{\begin{smallmatrix}1 &\ldots& 1\\0 & \ldots & 0\end{smallmatrix}}_{n\textrm{ times}};z,\tau\big)=\frac{1}{n!}\gamtt{1}{0}{z}{\tau}^n\,,
\eeq
with
%\beq \label{eq:Gamma_reg}
%\gamtt{1}{0}{z}{\tau} = \log z + \int_0^zdz'\,\left(g^{(1)}(z',\tau)-\frac{1}{z'}\right)\,.
%\eeq
\beq \label{eq:Gamma_reg}
\gamtt{1}{0}{z}{\tau} = \log\left(1-e^{2\pi i z}\right) - 2\pi i z + \int_0^zdz'\,\left(g^{(1)}(z',\tau)-\frac{2\pi i}{e^{2\pi i z'}-1}\right)\,.
\eeq
The reason for this rather complicated-looking formula lies in the fact that we want eMPLs to have a nice $q$-expansion (see Appendix~\ref{app:q_exp}), while at the same time preserving the shuffle algebra structure.

The integration kernels in eq.~\eqref{eq:gamt_def} are defined through a generating series known as the \emph{Eisenstein-Kronecker series},
\beq\label{eq:Eisenstein-Kronecker}
F(z,\alpha,\tau) = \frac{1}{\alpha}\,\sum_{n\ge0}g^{(n)}(z,\tau)\,\alpha^n = \frac{\theta'_1(0,\tau)\,\theta_1(z+\alpha,\tau)}{\theta_1(z,\tau)\,\theta_1(\alpha,\tau)}\,,
\eeq
where $\theta_1$ is the odd Jacobi theta function, and $\theta'_1$ is its derivative with respect to its first argument. The functions $g^{(n)}$ appearing in eq.~\eqref{eq:Eisenstein-Kronecker} have various nice properties, cf. refs.~\cite{MatthesThesis,Broedel:2014vla,BrownLevin}. In particular, they have definite parity
\beq
g^{(n)}(-z,\tau) = (-1)^n\,g^{(n)}(z,\tau)\,,
\eeq
and satisfy the Fay identity~\cite{BrownLevin},
\beq\bsp\label{eq:Fay}
g^{(m)}(z_1,\tau)\,g^{(n)}(z_2,\tau)&\, = -(-1)^n\,g^{(m+n)}(z_1-z_2,\tau)\\
&\, + \sum_{r=0}^n\binom{m+r-1}{m-1}\,g^{(n-r)}(z_2-z_1,\tau)\,g^{(m+r)}(z_1,\tau)\\
&\,+ \sum_{r=0}^m\binom{n+r-1}{n-1}\,g^{(m-r)}(z_1-z_2,\tau)\,g^{(n+r)}(z_2,\tau)\,.
\esp\eeq
Seen as a function of $z$, the function $g^{(1)}(z,\tau)$ has a simple pole with unit residue at every point of the lattice $\Lambda_{\tau}$. For $n>1$, $g^{(n)}(z,\tau)$ has a simple pole only at those lattice points that do not lie on the real axis. As a consequence, the iterated integrals in eq.~\eqref{eq:gamt_def} have at most logarithmic singularities, and therefore define generalizations of multiple polylogarithms to elliptic curves.

At this point we have to make a comment about the integration kernels $g^{(n)}$, and the ensuing iterated integrals $\widetilde{\Gamma}$. Every well-defined function on the torus $\mathbb{C}/\Lambda_{\tau}$ must be invariant under translations by elements of the lattice $\Lambda_{\tau}$. It turns out that the integration kernels defined by the Eisenstein-Kronecker series in eq.~\eqref{eq:Eisenstein-Kronecker} are not invariant under translations by $\tau$. For example, we have
\beq\label{eq:g1_periodicity}
g^{(1)}(z+1,\tau) = g^{(1)}(z,\tau) \textrm{~~and~~} g^{(1)}(z+\tau,\tau) = g^{(1)}(z,\tau)-{2\pi i}\,.
\eeq
More generally, at the level of the generating function, we have
\beq
F(z+1,\alpha,\tau) = F(z,\alpha,\tau) \textrm{~~and~~} F(z+\tau,\alpha,\tau) = e^{-2\pi i\alpha}\,F(z,\alpha,\tau)\,.
\eeq
Hence, strictly speaking, the $g^{(n)}$ are not well-defined functions on the torus.
One can show that it is \emph{not} possible to find a set of independent integration kernels that are at the same time holomorphic, periodic and have at most simple poles at the lattice points. Instead, one has to give up either holomorphicity or periodicity in order to define generalizations of MPLs to elliptic curves. While in our applications the holomorphic, but non-periodic, version of eMPLs defined in eq.~\eqref{eq:gamt_def} is most appropriate, we mention that in the mathematics and string theory literature it is customary to consider iterated integrals defined through integration kernels that are periodic, at the expense of giving up holomorphicity~\cite{MatthesThesis,Broedel:2014vla,BrownLevin},
\beq\label{eq:periodic_eMPLs}
\gam{n_1 &\ldots& n_k}{z_1 & \ldots & z_k}{z,\tau} = \int_0^zdz'\,f^{(n_1)}(z'-z_1,\tau)\,\gam{n_2 &\ldots& n_k}{z_2 & \ldots & z_k}{z',\tau}\,,
\eeq
where the functions $f^{(n)}$ are defined by the generating series
\beq\label{eq:Omega_def}
\Omega(z,\alpha,\tau) = \frac{1}{\alpha}\sum_{n\ge 0}f^{(n)}(z,\tau)\,\alpha^n = \exp\left[2\pi i\,\alpha\,\frac{\textrm{Im}\,z}{\textrm{Im} \,\tau}\right]\,F(z,\alpha,\tau)\,.
\eeq
The functions $f^{(n)}(z,\tau)$ are periodic with respect to translations in both the real and $\tau$ directions, but they depend explicitly on the antiholomorphic variable $\bar{z}$ through the exponential factor in the right-hand side of eq.~\eqref{eq:Omega_def}.

\subsection{The total differential and the symbol of eMPLs}

In this section we propose a generalization of the notion of symbols from ordinary MPLs to eMPLs. Our starting point is the recursive definition of the symbol in eqs.~\eqref{eq:MPL_d_F} and~\eqref{eq:SF_rec}. In order to apply the recursion, we need a closed form for the total differential of eMPLs, similar to eq.~\eqref{eq:MPL_tot_diff}. We also need to modify the recursive definition slightly, because in the case of eMPLs not all the basic integration kernels have logarithmic divergences, but some of them are regular everywhere. In the next subsection we give the general formula for the total differential and the resulting symbol map, and we discuss some of its basic properties.

If we introduce the following shorthand for the arguments of eMPLs
\beq\bsp 
A_i^{[r]}\,\equiv\, \left( \begin{smallmatrix}
	n_i+r \\ z_i
\end{smallmatrix}\right) {\rm~~and~~} A_i^{[0]}\,\equiv\,A_i\,,
\esp\eeq
the total differential of an eMPL takes the form
%\beq\bsp
%%\label{eq:gamma_differential}
%d\gamtt{n_k&\ldots& n_1}{z_k&\ldots& z_1}{z}{\tau} &\,= \sum_{p=1}^{k-1}(-1)^{n_{p+1}}\,\gamtt{n_k&\ldots&0&\ldots& n_1}{z_k&\ldots&0&\ldots& z_1}{z}{\tau}\,\omega_{p,p+1}^{(n_p+n_{p+1})}\\
%&\,+\sum_{p=1}^{k}\sum_{r=0}^{n_p+1}\Bigg[\binom{n_{p-1}+r-1}{n_{p-1}-1}\,\gamtt{n_k&\ldots&n_{p-1}+r&\ldots& n_1}{z_k&\ldots&z_{p-1}&\ldots& z_1}{z}{\tau}\,\omega_{p,p-1}^{(n_p-r)}\\
%&\,\phantom{\sum_{p=1}^{k}\sum_{r=0}^{n_p+1}\Big[}
%-\binom{n_{p+1}+r-1}{n_{p+1}-1}\,\gamtt{n_k&\ldots&n_{p+1}+r&\ldots& n_1}{z_k&\ldots&z_{p+1}&\ldots& z_1}{z}{\tau}\,\omega_{p,p+1}^{(n_p-r)}\Bigg]\,.
%\esp\eeq
\beq\bsp\label{eq:gamma_differential}
d\widetilde{\Gamma}&\left(A_1\cdots A_k;z,\tau \right) = \sum_{p=1}^{k-1}(-1)^{n_{p+1}}\,\widetilde{\Gamma}\left(A_1\cdots A_{p-1}\; ^0 _0 \; A_{p+2}\cdots A_k;z,\tau \right)\,\omega_{p,p+1}^{(n_p+n_{p+1})}\\
&\,+\sum_{p=1}^{k}\sum_{r=0}^{n_p+1}\Bigg[\binom{n_{p-1}+r-1}{n_{p-1}-1}\,\widetilde{\Gamma}\left(A_1\cdots A_{p-1}^{[r]} \; \hat{A}_{p} \; A_{p+1}\cdots A_k;z,\tau \right)\,\omega_{p,p-1}^{(n_p-r)}\\
&\,\phantom{\sum_{p=1}^{k}\sum_{r=0}^{n_p+1}\Big[}
-\binom{n_{p+1}+r-1}{n_{p+1}-1}\,\widetilde{\Gamma}\left(A_1\cdots A_{p-1} \; \hat{A}_{p} \; A_{p+1}^{[r]}\cdots A_k;z,\tau \right)\,\omega_{p,p+1}^{(n_p-r)}\Bigg]\,,
\esp\eeq
where similarly to the case of MPLs, the hat indicates that the corresponding argument is absent.
In the previous equation, we let $(z_0,z_{k+1})=(z,0)$ and $(n_0,n_{k+1})=(0,0)$, and we use the convention that the binomial number $\binom{-1}{-1}$ is $1$. The differential one-forms in this formula are, for $n\ge 0$,
\beq\bsp\label{eq:empl_letter}
\omega_{ij}^{(n)} &\,= d\gamtt{n}{z_i}{z_j}{\tau}-(-1)^n\,d\gamtt{n}{0}{z_i}{\tau} - \frac{n\,d\tau}{2\pi i}\,G_{n+1}(\tau) \\
&\,= (dz_j-dz_i)\,g^{(n)}(z_j-z_i,\tau) + \frac{n\,d\tau}{2\pi i}\,g^{(n+1)}(z_j-z_i,\tau)\,,
\esp\eeq
where $G_{2m+1}(\tau)=0$ and $G_{2m}(\tau)$ are the Eisenstein series\footnote{We assume the standard regularization for the case $m=1$.}
\beq\label{eq:Eisenstein_series}
G_{2m}(\tau) =\sum_{\substack{(\alpha,\beta)\in \mathbb{Z}^2\\ (\alpha,\beta)\neq(0,0)}}\frac{1}{(\alpha+\beta\tau)^{2m}}\,.
\eeq
For $n=-1$, we define
\beq\label{eq:empl_letter_m1}
\omega_{ij}^{(-1)} = -\frac{d\tau}{2\pi i}\,.
\eeq
The proof of eq.~\eqref{eq:gamma_differential} is given in Appendix~\ref{app:proof}. The structure of the total differential is very similar to the total differential for ordinary MPLs in eq.~\eqref{eq:MPL_tot_diff}. There are two main differences between eq.~\eqref{eq:MPL_tot_diff} and~\eqref{eq:gamma_differential}. First, the terms involving $\gamtt{\ldots & 0& \ldots}{\ldots& 0&\ldots}{z}{\tau}$ in the first line are absent in the case of ordinary MPLs, because there are no non-trivial abelian differentials of the first kind on curves of genus zero. Second, the terms proportional to non-trivial binomial coefficients are absent in eq.~\eqref{eq:MPL_tot_diff}. These arise from the application of the Fay identity in eq.~\eqref{eq:Fay}, which generalizes partial fractioning to curves of genus one. A very similar formula for the differential of (twisted) elliptic multi-zeta values was obtained in refs.~\cite{EnriquezEMZV,Broedel:2015hia,Broedel:2017jdo}. Note that the right-hand side of eq.~\eqref{eq:gamma_differential} only involves eMPLs of length $k-1$. In other words, the action of the differential strictly lowers the length of an eMPL, i.e., the number of integrations. The weight is not preserved, and the right-hand side involves eMPLs of different weights. Finally, we stress that we can only prove this formula for the holomorphic, non-periodic version of eMPLs in eq.~\eqref{eq:periodic_eMPLs}. The proof relies on the relations~\cite{BrownLevin},
\beq\bsp\label{eq:proof_elements}
\partial_{z_i}g^{(n)}(z'-z_i,\tau) &\,= -\partial_{z'}g^{(n)}(z'-z_i,\tau)\,,\\
 2\pi i\,\partial_{\tau}g^{(n)}(z'-z_i,\tau) &\,= n\,\partial_{z'}g^{(n+1)}(z'-z_i,\tau)\,,
\esp\eeq
which can be used to turn all partial derivatives into derivatives in the integration variable $z'$. The derivatives with respect to $z'$ can be integrated away using integration by parts. This last step fails, at least naively, in the case of non-holomorphic functions, preventing us from extending the proof to the non-holomorphic eMPLs of eq.~\eqref{eq:periodic_eMPLs}.

Since the length of an eMPL is strictly lowered by the differential, we see that eMPLs satisfy a differential equation without homogeneous part.
We can then immediately write down a recursive formula for the symbol of eMPLs in the same spirit as in eq.~\eqref{eq:SF_rec},
%\beq\bsp\label{eq:empl_symbol_rec}
%\cS\Big(&\gamtt{n_k&\ldots& n_1}{z_k&\ldots& z_1}{z}{\tau}\Big) = \sum_{p=1}^{k-1}(-1)^{n_{p+1}}\,\cS\Big(\gamtt{n_k&\ldots&0&\ldots& n_1}{z_k&\ldots&0&\ldots& z_1}{z}{\tau}\Big)\otimes\omega_{p,p+1}^{(n_p+n_{p+1})}\\
%&\,+\sum_{p=1}^{k}\sum_{r=0}^{n_p+1}\Bigg[\binom{n_{p-1}+r-1}{n_{p-1}-1}\,\cS\Big(\gamtt{n_k&\ldots&n_{p-1}+r&\ldots& n_1}{z_k&\ldots&z_{p-1}&\ldots& z_1}{z}{\tau}\Big)\otimes\omega_{p,p-1}^{(n_p-r)}\\
%&\,\phantom{\sum_{p=1}^{k}\sum_{r=0}^{n_p+1}\Big[}
%-\binom{n_{p+1}+r-1}{n_{p+1}-1}\,\cS\Big(\gamtt{n_k&\ldots&n_{p+1}+r&\ldots& n_1}{z_k&\ldots&z_{p+1}&\ldots& z_1}{z}{\tau}\Big)\otimes\omega_{p,p+1}^{(n_p-r)}\Bigg]\,.
%\esp\eeq
\begin{align}\nonumber
&\cS\Big(\widetilde{\Gamma}\left(A_1\cdots A_k;z,\tau \right)\!\!\Big) =\! \sum_{p=1}^{k-1}(-1)^{n_{p+1}}\,\pi_k\!\left[\cS\Big(\widetilde{\Gamma}\left(A_1\cdots A_{p-1}\; ^0 _0 \; A_{p+2}\cdots A_k;z,\tau \right)\!\!\Big)\otimes\omega_{p,p+1}^{(n_p+n_{p+1})}\right]\\
\label{eq:empl_symbol_rec}&\,+\sum_{p=1}^{k}\sum_{r=0}^{n_p+1}\pi_k\Bigg[\binom{n_{p-1}+r-1}{n_{p-1}-1}\,\cS\Big(\widetilde{\Gamma}\left(A_1\cdots A_{p-1}^{[r]} \; \hat{A}_{p} \; A_{p+1}\cdots A_k;z,\tau \right)\!\!\Big)\otimes\omega_{p,p-1}^{(n_p-r)}\\
\nonumber&\,\phantom{\sum_{p=1}^{k}\sum_{r=0}^{n_p+1}\Big[}
-\binom{n_{p+1}+r-1}{n_{p+1}-1}\,\cS\Big(\widetilde{\Gamma}\left(A_1\cdots A_{p-1} \; \hat{A}_{p} \; A_{p+1}^{[r]}\cdots A_k;z,\tau \right)\!\!\Big)\otimes\omega_{p,p+1}^{(n_p-r)}\Bigg]\,,
\end{align}
where $\pi_k$ denotes the projection onto tensors of length $k$. The role of this projection will be clarified below.
Note that we have mildly generalized eq.~\eqref{eq:SF_rec}, and we interpret the entries in the symbol as one-forms rather than rational or algebraic functions.
We emphasize that all the usual computational rules for symbols remain valid. In particular, the symbol map $\cS$ is linear and maps a product of eMPLs and/or MPLs to the shuffle product of their symbols, cf. eq.~\eqref{eq:symbol_shuffle}. Finally, let us comment on the role of the projection $\pi_k$. In the case of generic arguments, this projection can be ignored, because all tensors will always have length $k$. In the case of non-generic arguments, however, (which is the situation most commonly encountered in applications), it can happen that the length of the eMPLs in the right-hand side is strictly less than $k$, because some eMPLs of weight zero evaluate to rational numbers, e.g., $\gamtt{0}{0}{1}{\tau} = 1$. If we want our definition of the symbol of an eMPL to be consistent with specializations of the arguments, we need to restrict ourselves to the part of the tensor that has maximal length. Note that in the case of ordinary MPLs this issue never shows up, because for ordinary MPLs the length is always equal to the weight.

From eq.~\eqref{eq:empl_symbol_rec} we can also read off the symbol alphabet associated with eMPLs: it is precisely the set of differential one-forms $\omega_{ij}^{(n)}$ in eqs.~\eqref{eq:empl_letter} and~\eqref{eq:empl_letter_m1}. Conversely, if we start from a generic tensor made out of the letters $\omega_{ij}^{(n)}$, then this tensor is the symbol of a function if and only if the tensor satisfies an integrability condition very similar to the case of ordinary MPLs in eq.~\eqref{eq:MPL_integrability}, with the product of differential forms $d\log R_i\wedge d\log R_j$ replaced by $\omega_{ij}^{(n)}\wedge \omega_{kl}^{(m)}$.

At this point, let us make a technical comment on the integrability of the symbol. The symbol of an eMPL is always by construction an integrable tensor, and the integrals in eq.~\eqref{eq:gamt_def} are homotopy-invariant iterated integrals over the kernels $g^{(n)}(z,\tau)$ (see Appendix~\ref{app:iterated_integrals} for a review of homotopy-invariant iterated integrals).
This should be contrasted with the non-holomorphic, doubly-periodic iterated integrals over the $f^{(n)}(z,\tau)$ kernels defined in eq.~\eqref{eq:periodic_eMPLs}. The $g^{(n)}(z,\tau)$ and  $f^{(n)}(z,\tau)$ are related to each other, and the $g^{(n)}(z,\tau)$ can be obtained from the $f^{(n)}(z,\tau)$ by formally setting $\textrm{Im}(z)$ to zero.
However, if one takes an integrable symbol in the letters $g^{(n)}(z,\tau)$ and simply replaces all occurrences of $g^{(n)}(z,\tau)$ by $f^{(n)}(z,\tau)$, then the resulting symbol will in general \emph{not} be integrable. To restore the integrability, an additional non-holomorphic letter needs to be included~\cite{BrownLevin}.
This is no contradiction: Our eMPLs are not doubly-periodic, and so they do not define genuine iterated integrals on the torus but only on its universal cover. Non-holomorphic iterated integrals, in contrast, are doubly-periodic and therefore define iterated integrals on the torus. The two classes of functions are similar and can be related to one another, but they are \emph{not} equivalent.

Equation~\eqref{eq:empl_symbol_rec} allows us to define symbols for elliptic multi-zeta values (eMZVs). Elliptic MZVs are defined as~\cite{EnriquezEMZV} (see also refs.~\cite{Broedel:2014vla,MatthesThesis}),
\beq
\omega(n_k,\ldots,n_1;\tau) \equiv \gamtt{n_1&\ldots& n_k}{0&\ldots& 0}{1}{\tau}\,.
\eeq
At this point we have to make a comment. In principle, eMZVs are defined in terms of the non-holomorphic, periodic version of eMPLs in eq.~\eqref{eq:periodic_eMPLs}. However, in the context of the so-called $A$-type eMZVs, the integration contour is the real segment $[0,1]$. On this integration contour the holomorphic and non-holomorphic versions of eMPLs agree, and we can define $A$-type eMZVs in terms of the holomorphic, non-periodic eMPLs. If we specialize eq.~\eqref{eq:gamma_differential} to eMZVs, we find,
%\begin{align}
%\nonumber
%2\pi i&\,d\omega(n_k,\ldots, n_1;\tau) = d\tau\sum_{p=1}^{k-1}(-1)^{n_{p+1}}\,(n_p+n_{p+1})\,\omega(n_1\ldots 0\ldots n_k;\tau)\,G_{n_p+n_{p+1}+1}(\tau)\\
%\label{eq:eMZV_differential}&\,+d\tau\sum_{p=1}^{k}\sum_{r=0}^{n_p+1}\Bigg[\binom{n_{p-1}+r-1}{n_{p-1}-1}\,(n_p-r)\,\omega(n_1\ldots n_{p-1}+r\ldots n_k;\tau)\,G_{n_p-r+1}(\tau)\\
%\nonumber&\,\phantom{d\tau\sum_{p=1}^{k}\sum_{r=0}^{n_p+1}\Big[}
%-\binom{n_{p+1}+r-1}{n_{p+1}-1}\,(n_p-r)\,\omega(n_1\ldots n_{p+1}+r\ldots n_k;\tau)\,G_{n_p-r+1}(\tau)\Bigg]\,.
%\end{align}
%%
\begin{align}
\nonumber
2\pi i&\,d\omega(n_k,\ldots, n_1;\tau) = \\
\nonumber
&-d\tau\sum_{p=1}^{k-1}(-1)^{n_{p+1}}\,(n_p+n_{p+1})\,\omega(n_k,\ldots,n_{p+2},0,n_{p-1},\ldots n_1;\tau)\,G_{n_p+n_{p+1}+1}(\tau)\\
\nonumber
&\,-d\tau\sum_{p=1}^{k}\sum_{r=0}^{n_p+1}\Bigg[\binom{n_{p-1}+r-1}{n_{p-1}-1}\,(n_p-r)\,\omega(n_k,\ldots,n_{p+1},\hat{n}_p,n_{p-1}^{[r]},\ldots n_1;\tau)\,G_{n_p-r+1}(\tau)\\
\label{eq:eMZV_differential}&\,\phantom{d\tau\sum_{p=1}^{k}\sum_{r=0}^{n_p+1}\Big[}
-\binom{n_{p+1}+r-1}{n_{p+1}-1}\,(n_p-r)\,\omega(n_k,\ldots,n_{p+1}^{[r]},\hat{n}_p,n_{p-1},\ldots n_1;\tau)\,G_{n_p-r+1}(\tau)\Bigg]\,.
\end{align}
where the hat denotes omission of the corresponding argument and
$n_{i}^{[r]}=n_i+r$. This formula agrees with the one given in
refs.~\cite{EnriquezEMZV,Broedel:2015hia}. Hence, we obtain the following
recursive formula for the symbol,
%%
%\begin{align}
%\nonumber
%&\cS(\omega(n_k,\ldots, n_1;\tau)) = \sum_{p=1}^{k-1}(-1)^{n_{p+1}}\,(n_p+n_{p+1})\,\pi_k\!\left[\cS(\omega(n_1\ldots 0\ldots n_k;\tau))\otimes\gamma_{n_p+n_{p+1}+1}\right]\\
%&\,+d\tau\sum_{p=1}^{k}\sum_{r=0}^{n_p+1}\pi_k\!\Bigg[\binom{n_{p-1}+r-1}{n_{p-1}-1}\,(n_p-r)\,\cS(\omega(n_1\ldots n_{p-1}+r\ldots n_k;\tau))\otimes \gamma_{n_p-r+1}\\
%\nonumber&\,\phantom{d\tau\sum_{p=1}^{k}\sum_{r=0}^{n_p+1}\Big[}
%-\binom{n_{p+1}+r-1}{n_{p+1}-1}\,(n_p-r)\,\cS(\omega(n_1\ldots n_{p+1}+r\ldots n_k;\tau))\otimes\gamma_{n_p-r+1}\Bigg]\,,
%\end{align}
%%
\begin{align}
\nonumber
&\cS(\omega(n_k,\ldots, n_1;\tau)) = \\
\nonumber &-\sum_{p=1}^{k-1}(-1)^{n_{p+1}}\,(n_p+n_{p+1})\,\pi_k\!\left[\cS(\omega(n_k,\ldots,n_{p+2},0,n_{p-1},\ldots n_1;\tau))\otimes\gamma_{n_p+n_{p+1}+1}\right]\\
\nonumber&\,-d\tau\sum_{p=1}^{k}\sum_{r=0}^{n_p+1}\pi_k\!\Bigg[\binom{n_{p-1}+r-1}{n_{p-1}-1}\,(n_p-r)\,\cS(\omega(n_k,\ldots,n_{p+1},\hat{n}_p,n_{p-1}^{[r]},\ldots n_1;\tau))\otimes \gamma_{n_p-r+1}\\
&\,\phantom{d\tau\sum_{p=1}^{k}\sum_{r=0}^{n_p+1}\Big[}
-\binom{n_{p+1}+r-1}{n_{p+1}-1}\,(n_p-r)\,\cS(\omega(n_k,\ldots,n_{p+1}^{[r]},\hat{n}_p,n_{p-1},\ldots n_1;\tau))\otimes\gamma_{n_p-r+1}\Bigg]\,,
\end{align}
where $\gamma_n$ denotes the one-form
\beq
\gamma_n=\frac{d\tau}{2\pi i}\,G_n(\tau)\,.
\eeq
We see that the letters of the symbol alphabet of eMZVs are Eisenstein series. 
The symbol of eMZVs defined above is in fact closely related, though not completely equivalent, to the decomposition map $\psi^A$ of ref.~\cite{Matthes:Eisenstein2}. $\psi^A$ associates to each $A$-type eMZV a non-commutative word in letters ${\bf e}^\vee_n$, one for each Eisenstein series. More precisely, if we identify the word ${\bf e}^\vee_{n_{i_m}}\ldots {\bf e}^\vee_{n_{i_1}}$ with the tensor $\gamma_{n_{i_m}}\otimes\ldots\otimes\gamma_{n_{i_1}}$, then  the symbol of an $A$-type eMZV corresponds to the maximal length part of the image under the map $\psi^A$ (possibly up to reversal  of words),
\beq
\cS(\omega(n_k,\ldots, n_1;\tau)) = \pi_k(\psi^A(\omega(n_k,\ldots, n_1;\tau)))\,.
\eeq

Just like in the case of ordinary MPLs, the symbol of eMPLs has some obvious drawbacks, because it looses a lot of information about the function. In the remainder of this paper, we define (a variant of) a coaction on MPLs, which allows us to recover some of the missing information.

% !TEX root = elliptic_symbols.tex

\section{A general construction of a coaction}
\label{sec:de-Rham-symbols}

In the previous section we have extended the symbol map from ordinary to elliptic MPLs. While the symbol is extremely useful in practice, 
it has the drawback that it maps to zero all constants, and a lot of information is lost when passing from a function to its symbol.
In the case of MPLs, it can be useful to work with the coaction rather than the symbol, because the coaction allows one to 
recover some of the lost information (see Section~\ref{sec:functional_eqs}). The purpose of this section is to review a general construction due to Brown~\cite{Brown:coaction} of a coaction on a larger class of integrals than just MPLs -- in particular on (certain classes of) \emph{periods}, i.e., integrals of rational (or algebraic) functions over a domain specified by rational (or algebraic) inequalities. We do not aim at mathematical rigor in this section, and we content ourselves to introduce the main points needed to understand the construction of this coaction.
In a nutshell, we will construct a coaction on some classes of periods where the first factor is itself a period, while the second factor is a symbol of some sort. 
At the end of Section~\ref{sec:functional_eqs} we argued that such a coaction is sufficient for all known applications in physics, and we therefore obtain a generalization of the techniques of refs.~\cite{Duhr:2012fh,Duhr:2014woa} beyond ordinary MPLs. We first review the general construction of ref.~\cite{Brown:coaction} in this section, before we apply it to eMPLs in the next section. 

\subsection{Periods and the motivic coaction}
\label{eq:motivic_coaction}
Our starting point for constructing a coaction on eMPLs is the \emph{motivic coaction} of ref.~\cite{Brown:coaction}. Very loosely speaking, the motivic coaction provides a general framework to define a coaction on arbitrary periods, and it contains the coaction on MPLs in eq.~\eqref{eq:Delta_MPL} as a special case.
The formula for the {motivic coaction} $\Delta^{\mathfrak{m}}$ can (schematically) be written as\footnote{We have to make a comment about our notations, which differ from ref.~\cite{Brown:coaction}. In ref.~\cite{Brown:coaction} motivic ($\mathfrak{m}$) periods are triplets $[M,\gamma,\omega]^{\mathfrak{m}}$ rather than pairs, where $M$ keeps track of the underlying `geometric space', while de Rham ($\mathfrak{dr}$) periods are triplets $[M,\omega_i^\vee,\omega]^{\mathfrak{dr}}$, where $\omega_i^\vee$ is a dual differential form. The dual of a differential form is defined in the following way: if $\{\omega_i\}$ is a basis for  the vector space of differential forms, then the dual basis $\{\omega_i^\vee\}$ is defined by $\omega_i^\vee(\omega_j)=\delta_{ij}$. The relationship between differential forms and their duals is the same as the one between kets and bras. In order to keep the notation as light as possible we denote the motivic period $[M,\gamma,\omega]^{\mathfrak{m}}$ simply by $[\gamma,\omega]$, and the de Rham period $[M,\omega_i^\vee,\omega]^{\mathfrak{dr}}$ by $[\omega_i,\omega]$.}
\beq\label{eq:Delta_m_def}
\Delta^{\mathfrak{m}}([\gamma,\omega]) = \sum_i\,[\gamma,\omega_i]\otimes [\omega_i,\omega]\,.
\eeq

Strictly speaking, the motivic coaction is not directly defined on integrals, but rather on their `motivic versions'. For our purposes, it is sufficient to think of a \emph{motivic period} as a pair $[\gamma,\omega]$, where $\gamma$ is an integration contour and $\omega$ is a differential form that can be integrated over $\gamma$. The motivic period $[\gamma,\omega]$ can thus be thought of a `motivic avatar' of the integral $\int_{\gamma}\omega$. Motivic periods satisfy the same basic relations as integrals, namely they are linear in both the contour and the integrand, and two motivic periods are considered identical if they are related by a change of variables. We can multiply integrals in an obvious way, and so we can equip the vector space of all motivic periods with an algebra structure. There is a natural algebra homomorphism, the \emph{period map}, which associates to a motivic period the corresponding integral,
\beq
\textrm{per}:\, [\gamma,\omega] \mapsto \int_{\gamma}{\omega}\,.
\eeq
A folklore conjecture states that the period map is injective, i.e., no information is lost when passing from its motivic avatar to the actual integral. For this reason we do not distinguish the motivic period $[\gamma,\omega]$ and the integral $\int_{\gamma}\omega$, and we only talk about the integral. In physics parlance, the sum in eq.~\eqref{eq:Delta_m_def} runs over a basis $[\gamma,\omega_i]$ of master integrals associated to the integral family to which $[\gamma,\omega]$ belongs.

The object in the second factor in the tensor product on the right-hand side of eq.~\eqref{eq:Delta_m_def} is more mysterious, as it does not directly correspond to an integral, unlike what is suggested by the polylogarithmic case in eq.~\eqref{eq:Delta_MPL}. A \emph{de Rham period} is a pair $[\omega_i,\omega]$ of differential forms $\omega_i$ and  $\omega$. A hand-wavy argument why the second factor in the coaction cannot be interpreted as an integral in a straightforward way goes as follows. The integrals we want to consider usually define multivalued functions, and the analytic continuation and the discontinuities of the integrals are tightly connected to non-trivial deformations of the integration contour $\gamma$. Since discontinuities only act on the first factor of the coaction~\cite{Brown:coaction} (cf. eq.~\eqref{eq:disc_coproduct}), the second factor should be invariant under deformations of the contour. Hence, it cannot take the form of an ordinary integral, but it can at best be an `integral defined up to branch cuts'.\footnote{Alternatively, we could represent de Rham periods as single-valued objects~\cite{Brown:coaction}. We do not explore this possibility here.} In the case of MPLs, discontinuities are always proportional to powers of $i\pi$, and so in the case of MPLs we can interpret de Rham periods as `MPLs defined modulo $i\pi$'. This hand-wavy argument can be made mathematically rigorous~\cite{Brown:coaction}, resulting in the form of the coaction on MPLs given in eq.~\eqref{eq:Delta_MPL}. 

In more general applications, and in particular in the case of integrals of elliptic type, not all discontinuities are proportional to powers of $i\pi$, e.g., 
\beq
\textrm{Disc}_{\lambda}\textrm{K}(\lambda) =  \textrm{K}(\lambda+i\varepsilon)-\textrm{K}(\lambda-i\varepsilon) = \theta(\lambda-1)\,\frac{2}{\sqrt{\lambda}}\,\textrm{K}(1-1/\lambda)\,,
\eeq
where $\theta$ denotes the Heaviside step function, and $\textrm{K}(\lambda)$ is the complete elliptic integral of the first kind,
\beq
\textrm{K}(\lambda) = \int_0^1\frac{dt}{\sqrt{(1-t^2)(1-\lambda t^2)}}\,.
\eeq
Hence, if we go beyond the case of MPLs, we cannot identify de Rham periods, i.e., pairs of differential forms, with `integrals defined modulo $i\pi$'.

From the previous discussion, it is not immediately clear how to make eq.~\eqref{eq:Delta_m_def} explicit in the case of eMPLs because there is no immediate way to work with de Rham periods as integrals defined modulo $i\pi$, as we did for ordinary MPLs. In the following we review a construction of ref.~\cite{Brown:coaction}, where one associates symbols to de Rham periods. 
Since de Rham periods are \emph{pairs} of differential forms, at least naively this notion of symbols cannot be defined immediately through the recursive definition in terms of the total differential of a transcendental function, which is defined via a \emph{single} differential form integrated over a contour. In the next subsection we review how to define symbols for pairs of differential forms.

 %%%%%%%%%%%%%%%%%%%%%%%%%%%%%%%%%%%
 
 \subsection{Unipotent differential equations and symbols}
 
In this section we review the construction of ref.~\cite{Brown:coaction} of symbols for (certain classes of) de Rham periods, i.e., pairs of differential forms. 
We have already pointed out in Section~\ref{sec:MPL_symbol} that not every function admits a symbol, but we can only define the symbol of a function that satisfies a first-order differential equation without a homogeneous part.
However, in applications one often encounters integrals that satisfy a differential equation with a non-trivial homogeneous part. For example, it is known that Feynman integrals can lead to expressions that involve complete elliptic integrals, cf.~refs.~\cite{Remiddi:2016gno,Bonciani:2016qxi}, and the latter satisfy a set of coupled homogeneous first order differential equations,
 \beq\bsp\label{eq:KE_diff_eq}
 \partial_{\lambda}\textrm{K}(\lambda) &\,= -\frac{1}{2\lambda}\,\textrm{K}(\lambda)+\frac{1}{2\lambda(1-\lambda)}\,\textrm{E}(\lambda)\,,\\
 \partial_{\lambda}\textrm{E}(\lambda) &\,= -\frac{1}{2\lambda}\,\textrm{K}(\lambda)+\frac{1}{2\lambda}\,\textrm{E}(\lambda)\,,
 \esp\eeq
 where $\textrm{E}(\lambda)$ is the complete elliptic integral of the second kind,
 \beq
 \textrm{E}(\lambda) = \int_0^1 dt\,\frac{(1-\lambda t^2)}{\sqrt{(1-t^2)(1-\lambda t^2)}}\,.
 \eeq
 As these functions do not satisfy differential equations without homogeneous parts, they do not have the right properties to admit a symbol.
In the remainder of this section we therefore restrict our attention to a subclass of integrals that satisfy a system of first order differential equations with a trivial homogeneous part. The corresponding class of functions is closely related, though not identical, to the notion of `pure functions' encountered in the physics literature~\cite{ArkaniHamed:2010gh,Henn:2013pwa}. We will comment more precisely on this connection at the end of this section.
 
Consider the vector of integrals
\beq\label{eq:I_vec_def}
I = \left(\int_\gamma\xi_1,\ldots, \int_{\gamma}\xi_n\right)^T\,.
\eeq 
We assume that the components of $I$ form a complete and independent set of integrals with respect to integration-by-parts identities. In other words, we assume that $I$ is a vector of master integrals. We also assume that $I$ satisfies a linear differential equation of first order of the form
\beq\label{eq:diff_eq}
dI = A\,I\,,\quad A = \sum_iA_i\,\omega_i\,,
\eeq
where the $A_i$ are $n\times n$ matrices with rational numbers as entries and the $\omega_i$ are one-forms. Without loss of generality, we may assume that the $\omega_i$ are all independent. The differential equation has a non-trivial solution only if $dA=A\wedge A$, and we assume from now on that this condition is satisfied. 

So far we have made no assumption on whether or not the components of $I$ satisfy a differential equation a with trivial homogeneous part. We say that the integrals in eq.~\eqref{eq:I_vec_def} and the differential equation~\eqref{eq:diff_eq} are \emph{unipotent} if the matrix $A$ is nilpotent. In that case we can find a change of basis such that in the new basis the matrix $A$ is strictly upper triangular, and from now on we assume that we are working in that basis. All the differential equations are then decoupled, and the homogeneous part for each component of $I$ is trivial. 

We now show how we can define a notion of symbols on pairs of differential forms $[\xi_i,\xi_j]$, where the differential forms $\xi_i$ are those that appear in eq.~\eqref{eq:I_vec_def}.
Consider the following matrix, 
\beq\label{eq:TA_series}
T_A = 1+ [A]^R + [A|A]^R + [A|A|A]^R+\ldots\,.
\eeq
This matrix is independent of the choice of the contour $\gamma$ used to define the basis in eq.~\eqref{eq:I_vec_def}.
The matrix multiplication corresponds to ordinary matrix multiplication, combined with the concatenation of words formed out of the one-forms $\omega_i$. In the following we denote words of one-forms by $[\omega_{i_1}|\ldots|\omega_{i_k}]$. 
The superscript $R$ denotes the operation that reverses words\footnote{In ref.~\cite{Brown:coaction} the map $R$ is absent from the definition of the matrix. This can be traced back to a difference in conventions between mathematics and physics.},
\beq 
[\omega_{i_1}|\ldots|\omega_{i_k}]^R = [\omega_{i_k}|\ldots|\omega_{i_1}]\,.
\eeq
 For example, consider the nilpotent matrix
\beq\label{eq:example_A_matrix}
A=\left(\begin{array}{ccc}
0&\omega_0&0\\
0&0&\omega_1\\
0&0&0
\end{array}\right)\,,
\eeq
where $\omega_0$ and $\omega_1$ are closed one-forms. Then we have
\beq
[A|A]^R=\left(\begin{array}{ccc}
0&0&[\omega_1|\omega_0]\\
0&0&0\\
0&0&0
\end{array}\right) \,,
\eeq
and
$[A|A|A] = 0$, so that
\beq
T_A=\left(\begin{array}{ccc}
1&[\omega_0]&[\omega_1|\omega_0]\\
0&1&[\omega_1]\\
0&0&1
\end{array}\right)\,.
\eeq
In general, the series in eq.~\eqref{eq:TA_series} always terminates if $A$ is nilpotent, in which case the matrix $T_A$ is unipotent (a square matrix is unipotent if it can be written as a sum of the identity matrix and a nilpotent matrix).
We define the \emph{symbol of the pair of periods} $[\xi_i,\xi_j]$ as the matrix elements of the matrix $T_A$~\cite{Brown:coaction},
\beq\label{eq:de_Rham symbol}
\symb([\xi_i,\xi_j]) \equiv \langle\xi_i|T_A|\xi_j\rangle =(T_A)_{ji}\,.
\eeq

The matrix $T_A$ has a very simple interpretation: It is an `avatar' for the fundamental solution matrix (i.e., the Wronskian) of the unipotent differential equation $dI=AI$. More precisely, let us choose a base point $x_0$ and a path $\gamma$ (not necessarily identical to the path in eq.~\eqref{eq:I_vec_def}) with endpoints $x_0$ and $x$. Then the solution $I(x)$ with initial condition $I(x=x_0)=I_0$ can be written in the form
\beq
I(x) = \cW\,I_0\,,
\eeq
where $\cW\equiv \int_{\gamma}T_A$ is a matrix of iterated integrals. If $t\in[0,1]$ is a local coordinate parametrizing the path $\gamma$ and we write $\omega_i=dt\,f_i(t)$, then 
\beq
\int_\gamma[\omega_{i_1}|\ldots|\omega_{i_n}] \equiv \int_{0\le t_1\le\ldots\le t_n\le 1}dt_1\,f_{i_1}(t_1)\ldots dt_n\,f_{i_n}(t_n)\,.
\eeq
We have collected some background material on iterated integrals in Appendix~\ref{app:iterated_integrals}. 
In order for $I(x)$ to be a well-defined function of $x$, it must be independent of the details of the path $\gamma$ and it can only depend on its endpoint $x$. Said differently, the iterated integrals must be homotopy-invariant. A general criterion due to Chen for an iterated integral to be homotopy-invariant -- the \emph{integrability condition} -- is reviewed in Appendix~\ref{app:iterated_integrals}. Here it suffices to say that in our case this criterion is always satisfied, provided that $dA=A\wedge A$. In other words, the entries in the matrix $T_A$ are always \emph{integrable words of one-forms}.

%In addition, all entries of $T_A$ are integrable words. Indeed, we have
%\beq\bsp
%DT_A&\, = D_1T_A+D_2T_A \\
%&\,= -R[dA] -R[dA|A]-R[A|dA]+R[A\wedge A] +R[A\wedge A|A] + R[A|A\wedge A] +\ldots\\
%&\, = R[A\wedge A-dA] +R[A\wedge A-dA|A] + R[A|A\wedge A-dA] + \ldots\\
%&\,=0\,.
%\esp\eeq

Let us illustrate this definition on the example in eq.~\eqref{eq:example_A_matrix}. This choice of $A$ corresponds to the basis
\beq\label{eq:example_I_basis}
I=(I_1,I_2,I_3)^T = \left(\int_{\gamma}\xi_1,\int_{\gamma}\xi_2,\int_{\gamma}\xi_3\right)^T=\left(\int_{\gamma}[\omega_1|\omega_0],\int_{\gamma}\omega_1,1\right)^T\,,
\eeq
for some contour $\gamma$ and where we used that fact the the integral over the empty word is equal to unity (by definition). It is easy to check that $I$ satisfies the differential equation $dI=AI$. We then find for example
\beq\label{eq:example_symbol}
\symb([\xi_3,\xi_1]) = \langle\xi_3|T_A|\xi_1\rangle = [\omega_1|\omega_0]\,.
\eeq

Let us discuss some of the properties of the symbol map $\symb$. First, $\symb$ satisfies all the well-known properties of a symbol. In particular, since $T_A$ is a matrix of integrable words, symbols are always integrable. Moreover, the symbol map $\symb$ is linear and maps a product of two pairs of differential forms to the shuffles of their symbols,
\beq
\symb([\xi_i,\xi_j]\cdot [\xi_k,\xi_l]) = \symb([\xi_i,\xi_j])\sha \symb([\xi_k,\xi_l])\,.
\eeq
%Moreover, differential operators act on symbols by clipping off entries from the right,
%\beq\label{eq:dr_symbol_der}
%\symb([\xi_i,\partial_z\xi_j]) = \sum_kA_{jk}\,\symb([\xi_i,\xi_k])\,,
%\eeq
%while the operation of taking discontinuities clips off symbol entries from the left. 
In practice, it is not necessary to construct the entire matrix $T_A$ explicitly. Indeed, if $i=j$, we have $\symb([\xi_i,\xi_j])=1$.  For $i\neq j$, the symbol of a pair of 
differential forms can be computed recursively from the knowledge of the matrix $A$,
\beq\label{eq:dr_rec}
\symb([\xi_i,\xi_j]) = \sum_k\left[\symb([\xi_i,\xi_k])\Big| A_{jk}\right]\,.
\eeq
Indeed, we have, for $i\neq j$ (repeated indices are summed),
\beq\bsp
\symb([\xi_i,\xi_j]) &\,= [A_{ji}] + [A_{jk}|A_{ki}]^R + [A_{jl}|A_{lk}|A_{ki}]^R+\ldots\\
&\,= [A_{ji}] + [A_{ki}|A_{jk}] + [A_{li}|A_{kl}|A_{jk}]+\ldots\\
&\,= \left[\delta_{ki} + [A_{ki}] + [A_{li}|A_{kl}]+\ldots\big|A_{jk}\right]\\
&\,=\left[\symb([\xi_i,\xi_k])\big| A_{jk}\right]\,.
\esp\eeq
The form of this recursion is very reminiscent of the recursion satisfied by the symbol of a transcendental function in eq.~\eqref{eq:SF_rec}.

From the definition in eq.~\eqref{eq:de_Rham symbol}, it looks like the symbol is tightly connected to a specific choice for the basis of master integrals, because the right-hand side of eq.~\eqref{eq:de_Rham symbol} depends explicitly on the the matrix $A$, which itself depends on the choice of basis in eq.~\eqref{eq:example_I_basis}. If we change basis according to $I' = MI$, for some invertible matrix $M$, then the differential equation, and thus the matrix $T_A$, changes. Hence, it is a priori not clear that symbols computed in two different bases will agree. In ref.~\cite{Brown:coaction} it was shown that the definition in eq.~\eqref{eq:de_Rham symbol} is independent of the choice of the basis provided that certain relations among words are taken into account. In the case that the forms $\omega_i$ are closed (which is the situation most often encountered in applications), these relations read,
\begin{align}\label{eq:symbol_ibp}
\nonumber[\omega_1|\dots|\omega_k|df|\omega_{k+1}|\ldots|\omega_n] &\,= [\omega_1|\dots|\omega_k|f\,\omega_{k+1}|\ldots|\omega_n]-[\omega_1|\dots|\omega_k\,f|\omega_{k+1}|\ldots|\omega_n]\,,\\
[df|\omega_{1}|\ldots|\omega_n] &\,= [f\,\omega_1|\dots|\omega_n]-f\,[\omega_1|\dots|\omega_n]\,,\\
\nonumber[\omega_{1}|\ldots|\omega_n|df] &\,= f\,[\omega_1|\dots|\omega_n]-[\omega_1|\dots|\omega_n\,f]\,.
\end{align}
Equation~\eqref{eq:symbol_ibp} implements integration by parts at the level of the symbol. We can illustrate this on our example with the matrix $A$ in eq.~\eqref{eq:example_A_matrix} and the choice of basis of master integrals $I$ in eq.~\eqref{eq:example_I_basis} (see also Example 9.5 in ref.~\cite{Brown:coaction}). Let us choose a different basis of master integrals $I'$ by
\beq
I' = MI\,\quad\textrm{with } M=\left(\begin{array}{ccc}
1&f&0\\
0&1&0\\
0&0&1
\end{array}\right) \,,
\eeq
where $f$ is some algebraic function. It is easy to check that
\beq
dI'=A'I'\,,\quad\textrm{with }A'=\left(\begin{array}{ccc}
0&\omega_0+df&f\,\omega_1\\
0&0&\omega_1\\
0&0&0
\end{array}\right)\,.
\eeq
We see that $I'$ still satisfies a unipotent differential equation with a matrix $A'$ that is strictly upper triangular.
The matrix $T_{A'}$ is given by
\beq
T_{A'}=\left(\begin{array}{ccc}
1&[\omega_0+df]&[\omega_1|\omega_0+df]+[f\omega_1]\\
0&1&[\omega_1]\\
0&0&1
\end{array}\right)\,.
\eeq
We then obtain
\beq\bsp
\symb([\xi_3,\xi_1])&\,  =  \langle\xi_3|T_{A'}|\xi_1\rangle\\
&\,=\langle\xi'_3|T_{A'}|\xi'_1\rangle -  f \langle\xi'_3|T_{A'}|\xi'_2\rangle\\
&\,=[\omega_1|\omega_0+df] + [f\,\omega_1] -f\,[\omega_1]\\
&\,=\langle\xi_3|T_{A}|\xi_1\rangle + [\omega_1|df] + [f\,\omega_1] -f\,[\omega_1]\,.
\esp\eeq
Comparing the previous equation to eq.~\eqref{eq:example_symbol}, we see that the two expressions agree once eq.~\eqref{eq:symbol_ibp} is imposed,
\beq
[\omega_1|df] =  f\,[\omega_1] -[f\,\omega_1]\,.
\eeq

Let us conclude this section with a comment on the relationship between unipotent and pure functions. If we choose $\omega_a=d\log(a-z)$ and $\gamma$ the straight line from $0$ to $z$, then $I=(G(0,1;z),G(1;z),1)^T$. After changing the basis with $f(z)=1/z$, say, we find $I' =(G(0,1;z)+G(1;z)/z,G(1;z),1)$. 
The vector $I$ is what is usually called a `pure function' in the physics literature. The vector $I'$ is not pure because it involves MPLs multiplied by rational functions. The previous analysis shows that both $I$ and $I'$ are unipotent, because they both satisfy a differential equation with a strictly upper triangular matrix. The definition of symbols of ref.~\cite{Brown:coaction} (and reviewed here) assigns a matrix of symbols not just to pure objects, but more generally to unipotent objects. This may have implications when thinking about canonical differential equations for Feynman integrals~\cite{Henn:2013pwa}, because unipotency seems to be a more general concept than purity.

%%%%%%%%%%%%%%%%%%%%%%%%%%%%%%%%%

\section{Elliptic symbol calculus}
\label{sec:elliptic_symbol_calculus}

\subsection{Unipotent and semi-simple periods}

In the previous section we have reviewed how to assign a symbol to pairs of differential forms. We can use this definition and combine it with the motivic coaction in eq.~\eqref{eq:Delta_m_def} to give a meaning to the pairs of differential forms that appear in the second factor in eq.~\eqref{eq:Delta_m_def}. More precisely, following ref.~\cite{Brown:coaction}, we define a map $\Delta$ by composing the motivic coaction $\Delta^{\mathfrak{m}}$ with the symbol map,
\beq\label{eq:Delta_def}
\Delta \equiv (\textrm{id}\otimes \symb)\,\Delta^{\mathfrak{m}}\,.
\eeq
One can check that $\Delta$ satisfies all the axioms of a coaction (see Appendix~\ref{app:algebras}).
At this point we should make a comment about some differences between the motivic coaction in eq.~\eqref{eq:Delta_m_def} and the coaction in eq.~\eqref{eq:Delta_def}. The motivic coaction $\Delta^{\mathfrak{m}}$ is defined for arbitrary (motivic) periods. The symbol map $\symb$, which appears in the definition of $\Delta$ in eq.~\eqref{eq:Delta_def}, 
is only defined for unipotent quantities. As a consequence, unlike $\Delta^{\mathfrak{m}}$, $\Delta$ is only defined for unipotent quantities. MPLs are unipotent (see eq.~\eqref{eq:MPL_tot_diff}), and so we can apply this construction to ordinary MPLs, and recover in this way (a version of) the coaction on MPLs from Section~\ref{sec:mpls}. Details can be found in Appendix~\ref{sec:Brown_MPLs}.
%\footnote{Strictly speaking, the coaction used here does not correspond to $\Delta^{\mathfrak{m}}$ of Section~\ref{eq:motivic_coaction}, but rather to the coaction $\Delta^{\mathfrak{u}}$ associated to the unipotent part of the de Rham group, cf. ref.~\cite{Brown:coaction}, which leaves semi-simple period invariant. We do not dwell on this distinction here.}. 

In general, one can show that every (motivic) period $x$ can be decomposed into a linear combination of products of a unipotent period $u_i$ and a \emph{semi-simple} period $s_i$~\cite{Brown:coaction},
\beq\label{eq:ss_unipotent}
x = \sum_{i}s_i\,u_i\,,
\eeq
We define $\Delta$ in such a way that it acts trivially on semi-simple periods,
\beq
\Delta(s_i) = s_i\otimes 1\,.
\eeq
From eq.~\eqref{eq:Delta_ipi} we see that $i\pi$ is semi-simple. 
Other semi-simple periods that will appear in our context are related to the periods $\omega_i$ and the quasi-periods $\eta_i$ of the elliptic curve. In the following we assume without loss of generality that $\textrm{Im }\omega_2/\omega_1>0$. The periods and quasi-periods are not independent, but they satisfy the Legendre relation,
\beq
\eta_2\,\omega_1-\eta_1\omega_2 = -i\pi\,.
\eeq
If we combine the periods and quasi-periods into a matrix,
\beq
P=\left(\begin{array}{cc}
\omega_1&\omega_2\\
\eta_1&\eta_2
\end{array}\right)\,,
\eeq
then we can decompose this matrix into a product of a semi-simple\footnote{For our purposes, it is sufficient to think of a semi-simple matrix as a diagonalizable matrix. Note that the only diagonalizable unipotent matrix is the identity.} and a unipotent matrix, 
\beq\label{eq:period_splitting}
P=SU\,,
\eeq
with
\beq
S=\left(\begin{array}{cc}
\omega_1&0\\
\eta_1&-i\pi/\omega_1
\end{array}\right)
\textrm{~~and~~}
U=\left(\begin{array}{cc}
1&\tau\\
0&1
\end{array}\right)\,,
\eeq
and $\tau=\omega_2/\omega_1$. 
The matrix $U$ satisfies a unipotent differential equation, 
\beq
dU = AU\,,\textrm{~~with~~} A=\left(\begin{array}{cc}
0&d\tau\\
0&0
\end{array}\right)\,,
\eeq
and so $\tau$ itself is unipotent. Hence, the coaction $\Delta$ acts non-trivially on $\tau$,
\beq
\Delta(\tau) = \tau\otimes 1+ 1\otimes [d\tau]\,.
\eeq
The elements of $S$, instead, are semi-simple, and so $\Delta$ acts trivially on them. In particular, we have,
\beq
\Delta(\omega_1) = \omega_1\otimes 1 \textrm{~~and~~} \Delta(\eta_1) = \eta_1\otimes 1\,.
\eeq
It is easy to see that the previous relations imply
\beq
\Delta(\omega_2) = \omega_2\otimes 1 +\omega_1\otimes [d\tau] \textrm{~~and~~} \Delta(\eta_2) = \eta_2\otimes 1+\eta_1\otimes [d\tau]\,.
\eeq

Let us make a comment. The separation of the periods and quasi-periods into semi-simple and unipotent quantities in eq.~\eqref{eq:period_splitting} depends on a choice. For example, we could have swapped the roles of $\omega_1$ and $\omega_2$, or chosen any other linear combination of $(\omega_1,\omega_2)$ as a basis for the periods. While other choices would be equally valid, in many applications there is a natural choice: we often have to deal with elliptic curves defined over the real numbers, in which case we can choose a basis of periods such that $\omega_1$ is real and positive, and $\omega_2$ is purely imaginary with positive imaginary part (at least in some region of the space of external parameters). In the following we always assume that we work with this basis.

The coaction $\Delta$ has all the properties known from the coaction on MPLs. In particular, it preserves the multiplication,
\beq
\Delta(x_1\,x_2) = \Delta(x_1)\,\Delta(x_2)\,.
\eeq
The coaction also interacts with the operations of taking derivatives and discontinuities in ways very similar to eqs.~\eqref{eq:disc_coproduct} and~\eqref{eq:der_coproduct}. In particular, discontinuities only act in the first factor of the coaction,
\beq
\Delta\left(\textrm{Disc}\,x\right) = (\textrm{Disc}\otimes\textrm{id})\Delta(x)\,.
\eeq
The corresponding formula for derivatives looks slightly different from eq.~\eqref{eq:der_coproduct}. If $x$ is decomposed into semi-simple and unipotent periods according to eq.~\eqref{eq:ss_unipotent}, then we have
\beq\label{eq:der_coaction}
\Delta(\partial_zx) = \sum_i\left[\Delta(u_i)\,\partial_zs_i + s_i\,(\textrm{id}\otimes\partial_z)\Delta(u_i)\right]\,.
\eeq
We can easily recover eq.~\eqref{eq:der_coproduct} from eq.~\eqref{eq:der_coaction}, because for MPLs the only semi-simple quantities of non-zero weight are powers of $i\pi$, for which the first term in eq.~\eqref{eq:der_coaction} vanishes. In the case of eMPLs there are semi-simple objects with non-trivial functional dependence and discontinuities, namely the (real) period $\omega_1$ and quasi-period $\eta_1$.

Finally, let us make a comment about the meaning of the coaction $\Delta$. The recursive property of the symbol for pairs of differential forms in eq.~\eqref{eq:dr_rec} implies that $\Delta$ encodes all the iterated differentials of a given (unipotent) period. In particular, the coaction $\Delta$ can be computed in a recursive manner similar to the symbol. More precisely, if $I = (\int_\gamma\xi_a)_{a\in S}$, with $S$ a set of labels of the words of one-forms, is a vector of unipotent integrals satisfying the differential equation $dI=A\,I$, with $A$ strictly upper triangular, then we have
\beq
\Delta(I_a) = I_a\otimes 1 + \sum_{b\neq a}\left[\Delta(I_b)\big| A_{ab}\right]\,.
\eeq
Indeed, inserting the recursion for the symbol in eq.~\eqref{eq:dr_rec} into the formula for the coaction \eqref{eq:Delta_def}, we find,
\beq\bsp
\Delta(I_a) &\,= \sum_{c}I_c\otimes \symb([\xi_c,\xi_a]) \\
&\,= I_a\otimes \symb([\xi_a,\xi_a]) + \sum_{c\neq a}I_c\otimes \symb([\xi_c,\xi_a])\\
&\, = I_a\otimes 1 + \sum_{b,c\neq a}I_c\otimes \left[\symb([\xi_c,\xi_b])\Big|A_{ab}\right]\\
&\,=I_a\otimes 1 + \sum_{b\neq a}\left[\Delta(I_b)\big| A_{ab}\right]\,.
\esp\eeq
In this sense the computation of the coaction is similar to the computation of the symbol. However, the information contained in $\Delta$ is less coarse than the information in the symbol because the coaction retains additional information on constants that are mapped to zero by the symbol. These constants can provide valuable information when using the coaction to work out relations among functions, cf. Section~\ref{sec:functional_eqs}.

%%%%%%%%%%%%%%%%%%%%%%%%%%%

% !TEX root = elliptic_symbols.tex

\subsection{A coaction on elliptic multiple polylogarithms}
\label{sec:empls_coaction}

In this section we give the explicit form of the coaction $\Delta$ for the eMPLs of Section~\ref{sec:empls}. 
Our starting point is the formula~\eqref{eq:gamma_differential} for the total differential of eMPLs. For simplicity, we restrict the discussion to the generic case where all the arguments of the eMPLs are distinct. Since the eMPLs in the right-hand side of eq.~\eqref{eq:gamma_differential} all have length $k-1$, we see that eMPLs are unipotent.
Consider the function $\widetilde{\Gamma}(\vec A;z,\tau)$, with $\vec A=\big(A_1\begin{smallmatrix}\ldots\\ \end{smallmatrix} A_k\big) = \left(\begin{smallmatrix}n_1&\ldots& n_k\\ z_1&\ldots& z_k\end{smallmatrix}\right)$, and the vector
 \beq
 J = \left(J_{\vec B}\right)_{\vec B\in S}^T = \left(\int_0^z\Omega_{\vec B}\right)_{\vec B\in S}^T\,.
\eeq
The elements of $J$ are labeled by the set $S$, whose elements label the eMPLs obtained by iterating the differential on $\widetilde{\Gamma}(\vec A;z,\tau)$.
While the explicit form of $S$ is very easy to work out in concrete cases, its general structure is quite complicated and we refrain from giving an explicit representation for it.
The integrands $\Omega_{\vec B}$ are sequences of one-forms,
\beq
\Omega_{\vec B} = \big[dt\,g^{(n_l)}(t-z_l,\tau)\big|\ldots\big|dt\,g^{(n_1)}(t-z_1,\tau)\big]\,,\quad 
\vec B = \left(\begin{smallmatrix}n_1&\ldots& n_l\\ z_1&\ldots& z_l\end{smallmatrix}\right)\,.
\eeq
The vector $J$ satisfies a unipotent differential equation of the form $dJ=MJ$, where $M$ is a strictly upper triangular matrix\footnote{In order for the matrix $M$ to be strictly upper triangular, the elements in the vector $J$ need to be ordered by descending length of the sets $B$.} whose entries are $\mathbb{Q}$-linear combinations of the one-forms in eqs.~\eqref{eq:empl_letter} and~\eqref{eq:empl_letter_m1}.

The symbol of a pair $[\Omega_{\vec B},\Omega_{\vec A}]$ is obtained from the recursion in eq.~\eqref{eq:dr_rec},
\beq\label{eq:S_dR_dMPL}
\symb([\Omega_{\vec B},\Omega_{\vec A}]) = \sum_{\vec C\in S}\left[\symb([\Omega_{\vec B},\Omega_{\vec C}])\Big| M_{\vec A\vec C}\right]\,.
\eeq
The coaction $\Delta$ from eq.~\eqref{eq:Delta_def} then acts on eMPLs via
\beq\bsp\label{eq:Delta_u_eMPL}
\Delta(\widetilde{\Gamma}(\vec A;z,\tau)) &\,= \sum_{\vec B\in S} \widetilde{\Gamma}(\vec B;z,\tau)\otimes \symb([\Omega_{\vec B},\Omega_{\vec A}])
\,.
\esp\eeq
The previous equation is the final form of our coaction on eMPLs. 
We are currently unaware of a nice closed formula that allows us to write the symbol in the second entry as the symbol of a function,\footnote{Note that there is always a function whose symbol is $\symb([\Omega_{\vec B},\Omega_{\vec A}])$, because the symbol is integrable. Hence, while we know that such a function exists and can be constructed on a case-by-case basis, we were not able to find a closed form that works in all cases.} unlike what was the case for ordinary MPLs (cf. eq.~\eqref{eq:Delta_MPL_decompose}). 
By specializing the previous equation to $z=1$ and $z_i=0$, we obtain a coaction on eMZVs.
%We recall at this point that the periods and quasi-periods of an elliptic curve are semi-simple objects, and $\Delta$ acts trivially on them,
%\beq\label{eq:Delta_omega}
%\Delta(\omega_i) = \omega_i\otimes 1 {\rm~~and~~} \Delta(\eta_i) = \eta_i\otimes 1\,.
%\eeq
%We also find it convenient to define
%\beq
%\cS(\omega_i) = \omega_i {\rm~~and~~} \cS(\eta_i) = \eta_i\,.
%\eeq

Let us illustrate this on an example. Consider the function
\beq
f = \gamtt{2&2}{z_1&z_2}{z}{\tau}\,.
\eeq
In order to avoid lengthy technical discussions about regularization, we assume that the points $z_1$, $z_2$ and $z$ are generic.
In a first step, we compute the iterated total differential of $f$ using eq.~\eqref{eq:gamma_differential}. We can then determine $J$ to be the eleven-component vector
\beq
\bsp
\label{eq:s_set}
J = \Big(&
\gamtt{2&2}{z_1&z_2}{z}{\tau},
\gamtt{5}{z_2}{z}{\tau},
\gamtt{5}{z_1}{z}{\tau},
\gamtt{4}{z_2}{z}{\tau},
\gamtt{4}{z_1}{z}{\tau},\\
&\gamtt{3}{z_2}{z}{\tau},
\gamtt{3}{z_1}{z}{\tau},
\gamtt{2}{z_2}{z}{\tau},
\gamtt{2}{z_1}{z}{\tau},
\gamtt{0}{0}{z}{\tau},
1
\Big)^T\,,
\esp
\eeq
from which we can also read off the set $S$. 
Next, we can determine the matrix $M$, by computing the total derivative of $J$ using eq.~\eqref{eq:gamma_differential}.
The non-zero entries of the matrix appear only in the first row or last column of the matrix and can be written in terms of the one-forms $\omega^{(n)}_{ij}$ defined in eqs.~\eqref{eq:empl_letter} and~\eqref{eq:empl_letter_m1}. The non-zero entries in the first row of the $11\times 11$ matrix $M$ are
\beq\bsp
M_{1,2}= -M_{1,3} &\,= -4\,\omega^{(-1)}_{1,2}\,,\\
M_{1,4}= \phantom{-}M_{1,5} &\,= -3\,\omega^{(0)}_{1,2}\,,\\
M_{1,6}= -M_{1,7} &\,= -2\,\omega^{(1)}_{1,2}\,,\\
M_{1,8}&\,= -\omega^{(2)}_{3,1}-\omega^{(2)}_{1,2}\,, \\
M_{1,9}&\,= -\omega^{(2)}_{1,2}-\omega^{(2)}_{2,0}\,,\\
M_{1,10}&\,=\omega^{(4)}_{1,2}\,,
\esp\eeq
%\beq
%\bsp
%M_{1,2}= \omega^{(-1)}_{1,2},                       \,\,
%M_{1,3}= 4\omega^{(-1)}_{1,2},                      \,\,
%M_{1,4}= 3\omega^{(0)}_{1,2},                       \,\,
%M_{1,5}= 3\omega^{(0)}_{1,2},                       \,\,
%M_{1,6}=-2\omega^{(1)}_{1,2},                       \\
%M_{1,7}= 2\omega^{(1)}_{1,2},                       \,\,
%M_{1,8}= \omega^{(2)}_{3,1}+\omega^{(2)}_{1,2},     \,\,
%M_{1,9}= \omega^{(2)}_{1,2}+\omega^{(2)}_{2,0},     \,\,
%M_{1,10}=-\omega^{(4)}_{1,2},                       \,\,
%M_{11,2} = \omega^{(5)}_{3,2}-\omega^{(5)}_{2,0},   \\
%M_{11,3} = \omega^{(5)}_{3,1}-\omega^{(5)}_{1,0},   \,\,
%M_{11,4} = \omega^{(4)}_{3,2}+\omega^{(4)}_{2,0},   \,\,
%M_{11,5} = \omega^{(4)}_{3,1}+\omega^{(4)}_{1,0},   \,\,
%M_{11,6} = \omega^{(3)}_{3,2}-\omega^{(3)}_{2,0},   \\
%M_{11,7} = \omega^{(3)}_{3,1}-\omega^{(3)}_{1,0},   \,\,
%M_{11,8} = \omega^{(2)}_{3,2}+\omega^{(2)}_{2,0},   \,\,
%M_{11,9} = \omega^{(2)}_{3,1}+\omega^{(2)}_{1,0},   \,\,
%M_{11,10} = \omega^{(0)}_{3,0}.
%\esp
%\eeq
where we have identified $z_3=z$ and $z_0=0$. Similarly, the non-zero elements of the last column are
\beq\bsp
M_{2,11} = \phantom{-}\omega^{(5)}_{3,2}-\omega^{(5)}_{2,0}\,,  \qquad & M_{3,11} =  \phantom{-}\omega^{(5)}_{3,1}-\omega^{(5)}_{1,0}\,, \\
M_{4,11} = -\omega^{(4)}_{3,2}-\omega^{(4)}_{2,0}\,,   \qquad & M_{5,11} = -\omega^{(4)}_{3,1}-\omega^{(4)}_{1,0}\,,   \\
M_{6,11} =  \phantom{-}\omega^{(3)}_{3,2}-\omega^{(3)}_{2,0}\,,   \qquad & M_{7,11} =  \phantom{-}\omega^{(3)}_{3,1}-\omega^{(3)}_{1,0}\,,   \\
M_{8,11} = -\omega^{(2)}_{3,2}-\omega^{(2)}_{2,0}\,,   \qquad & M_{9,11} = -\omega^{(2)}_{3,1}-\omega^{(2)}_{1,0}\,,   \\
M_{10,11} =  -\omega^{(0)}_{3,0}\,.\phantom{-\omega^{(2)}_{2,0}}\qquad 
\esp\eeq
The knowledge of the matrix $M$ is equivalent to the symbols of pairs of differential forms. Hence, we can immediately write down the coaction of $f$
\beq
\bsp
\label{eq:cop_f}
\Delta&\left(\gamtt{2&2}{z_1&z_2}{z}{\tau}\right) = \gamtt{2&2}{z_1&z_2}{z}{\tau}\otimes\,1
+\sum_{k=1}^4k\,\Big((-1)^k\gamtt{k+1}{z_1}{z}{\tau}-\gamtt{k+1}{z_2}{z}{\tau}\Big)\otimes\left[\omega^{(3-k)}_{1,2}\right]\\
%+3\Big(\gamtt{4}{z_1}{z}{\tau}+\gamtt{4}{z_2}{z}{\tau}\Big)\otimes[\omega^{(0)}_{1,2}]
%+2\Big(\gamtt{3}{z_1}{z}{\tau}-\gamtt{3}{z_2}{z}{\tau}\Big)\otimes[\omega^{(1)}_{1,2}]\\
&-\gamtt{2}{z_1}{z}{\tau}\otimes\left[\omega^{(2)}_{2,0}\right]
-\gamtt{2}{z_2}{z}{\tau}\otimes\left[\omega^{(2)}_{3,1}\right]
+\gamtt{0}{0}{z}{\tau}\otimes\left[\omega^{(4)}_{1,2}\right]
+1\otimes\mathbb{S}\,.
\esp
\eeq
The quantity $\mathbb{S}$ can be identified with the symbol of $f$, and it is given by
\beq
\bsp
\label{eq:symb_f}
\mathbb{S} &= \cS\left(\gamtt{2&2}{z_1&z_2}{z}{\tau}\right)\\
&\,=\left[\omega^{(2)}_{3,1}+\omega^{(2)}_{1,0}\Big|\omega^{(2)}_{2,0}\right]
+\left[\omega^{(2)}_{3,2}+\omega^{(2)}_{2,0}\Big|\omega^{(2)}_{3,1}\right]
-\left[\omega^{(0)}_{3,0}\Big|\omega^{(4)}_{1,2}\right]+\sum_{k=1}^4k\,\left[\widetilde{\omega}^{(k+1)}\Big|\omega^{(3-k)}_{1,2}\right]\,,
\esp
\eeq
%\beq
%\bsp
%\label{eq:symb_f}
%\cS &= 4\omm{5}{}{-1}{1,2}+3\omm{4}{}{0}{1,2}+2\omm{3}{}{1}{1,2}\\
%&+[\omega^{(2)}_{3,1}+\omega^{(2)}_{3,2}+\omega^{(2)}_{1,0}+\omega^{(2)}_{2,0},\omega^{(2)}_{1,2}]
%+[\omega^{(2)}_{3,1}+\omega^{(2)}_{1,0},\omega^{(2)}_{2,0}]+[\omega^{(2)}_{3,2}+\omega^{(2)}_{2,0},\omega^{(2)}_{3,1}]\,,
%\esp
%\eeq
where we have defined the shorthand
\beq
\widetilde{\omega}^{(n)} \equiv \omega^{(n)}_{3,1}+(-1)^n\omega^{(n)}_{3,2}+(-1)^n\omega^{(n)}_{1,0}+\omega^{(n)}_{2,0}\,.
\eeq

% !TEX root = elliptic_symbols.tex

\section{Identities among eMPLs and eMZVs}
\label{sec:applications}

In this section we illustrate how one can use the coaction and the symbol map on eMPLs to obtain identities among eMPLs and eMZVs.
Our methodology will follow closely the one for ordinary MPLs of ref.~\cite{Duhr:2012fh,Duhr:2014woa} (cf. Section~\ref{sec:functional_eqs}). We illustrate it on two simple examples.
As a first application, we will derive a 30-term identity involving eMPLs of length two. As a second application, we show how we can derive identities among eMZVs, and we reproduce in particular an identity among eMZVs of length four obtained in ref.~\cite{Broedel:2014vla}.\footnote{We have checked that we are able to reproduce all identities of ref.~\cite{Broedel:2014vla}. We limit ourselves here to present only one of these cases.}

\subsection{A 30-term identity among eMPLs}
Consider the function 
\beq
f = \gamtt{2&2}{z_1&z_2}{z}{\tau}\,.
\eeq
We assume that all the arguments are generic.
In the following we show how we can use the symbol map to obtain an expression for $f$ where the variable $z_2$ only appears in the form $\gamtt{\ldots}{\ldots}{z_2}{\tau}$.
Such identities are encountered frequently in applications, e.g., if one is interested in studying the behavior of the function $f$ close to $z_2=0$.
%We start by computing the symbol of $f$.
%\beq
%S= \cS(f) = \cS\big(\gamtt{2&3}{z_1&z_2}{z}{\tau}\big)\,.
%\eeq

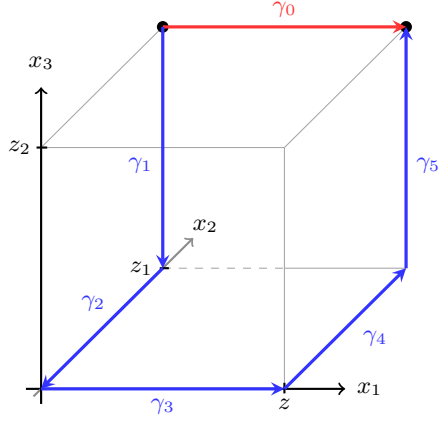
\begin{figure}[t]
    \center
    \begin{tikzpicture}[y=.4cm, x=.4cm, z=.2cm, font=\footnotesize]
        % cube
        \draw[gray!70] (0,8,0) -- (8,8,0);
        \draw[gray!70] (0,8,8) -- (8,8,8);
        \draw[gray!70] (8,0,0) -- (8,0,8);
        \draw[gray!70] (8,8,0) -- (8,0,0);
        \draw[gray!70] (8,8,0) -- (8,8,8);
        \draw[gray!70] (8,0,8) -- (8,8,8);
        \draw[gray!70] (0,8,0) -- (0,8,8);
        % cube dashed
        \draw[dashed,gray!70] (0,0,8) -- (0,4,8);
        \draw[gray!70] (0,4,8) -- (0,8,8);
        \draw[dashed,gray!70] (0,0,8) -- (4,0,8);
        \draw[gray!70] (4,0,8) -- (8,0,8);
        % axes
        \draw [->, thick]   (-0.5,0,0) -- coordinate (x1 axis mid) (10,0,0);
        \draw [->, gray!90, thick]  (0,0,-.5) -- coordinate (x2 axis mid) (0,0,10);
        \draw [->, thick]  (0,-.5,0) -- coordinate (x3 axis mid) (0,10,0);
        % axes labels
        \node at (10.8,0,0) {$x_1$};
	\node at (0,0,10.8) {$x_2$};
        \node at (0,10.8,0) {$x_3$};
        % ticks
	\draw [thick] (8,0.2,0) -- (8,-0.15,0);
        \draw [thick] (-0.1,0,8) -- (0.2,0,8);
        \draw [thick] (0.2,8,0) -- (-0.15,8,0);
	% tick labels
        \node [anchor=north] at (8,0,0) {$z$};
        \node [anchor=east] at (0,0,8) {$z_1$};
	\node [anchor=east] at (0,8,0) {$z_2$};
	% endpoints
	\fill[black] (8,8,8) circle (2.2pt);
	\fill[black] (0,8,8) circle (2.2pt);
        % gamma 0
        \draw[->,>=stealth,line width=1.2,red!80] (0,8,8) -- (8,8,8);
        % gamma 1
        \draw[->,>=stealth,line width=1.2,blue!80] (0,8,8) -- (0,0,8);
        % gamma 2
        \draw[->,>=stealth,line width=1.2,blue!80] (0,0,8) -- (0,0,0);
        % gamma 3
        \draw[->,>=stealth,line width=1.2,blue!80] (0,0,0) -- (8,0,0);
        % gamma 4
        \draw[->,>=stealth,line width=1.2,blue!80] (8,0,0) -- (8,0,8);
        % gamma 5
        \draw[->,>=stealth,line width=1.2,blue!80] (8,0,8) -- (8,8,8);
	% gamma labels
	\node [red!80,anchor=south] at (4,8,8) {$\gamma_0$};
	\node [blue!80,anchor=north east] at (0,4,8) {$\gamma_1$};
	\node [blue!80,anchor=south east] at (0.2,0,4.5) {$\gamma_2$};
	\node [blue!80,anchor=north] at (4,0,0) {$\gamma_3$};
	\node [blue!80,anchor=north west] at (8,0,4.5) {$\gamma_4$};
	\node [blue!80,anchor=north west] at (8,4,8) {$\gamma_5$};
    \end{tikzpicture}
    \caption{Choice of contours according to eq.~\eqref{eq:paths}.}
    \label{fig:paths}
\end{figure}

We have determined the coaction on $f$ and with that also the symbol of $f$ in the previous section in eqs.~\eqref{eq:cop_f} and~\eqref{eq:symb_f}.
%The symbol is rather lengthy, and we do not show it here {\bf [CD: So far my code only spits out symbols in terms of components, not in terms of $\omega_{ij}^{(n)}$. Maybe the symbol will be more compact in terms of these letters.]}.
The symbol is by construction integrable, and eMPLs are homotopy-invariant. As a consequence, we can deform the contour without changing the value of the integral. We start by interpreting $f$ as an integral in the four-dimensional space with coordinates $(x_1,\ldots, x_4)$, and the integral is over the contour $\gamma_0(t)= (t,z_1,z_2,\tau)$, $t\in [0,z]$. We can deform the contour  into a new one which consists of a product $\gamma_1\ldots\gamma_5$ of straight line segments, as shown in fig.~\ref{fig:paths}, 
\beq\bsp
\label{eq:paths}
\gamma_1(t) = (0,z_1,z_2-t,\tau) \,, \quad & t\in[0,z_2]\,,\\
\gamma_2(t) = (0,z_1-t,0,\tau) \,, \quad & t\in[0,z_1]\,,\\
\gamma_3(t) = (t,0,0,\tau) \,, \quad & t\in[0,z]\,,\\
\gamma_4(t) = (z,t,0,\tau) \,, \quad & t\in[0,z_1]\,,\\
\gamma_5(t) = (z,z_1,t,\tau) \,, \quad & t\in[0,z_2]\,.
\esp\eeq
%%%
For simplicity, we assume that the values of $(z,z_1,z_2,\tau)$ are such that
there is no singularity inside the hypercube with edges $\gamma_i$, in order to
avoid lengthy discussions about analytic continuation. Homotopy-invariance then
implies that 
\beq
\int_{\gamma_0}\cS(f) = \int_{\gamma_1\ldots\gamma_5}\cS(f) \,.
\eeq
We can use the path decomposition formula~\eqref{eq:path_composition} to reduce
the problem to the computation of integrals on the individual straight line
segments $\gamma_i$. The coordinate $x_3$ is non-zero only on $\gamma_1$ and
$\gamma_5$, where $z_2$ appears as an end-point of the integration. We
therefore obtain the desired representation for $f$,
\begin{align}
f&\,=\gamtt{2&2}{z_1&z_2}{z}{\tau} =-\gamtt{2}{0}{z_2}{\tau} \gamtt{2}{z}{z_1}{\tau}-2 \gamtt{2}{0}{z}{\tau} \gamtt{2}{z_1}{z_2}{\tau}\\
\nonumber&\,+\gamtt{2}{z_1}{z_2}{\tau} \gamtt{2}{z}{z_1}{\tau}+2 \gamtt{1}{z_1}{z_2}{\tau} \gamtt{3}{z}{z_1}{\tau}+\gamtt{0}{0}{z}{\tau} \gamtt{4}{z_1}{z_2}{\tau}\\
\nonumber&\,+3 \gamtt{0}{0}{z_2}{\tau} \gamtt{4}{z}{z_1}{\tau}+\gamtt{2}{0}{z_1}{\tau} \gamtt{2}{0}{z_2}{\tau}-\gamtt{2}{0}{z_1}{\tau} \gamtt{2}{z_1}{z_2}{\tau}\\
\nonumber&\,-2 \gamtt{3}{0}{z_1}{\tau} \gamtt{1}{z_1}{z_2}{\tau}-3 \gamtt{4}{0}{z_1}{\tau} \gamtt{0}{0}{z_2}{\tau}+2 \gamtt{2}{0}{z_1}{\tau} \gamtt{2}{0}{z}{\tau}\\
\nonumber&\,-\gamtt{2}{0}{z}{\tau} \gamtt{2}{z}{z_1}{\tau}-\gamtt{4}{0}{z_1}{\tau} \gamtt{0}{0}{z}{\tau}+6 \gamtt{0}{0}{z_1}{\tau} \gamtt{4}{0}{z}{\tau}+\gamtt{2 & 2}{0 & 0}{z}{\tau}\\
\nonumber&\,+\gamtt{2}{0}{z_2}{\tau} \gamtt{2}{0}{z}{\tau}-6 \gamtt{0}{0}{z_2}{\tau} \gamtt{4}{0}{z}{\tau}-2 \gamtt{1 & 3}{z_1 & z}{z_2}{\tau}+\gamtt{2 & 2}{z_1 & z}{z_2}{\tau}\\
\nonumber&\,+2 \gamtt{1 & 3}{z_1 & 0}{z_2}{\tau}-\gamtt{2 & 2}{z_1 & 0}{z_2}{\tau}-3 \gamtt{0 & 4}{0 & z}{z_1}{\tau}+2 \gamtt{1 & 3}{0 & z}{z_1}{\tau}-\gamtt{2 & 2}{0 & z}{z_1}{\tau}\\
\nonumber&\,+3 \gamtt{0 & 4}{0 & 0}{z_1}{\tau}-2 \gamtt{1 & 3}{0 & 0}{z_1}{\tau}+\gamtt{2 & 2}{0 & 0}{z_1}{\tau}+3 \gamtt{0 & 4}{0 & z}{z_2}{\tau}-3 \gamtt{0 & 4}{0 & 0}{z_2}{\tau}\,.
\end{align}

\subsection{Identities among eMZVs}
Our second example illustrates how to use the coaction $\Delta$ to study identities among eMZVs. 
As a warm up, consider an eMZV of length one. Its coaction is
\beq
\Delta(\omega(n;\tau)) = \omega(n;\tau)\otimes 1\,.
\eeq
Hence, we find
\beq
\Delta(\partial_\tau\omega(n;\tau)) = (\textrm{id}\otimes \partial_{\tau})\Delta(\omega(n;\tau))=0\,,
\eeq
and so $\omega(n;\tau)$ must be constant. In particular, it must be equal to its value at $\tau=i\infty$, and it is easy to show that~\cite{Broedel:2015hia} (See Appendix~\ref{app:q_exp})
\beq
\!\!\!\lim_{\tau\to i\infty}\omega(n_1,\ldots,n_k;\tau) \equiv \omega_0(n_1,\ldots,n_k) = \left\{\begin{array}{l}
\frac{1}{k!}\prod_{i=1}^k(-2\zeta_{n_i})\,, \textrm{ if all $n_i$ are even}\,,\\
0\,,\textrm{ if at least one $n_i$ is odd, all $n_i\neq 1$}\,.
\end{array}\right.
\eeq
Hence, we conclude that
\beq
\omega(n;\tau) = \omega_0(n) = \left\{\begin{array}{ll}
-2\zeta_{n}\,, &\textrm{ if $n$ even}\,,\\
0\,,&\textrm{ if $n$ odd}\,.
\end{array}\right.
\eeq
Using a very similar argument, one can show that
$\omega(n_1,n_2;\tau) = 2\zeta_{n_1}\zeta_{n_2}$, if $n_1$ and $n_2$ are both even and strictly greater than 1\,.
Both results are in agreement with ref.~\cite{Broedel:2014vla}. 

Next, let us discuss a less trivial example.
Consider the function
\beq
f(\tau) = 10\,\omega(0,0,0,5;\tau)+4\,\omega(0,0,3,2;\tau)+2\,\omega(0,2,0,3;\tau)-\omega(0,5;\tau)\,.
\eeq
Using the previous results up to length two, we can easily check that the symbol of $f$ vanishes, $\cS(f)=0$. However, it would be incorrect to conclude that $f$ is identically zero just because its symbol vanishes. Indeed, computing the full coaction, we find
\begin{align}
\Delta(f) &\,= f\otimes 1 + A\otimes\left[\frac{d\tau}{2\pi i}G_4(\tau)\right]+B\otimes\left[\frac{d\tau}{2\pi i}\right]\,,
\end{align}
with
\begin{align}
A&\,=-12\omega(0,0,2;\tau)-6\omega(0,2,0;\tau)\,,\\
\nonumber B&\,= 4\omega(0,3,3;\tau)-30\omega(0,0,6;\tau)-6\omega(0,2,4;\tau)-12\omega(0,4,2;\tau)-4\omega(3,0,3;\tau)-\frac{2\pi^6}{189}\,.
\end{align}
The structure of $\Delta(f)$ implies that both $A$ and $B$ must be constant. Indeed, if $A$ and $B$ were not constant, they would have a nontrivial image under $\Delta$, which, by coassociativity, would imply terms in $\Delta(f)$ with symbols of length two in the second factor. Since the latter are absent, we conclude that $A$ and $B$ must be constant, and equal to their value at $\tau=i\infty$. We find
\begin{align}
\Delta(f) &\,= f\otimes 1 +\pi^2\otimes\left[\frac{d\tau}{2\pi i}G_4(\tau)\right]-\frac{\pi^6}{45}\otimes\left[\frac{d\tau}{2\pi i}\right]\,.
\end{align}
From this we conclude that $f$ must have the form
\beq
f = f_0 + \pi^2 f_1\,,
\eeq
where $f_1$ is an eMZV of weight three (the total weight is five), and $f_0$ is such that $\Delta(f_0)=f_0\otimes 1$, which implies that $f_0$ must be constant. After some experimentation, we find,
\beq
f_0 = 0 \textrm{ and } f_1 = -\frac{1}{3}\omega(0,3;\tau)\,,
\eeq
in agreement with ref.~\cite{Broedel:2014vla}.

% !TEX root = elliptic_symbols.tex

\section{Symbols and iterated integrals of modular forms}
\label{sec:modular_forms}
So far we have used the construction of Section~\ref{sec:de-Rham-symbols} to extend the symbol and the coaction from ordinary to elliptic MPLs. The advantage of the construction in Section~\ref{sec:de-Rham-symbols} is that it is not restricted to polylogarithmic functions, but it applies more generally to arbitrary unipotent periods. In this section we apply the same construction to iterated integrals of modular forms~\cite{ManinModular,Brown:mmv}. In the next section, we show that these integrals arise naturally when evaluating eMPLs at certain special points.

\subsection{Modular forms}
We start by giving a lightning review of modular forms. We limit ourselves to the strict minimal mathematical background, and we refer to Appendix~\ref{app:modular_forms} and the literature for a more detailed and more rigorous discussion (cf., e.g., ref.~\cite{ZagierModular}).

In Section~\ref{sec:empls} we have seen that to every $\tau$ in the upper half-plane $\mathbb{H}$ we can associate a torus $\mathbb{C}/\Lambda_{\tau}$, and two points in $\mathbb{H}$ define the same torus if  and only if they are related by a modular transformation, cf. eq.~\eqref{eq:modular_trafo}.
Very loosely speaking, but sufficient for our purposes, a \emph{modular form of weight $n$} is a holomorphic function $f$ from the extended upper half-plane $\overline{\mathbb{H}}\equiv\mathbb{H}\cup \mathbb{Q}\cup\{i\infty\}$ into the complex numbers $\mathbb{C}$ which transforms nicely under modular transformations,
\beq\label{eq:weakly_modular}
f\left(\frac{a\tau+b}{c\tau+d}\right) = (c\tau+d)^n\,f(\tau)\,.
\eeq
For a more rigorous definition we refer to Appendix~\ref{app:modular_forms}.
Note that there are no modular forms of negative weight, because they would necessarily have a singularity, and so they cannot be holomorphic. Moreover, one can show that the only modular forms of weight zero are constants. 

In applications one is often not interested in functions that transform nicely under the full modular group $SL(2,\mathbb{Z})$, but only under a subgroup $\Gamma\subseteq SL(2,\mathbb{Z})$. Of particular interest are the so-called \emph{congruence subgroups of level $N$},
\beq\bsp\label{eq:congruence_subgroups}
\Gamma_0(N) &\,= \left\{\left(\begin{smallmatrix}a& b\\c& d\end{smallmatrix}\right)\in SL(2,\mathbb{Z}): c=0\!\!\!\!\mod N\right\}\,,\\
\Gamma_1(N) &\,= \left\{\left(\begin{smallmatrix}a& b\\c& d\end{smallmatrix}\right)\in SL(2,\mathbb{Z}): c=0\!\!\!\!\mod N\textrm{ and } a=d=1\!\!\!\!\mod N\right\}\,,\\
\Gamma(N) &\,= \left\{\left(\begin{smallmatrix}a& b\\c& d\end{smallmatrix}\right)\in SL(2,\mathbb{Z}): b=c=0\!\!\!\!\mod N\textrm{ and } a=d=1\!\!\!\!\mod N\right\}\,.
\esp\eeq
If $\Gamma$ is a congruence subgroup, then we denote the vector space of all modular forms of weight $n$ for $\Gamma$ by $\cM_n(\Gamma)$. Vector spaces of modular forms are always finite-dimensional. Modular forms form an algebra, and the product of two modular forms of weights $n_1$ and $n_2$ is a modular form of weight $n_1+n_2$. 

The prototypical modular forms are the Eisenstein series $G_{2m}(\tau)$ in eq.~\eqref{eq:Eisenstein_series}, which transform like modular forms of level $N=1$ and weight $2m$, except for $G_2(\tau)$, which transforms according to
\beq\label{eq:G2_trafo}
G_2\left(\frac{a\tau+b}{c\tau+d}\right) = (c\tau+d)^2\,G_2(\tau)-2\pi i\,c(c\tau+d)\,.
\eeq
$G_2(\tau)$ is an example of a \emph{quasi-modular form}. We will not discuss quasi-modular forms in the main text, though we include a short review in Appendix~\ref{app:modular_forms}.
As an algebra the space of all modular forms of level $N=1$ is generated by $G_4(\tau)$ and $G_6(\tau)$, and every modular form of level one is a polynomial in $G_4(\tau)$ and $G_6(\tau)$. In particular, there are no modular forms of level one and odd weight.

Every congruence subgroup of level $N$ contains the matrix $\left(\begin{smallmatrix}1&N\\ 0&1\end{smallmatrix}\right)$ which acts on the upper half-plane via translations by $N$. Equation~\eqref{eq:weakly_modular} then implies that a modular form of level $N$ is invariant under translations by $N$. Hence, every modular form is a periodic function, and thus admits a Fourier expansion, called \emph{q-expansion}, of the form
\beq
f(\tau) = \sum_{m=0}^\infty a_m \,e^{2\pi i m\tau/N} = \sum_{m=0}^\infty a_m\, q_N^m\,,
\eeq
with $q\equiv\exp(2\pi i\tau)$ and $q_N\equiv q^{1/N}$.

$SL(2,\mathbb{Z})$ acts separately on $\mathbb{H}$ and $\mathbb{Q}\cup\{i\infty\}$. Each congruence subgroup $\Gamma$ then partitions $\mathbb{Q}\cup\{i\infty\}$ into a set of distinct orbits, called the \emph{cusps of $\Gamma$}. A \emph{cusp form of weight $n$ for $\Gamma$} is an element of $\cM_n(\Gamma)$ that vanishes at all the cusps of $\Gamma$. We denote the vector space of cusp forms of weight $n$ by $\cS_n(\Gamma)$, and we have a decomposition
\beq
\cM_n(\Gamma) = \cE_n(\Gamma) \oplus \cS_n(\Gamma)\,,
\eeq
where $\cE_n(\Gamma)$ is the \emph{Eisenstein subspace of weight $n$ for $\Gamma$}. For $N=1$, the Eisenstein subspace $\cE_n(SL(2,\mathbb{Z}))$ is at most one-dimensional and generated by the Eisenstein series $G_{n}(\tau)$. The first non-zero cusp form for $N=1$ appears at weight 12.

%%%%%%%%%%%%%%%%%%%%%%%%%%%%%%%%%%%%%

\subsection{Iterated integrals of modular forms}

We now discuss iterated integrals of one-forms $d\tau\,f_i(\tau)$, where $f_i(\tau)$ is a modular form of a certain level and weight. More precisely, if $\tau_0$ and $\tau$ are elements of $\overline{\mathbb{H}}$, we define~\cite{ManinModular,Brown:mmv} (see also ref.~\cite{Adams:2017ejb})
\beq
I(f_{i_1},\ldots,f_{i_k};\tau,\tau_0) = \int_{\tau_0}^\tau d\tau'\,f_{i_1}(\tau')\,I(f_{i_2},\ldots,f_{i_k};\tau',\tau_0)\,,
\eeq
and the recursion starts with $I(;\tau,\tau_0)=1$. We call the number of integrations $k$ the \emph{length} of the integral. For simplicity, we always assume that the modular forms $f_{i_a}$ are all linearly independent (recall that modular forms form finite-dimensional vector spaces, so we can always choose a basis for a given space of modular forms).
All the concepts in this section carry over to iterated integrals of quasi-modular forms~\cite{Matthes:QuasiModular}. We only discuss modular forms here, and we refer to Appendix~\ref{app:modular_forms} for a summary of results in the quasi-modular case. 

Iterated integrals of modular forms satisfy all the basic properties of iterated integrals (see Appendix~\ref{app:iterated_integrals}). In particular, they form a shuffle algebra
\beq
I(f_{i_1},\ldots,f_{i_k};\tau,\tau_0)\,I(f_{i_{k+1}},\ldots,f_{i_{k+l}};\tau,\tau_0) = \sum_{\sigma\in \Sigma(k,l)}I(f_{i_{\sigma(1)}},\ldots,f_{i_{\sigma(k+l)}};\tau,\tau_0)\,.
\eeq
Although the integration kernels have nice behaviour under modular transformations, their iterated integrals usually do not exhibit any nice modular properties.

Let us now study the differential properties of iterated integrals of modular forms.
The total differential takes the form
\beq\bsp
dI(&f_{i_1},\ldots,f_{i_k};\tau,\tau_0) \\
&= d\tau\,f_{i_1}(\tau)\,I(f_{i_2},\ldots,f_{i_k};\tau,\tau_0) - I(f_{i_1},\ldots,f_{i_{k-1}};\tau,\tau_0)\,d\tau_0\,f_{i_k}(\tau_0)\,.
\esp\eeq
We see that the right-hand side only involves integrals of length $k-1$, and so iterated integrals of modular forms are unipotent. The symbol of $I(f_{i_1},\ldots,f_{i_k};\tau,\tau_0)$ can be obtained by the standard recursion,
\beq\bsp
\cS(I(&f_{i_1},\ldots,f_{i_k};\tau,\tau_0))\\
& = \cS(I(f_{i_2},\ldots,f_{i_k};\tau,\tau_0)) \otimes \hat{f}_{i_1}(\tau) -  \cS(I(f_{i_1},\ldots,f_{i_{k-1}};\tau,\tau_0)) \otimes \hat{f}_{i_k}(\tau_0)\,,
\esp\eeq
where we have introduced the shorthand $\hat{f}_i(\tau) = d\tau\,f_i(\tau)$. We see that, very naturally, the letters in the symbol alphabet are modular forms. This recursion can be solved explicitly, and we find a closed form for the symbol
\beq\bsp\label{eq:IEI_symbol_t0}
\cS(I(&f_{i_1},\ldots,f_{i_k};\tau,\tau_0))\\
& = \sum_{m=1}^{k+1}(-1)^{k-m+1}\,(\hat{f}_{i_m}(\tau_0)\otimes\ldots\otimes\hat{f}_{i_k}(\tau_0))\sha(\hat{f}_{i_{m-1}}(\tau)\otimes\ldots\otimes\hat{f}_{i_1}(\tau))\,.
\esp\eeq

There is a canonical choice for the base point of the integration, namely the cusp at infinity $\tau_0=i\infty$, and we will always work with this choice. We can use the path composition formula for iterated integrals in eq.~\eqref{eq:path_composition} to express integrals with a generic base point $\tau_0$ through integrals with this choice. In general, however, these integrals require regularization, because there may be divergences for $\tau_0\to i\infty$. This happens precisely if the modular forms do not vanish at the cusp at infinity, and the integral develops logarithmic divergences of the type $\log q_0$, with $q_0\equiv\exp(2\pi i\tau_0)\to 0$. We follow the regularization of ref.~\cite{Brown:mmv} (see also ref.~\cite{Adams:2017ejb}), and we define
\beq
I(f_{i_1},\ldots,f_{i_k};\tau) = \lim_{\tau_0\to i\infty} R\left[I(f_{i_1},\ldots,f_{i_k};\tau,\tau_0)\right]\,,
\eeq
where $R$ is the operator that sends $\log q_0$ to zero. This regularization has the feature that it preserves the shuffle algebra structure. Close to the cusp at infinity, every iterated integral of modular forms with the canonical choice of the base point admits a $q$-expansion of the type
\beq\label{eq:IIMF_log_q}
I(f_{i_1},\ldots,f_{i_k};\tau) = \sum_{l=0}^k\log^lq\,I_l(q)\,,
\eeq
where the $I_l(q)$ are functions that are holomorphic in a neighborhood of the cusp at infinity $q=0$.

Since iterated integrals of modular forms are unipotent, we can immediately use the construction of Section~\ref{sec:de-Rham-symbols} to define a coaction on iterated integrals of modular forms. For simplicity, we only discuss the case of the canonical base point $\tau_0=i\infty$. We have the unipotent differential equation
\beq
dJ = AJ\,,
\eeq
where 
\beq
J = \Bigg(\int_{i\infty}^\tau\Omega_1,\ldots, \int_{i\infty}^\tau\Omega_{k},1\Bigg)^T\,,
\eeq
with 
\beq
\Omega_a = \left[\hat{f}_{i_k}(\tau)|\ldots|\hat{f}_{i_a}(\tau)\right]\,.
\eeq
%By convention, we set $I(;\tau) = \int_{i\infty}^\tau\omega_{k+1} = 1$. 
The matrix $A$ has non-zero elements only on its superdiagonal,
\beq
A_{ab} = \delta_{b,a+1}\,\hat{f}_{a}(\tau)\,.
\eeq
We can easily construct the matrix $T_A$ and read off the symbol of a pair of differential forms,
\beq
\symb([\Omega_b,\Omega_a]) = \theta(b-a)\,\left[\hat{f}_{i_{b-1}}(\tau)|\ldots|\hat{f}_{i_a}(\tau)\right] = \theta(b-a)\,\cS(I(f_{i_a},\ldots,f_{i_{b-1}};\tau))\,,
\eeq
where $\theta$ denotes the Heaviside step function.
The coaction on iterated integral of modular forms then takes the compact form
\beq\bsp
\Delta(I(f_{i_1},\ldots,f_{i_k};\tau))& = \sum_{m=1}^{k+1}I(f_{i_m},\ldots,f_{i_k};\tau)\otimes \left[\hat{f}_{i_{m-1}}(\tau)|\ldots|\hat{f}_{i_1}(\tau)\right]\\ 
&= \sum_{m=1}^{k+1}I(f_{i_m},\ldots,f_{i_k};\tau)\otimes \cS(I(f_{i_1},\ldots,f_{i_{m-1}};\tau))\,.
\esp\eeq
%It is easy to check that also in this case we can identify the symbol with the maximal iteration fo the coaction.

% !TEX root = elliptic_symbols.tex

\section{Connecting eMPLs and iterated integrals of modular forms}
\label{sec:eMPLs_to_modular}

\subsection{A class of Eisenstein series from eMPLs}
In this section we discuss a close relationship between eMPLs evaluated at some special points and iterated integrals of (quasi-)modular forms. More precisely, we consider eMPLs of the form $\gamtt{n_1 & \ldots & n_k}{z_1&\ldots & z_k}{z_{k+1}}{\tau}$, where each $z_i$ is chosen to be a \emph{rational point}, i.e., a point $z_i = a_i+b_i\tau$, where $a_i$ and $b_i$ are rational numbers (such points are also known as \emph{torsion points} of the elliptic curve $\mathbb{C}/\Lambda_{\tau}$). Note that eMPLs evaluated at rational points cover in particular $A$-type eMZVs for $z_{k+1}=1$ and $z_i=0$, $1\le i\le k$, and more generally the twisted eMZVs of ref.~\cite{Broedel:2017jdo}.

eMPLs evaluated at rational points are functions of $\tau$. Our goal is to study the symbols (and the coaction) of these classes of functions. From the differential equation in eq.~\eqref{eq:gamma_differential} we can see that the symbol letters in eq.~\eqref{eq:empl_letter} reduce to
\beq
\omega_{ij}^{(n)} = d\tau\left[b_{ji}\,g^{(n)}(a_{ji}+b_{ji}\tau,\tau) + \frac{n}{2\pi i}g^{(n+1)}(a_{ji}+b_{ji}\tau;\tau)\right]\,,
\eeq
with $a_{ji}=a_j-a_i$ and $b_{ji}=b_j-b_i$. In the following, it will be convenient to write the coefficients $g^{(n)}$ of the Eisenstein-Kronecker series in eq.~\eqref{eq:Eisenstein-Kronecker} in terms of the periodic analogues $f^{(n)}$, cf. eq.~\eqref{eq:Omega_def}. It is easy to see that
\beq
g^{(n)}(z,\tau) = \sum_{k=0}^n\frac{1}{k!}\left[-2\pi i\frac{\textrm{Im }z}{\textrm{Im }{\tau}}\right]^k\,f^{(n-k)}(z,\tau)\,,
\eeq
and so, for rational $a$ and $b$,
\beq\label{eq:g_to_f}
g^{(n)}(a+b\tau,\tau) = \sum_{k=0}^n\frac{\left(-2\pi i b\right)^k}{k!}\,f^{(n-k)}(a+b\,\tau,\tau)\,.
\eeq
We see that for rational points, the functions $g^{(n)}$ and $f^{(n)}$ are linear combinations of each other, with coefficients that are rational polynomials in $2\pi i$. The letters of the symbol alphabet are then the functions $f^{(n)}$ evaluated at rational points. By periodicity, we may assume without loss of generality that $a$ and $b$ have the form
\beq
a=\frac{r}{N} {\rm~~and~~} b=\frac{s}{N}\,,\textrm{ with } 0\le r,s<N\,,
\eeq
and we find it convenient to introduce the following notation,\footnote{Note that the limit $\lim_{z\to 0}f^{(2)}(z,\tau)$ is not defined. By convention, we set $h_{N,0,0}^{(2)}(\tau) \equiv g^{(2)}(0,\tau) = -G_2(\tau)$.}
\beq
h^{(n)}_{N,r,s}(\tau)\equiv f^{(n)}\left(\frac{r}{N}+\tau\frac{s}{N},\tau\right)\,.  \label{eq:ftoh}
\eeq

One of the reasons to switch from $g^{(n)}$ to $f^{(n)}$ is that the latter have a nice behaviour under modular transformations~\cite{BrownLevin},
\beq
f^{(n)}\left(\frac{z}{c\tau+d},\frac{a\tau + b}{c\tau+d}\right) = (c\tau+d)^n\,f^{(n)}(z,\tau)\,.
\eeq
This implies the following transformation for the functions $h^{(n)}_{N,r,s}(\tau)$,\footnote{There is an exception to this rule, because $h_{N,0,0}^{(2)}(\tau) = -G_2(\tau)$ is the Eisenstein series of weight two, which is only quasi-modular.}
\beq\bsp\label{eq:trafo}
 h^{(n)}_{N,r,s}\left(\frac{a\tau + b}{c\tau+d}\right) &\,=(c\tau+d)^n\,h^{(n)}_{N,rd+sb,rc+sa}(\tau)\,.
\esp\eeq
Comparing eq.~\eqref{eq:trafo} to the transformation property of modular forms in eq.~\eqref{eq:weakly_modular}, we see that the $h^{(n)}_{N,r,s}$ transform like modular forms of weight $n$ for a certain subgroup $\Gamma\subseteq SL(2,\mathbb{Z})$, provided that for every matrix $\left(\begin{smallmatrix}a& b\\c&d\end{smallmatrix}\right)\in \Gamma$ we have
\beq\label{eq:constraint}
(s,r)\left(\begin{array}{cc}
a& b\\
c&d
\end{array}\right) = (s,r) \mod N\,.
\eeq
In the following we will not attempt to solve these constraints in full generality, but we focus on a special class of solutions with interesting properties. If we choose $\Gamma=\Gamma(N)$, then we have $\left(\begin{smallmatrix}a& b\\c&d\end{smallmatrix}\right)=\left(\begin{smallmatrix}1& 0\\0&1\end{smallmatrix}\right)\!\!\mod N$ (cf. eq.~\eqref{eq:congruence_subgroups}), and so eq.~\eqref{eq:constraint} is trivially satisfied. 
Hence, we see that the functions $h^{(n)}_{N,r,s}$ always transform like modular forms of weight $n$ for $\Gamma(N)$.
It is possible to make this statement more precise, and one can show that, for $n>1$, the functions $h^{(n)}_{N,r,s}$ are always Eisenstein series for $\Gamma(N)$, and they can be represented by a double sum very similar to eq.~\eqref{eq:Eisenstein_series}~\cite{ZagierBonn},
\beq
h^{(n)}_{N,r,s}(\tau) =-\!\!\!\!\!\!\!\sum_{\substack{(\alpha,\beta)\in \mathbb{Z}^2\\ (\alpha,\beta)\neq(0,0)}}\frac{e^{2\pi i(s\alpha-r\beta)/N}}{(\alpha+\beta\tau)^{2n}}\,.
\eeq
The previous equation does not hold for $n=1$. Indeed, seen as a function of $z$, the function $f^{(1)}(z,\tau)$ has a simple pole at every point of the lattice $\mathbb{Z}+\tau\mathbb{Z}$. As a consequence, the functions $h^{(1)}_{N,r,s}$ may have poles, in which case they do not define valid modular forms. The poles, however, always lie on the real axis, and so the functions $h^{(1)}_{N,r,s}$ define \emph{weakly holomorphic modular forms}. We conclude that the symbol alphabet of eMPLs evaluated at rational points $z_i=\frac{r_i}{N_i}+\tau\frac{s_i}{N_i}$ are Eisenstein series of weight $n>1$ and weakly holomorphic modular forms of weight one for $\Gamma(N)$, where $N$ is the least common multiple of the $N_i$. This is equivalent to saying that eMPLs evaluated at these rational points can always be written as iterated integrals of Eisenstein series of weight $n>1$ and weakly holomorphic modular forms of weight one for $\Gamma(N)$. 
This result generalizes and extends in a precise and natural way the fact that eMZVs, which are eMPLs evaluated at the rational points $z_i=0$ and $z_{k+1}=1$, can always be expressed as iterated integrals of the Eisenstein series $G_n(\tau)$~\cite{Broedel:2015hia,Matthes:Eisenstein2}. 
We find it convenient to introduce the following notation for iterated integrals of the functions $h_{N,r,s}^{(n)}$,
\beq\bsp
I\left(\begin{smallmatrix} n_1& N_1\\ r_1& s_1\end{smallmatrix}\big|\ldots\big|\begin{smallmatrix} n_k& N_k\\ r_k& s_k\end{smallmatrix};\tau\right) &\,\equiv I(h_{N_1,r_1,s_1}^{(n_1)},\ldots,h_{N_k,r_k,s_k}^{(n_k)};\tau)\\
&\,= \int_{i\infty}^{\tau}d\tau'\,h_{N_1,r_1,s_1}^{(n_1)}(\tau')\,I\left(\begin{smallmatrix} n_2& N_2\\ r_2& s_2\end{smallmatrix}\big|\ldots\big|\begin{smallmatrix} n_k& N_k\\ r_k& s_k\end{smallmatrix};\tau'\right)\,, \label{eq:itmod}
\esp\eeq
By convention, we set $h_{0,0,0}^{(0)}(\tau)\equiv 1$. 

Since our results imply that eMPLs evaluated at rational points are closely related to iterated integrals of Eisenstein series for $\Gamma(N)$, it is natural to ask whether the inverse is also true, i.e., if every iterated integral over Eisenstein series for $\Gamma(N)$ can be expressed in terms of eMPLs evaluated at some rational points. The answer to this question is negative, already in the simplest case $N=1$~\cite{Broedel:2015hia}. It can be shown that a necessary condition for an iterated integral over Eisenstein series of level $N=1$ to be expressible in terms of eMPLs is that its symbol is annihilated by the generators of a certain derivation algebra~\cite{EnriquezAssoc,Calaque,Hain1,Pollack}. It would be interesting to see if similar criteria on the symbols of iterated integrals of Eisenstein series of higher levels can be formulated.

So far we have argued that eMPLs evaluated at rational points can be written as iterated integrals of Eisenstein series for $\Gamma(N)$. In the case where $z_{k+1}\in \mathbb{Q}$, we can restrict the relevant spaces of iterated integrals of Eisenstein series even more. We have seen in eq.~\eqref{eq:IIMF_log_q} that in general iterated integrals of modular forms may have a logarithmic branch point at the cusp at infinity $\tau=i\infty$, or equivalently $q=0$. In the case of eMPLs evaluated at rational points with $z_{k+1}\in\mathbb{Q}$, however, this branch point must be absent. Indeed, we can start from the $q$-expansions for the integration kernels $g^{(n)}$ in Appendix~\ref{app:q_exp}, and integrate order by order in $z$ in the $q$-expansion using the definition of eMPLs in eq.~\eqref{eq:gamt_def}. As all integrations are in $z$ and not in $q$, and the upper integration limit $z_{k+1}$ does not depend on $q$, we never generate logarithmic singularities at $q=0$. Hence, eMPLs at rational points with $z_{k+1}\in\mathbb{Q}$ always evaluate to Taylor series in $q_N$, or equivalently to Fourier series in $\tau$ (if $z_{k+1}$ depends on $\tau$, the upper integration limit can introduce additional logarithmic divergences at $q_N=0$). Following ref.~\cite{Matthes:Eisenstein2,MatthesThesis} we define the \emph{Fourier subspace of iterated integrals of modular forms} as the vector space of iterated integrals of modular forms that admit a Fourier series in $\tau$. Hence, eMPLs evaluated at rational points with $z_{k+1}\in\mathbb{Q}$ always lie in the Fourier subspace. It is easy to check that the Fourier space is closed under multiplication, and it is generated precisely by linear combinations of the form
\beq
\label{eq:IFourier}
I_F(f_1,\ldots,f_k;\tau) \equiv I(f_1,\ldots,f_k;\tau) - f_k(i\infty)\,I(f_1,\ldots,f_{k-1},1;\tau)\,.
\eeq
Note that $f_k(i\infty)$ always exists. We have, for $n>1$~\cite{Broedel:2017jdo},
\beq
h_{N,r,s}^{(n)}(i\infty) = \frac{(2\pi i)^n}{n!}\,B_n\left(\frac{s}{N}\right)\,,
\eeq
where $B_n(x)$ are the Bernoulli polynomials,
\beq
B_n(x) = \sum_{k=0}^n\binom{n}{k}\,B_{n-k}\,x^k\,,
\eeq
with $B_n\equiv B_n(0)$ the Bernoulli numbers. For $n=1$, we have
\beq
h_{N,r,s}^{(1)}(i\infty) = \left\{\begin{array}{ll}
2\pi i\left(\frac{s}{N}-\frac{1}{2}\right)\,,& s\neq 0\,,\\
\pi\,\cot\frac{\pi\,r}{N}\,,&s=0, \textrm{  and  }r\neq 0\,.
\end{array}\right.
\eeq

We can translate the condition to lie in the Fourier subspace into a \emph{first entry condition} on the symbol: a linear combination of iterated integrals of modular forms lies in the Fourier subspace if and only if the first entries in its symbol all have the form $d\tau(f_j(\tau)-f_j(i\infty))$, for some modular forms $f_j$. For example, we have
\beq\label{eq:Fourier_1st_entry}
\cS(I_F(f_1,\ldots,f_k;\tau)) = \big[\hat{f}_k(\tau)-d\tau{f}_k(i\infty)|\hat{f}_{k-1}(\tau)|\ldots|\hat{f}_1(\tau)\big]\,.
\eeq
Note that the Fourier subspace is `closed' under the coaction (cf. ref.~\cite{Matthes:Eisenstein2,MatthesThesis} for a similar statement for $N=1$),
\beq\bsp
\Delta(I_F(f_{i_1},\ldots,f_{i_k};\tau))& = \sum_{m=1}^{k+1}I_F(f_{i_k},\ldots,f_{i_m};\tau)\otimes \cS(I(f_{i_1},\ldots,f_{i_{m-1}};\tau))\,.
\esp\eeq

Let us conclude with the question as to whether all Eisenstein series of a given weight can appear in the symbol alphabet of eMPLs evaluated at rational points, i.e., if every Eisenstein series of level $N$ and weight $n>1$ can be expressed in terms of the functions $h_{N,r,s}^{(n)}$.
For fixed values of $n$ and $N$, there are $N^2$ distinct functions $h_{N,r,s}^{(n)}$. The dimensions of the Eisenstein subspaces $\cE_n(\Gamma(N))$ of weight $n$  are known, and for $n>2$ they are independent of the weight $n$ (for $n=2$, the Eisenstein series $G_2(\tau)$ is missing). In addition, we always have $\textrm{dim }\cE_n(\Gamma(N))<N^2$. Hence, not all the functions $h_{N,r,s}^{(n)}$ are independent, but there must be linear relations among them. First, there is a \emph{reflection identity}, relating the functions associated to the values $(r,s)$ and $(N-r,N-s)$.
\beq\label{eq:reflect}
h_{N,r,s}^{(n)}(\tau) = (-1)^n\,h_{N,N-r,N-s}^{(n)}(\tau)\,.
\eeq
Second, for every $d|N$ and $0\le \rho, \sigma<d$ there is a \emph{distribution identity} of the form~\cite{ZagierBonn}
\beq\label{eq:latticesum}
\sum_{\frac{1}{N}(r,s) \in \frac{1}{N}(\rho,\sigma) + \Lambda_{N/d}^F} h^{(n)}_{N,r,s}(\tau) = \left(\frac{d}{N}\right)^{n-2}\,
h^{(n)}_{d,\rho,\sigma}(\tau)\,,
\eeq
where we used the shorthand notation $(r,s)\equiv {r}+\tau\,{s}$, and we defined
\beq
\Lambda_{N}^F = \left\{\frac{r}{N}+\tau\frac{s}{N}: 0\le r,s<N\right\}\,.
\eeq
Using {\tt Sage}~\cite{sage}, we have checked up to level $N=39$ that there are no other linear relations for weights $n>1$, and the dimension of the solution space agrees with the dimension of $\cE_n(\Gamma(N))$. Hence, the functions $h_{N,r,s}^{(n)}$ provide a spanning set for $\cE_n(\Gamma(N))$ for $n>1$. Moreover, we have been able to identify the following basis for $\cE_n(\Gamma(N))$,
\beq\label{eq:Eisenstein_Basis_2}
\cE_n(\Gamma(N)) = \Big\langle h_{N,r,s}^{(n)}: (r,s) \in B_N\Big\rangle_{\mathbb{C}}\,,
\eeq
with $B_N = B_{N,1}\cup B_{N,2}\cup B_{N,3}$ and
\begin{align}\label{eq:Eisenstein_Basis}
B_{N,1}&\, = \{(r,s): 0\le r<N\textrm{ and } 0<s\le\lceil N/2-1\rceil\textrm{ and } \textrm{gcd}(N,r,s) = 1\}\,, \\
\nonumber B_{N,2}&\, = \{(r,0): 1\le r\le N/2\textrm{ and } \textrm{gcd}(N,r) = 1\}\,, \\
\nonumber B_{N,3}&\, = \left\{\begin{array}{ll}\displaystyle \{(r,N/2): 1\le r\le N/2\textrm{ and } \textrm{gcd}(N,r,N/2) = 1\}\,,& \textrm{ $N$ even}\,,\\
\emptyset\,, &\textrm{ $N$ odd}\,.
\end{array}\right.
\end{align}
Note that this basis is not valid for small levels and weights, because in that case some basis elements may be absent (e.g., the Eisenstein series $G_2(\tau)=-h_{N,0,0}^{(2)}(\tau)$ is not a modular form, and there are no Eisenstein series of level one or two and odd weight). 

Finally, let us make a comment about other congruence subgroups. From eq.~\eqref{eq:congruence_subgroups} we see that there is an obvious inclusion $\Gamma(N)\subseteq \Gamma_1(N)$, and so $\cE_n(\Gamma_1(N))\subseteq \cE_n(\Gamma(N))$. Hence, every element of $\cE_n(\Gamma_1(N))$ can be written as a linear combination of the basis of $\Gamma(N)$ in eq.~\eqref{eq:Eisenstein_Basis_2}. In Appendix~\ref{app:modular_forms}, we show this linear combination and present an explicit basis elements for $\cE_n(\Gamma_1(N))$ in terms of certain `cyclic sums' of the basis for $\cE_n(\Gamma(N))$ in eq.~\eqref{eq:Eisenstein_Basis_2} .

\subsection{A worked out example}
In this section we illustrate on a simple example how one can express eMPLs evaluated at rational points in terms of iterated integrals of Eisenstein series. For concreteness, we analyze the function
\beq
f(\tau) = \gamtt{0&2}{0& 2\tau/3}{1}{\tau}\,.
\eeq
The coaction takes the form
\beq\bsp\label{eq:coaction_example_rational_1}
\Delta(f(\tau)) &\,= f(\tau)\otimes 1 -\frac{2}{3}\, \gamtt{2}{2\tau/3}{1}{\tau}\otimes[d\tau] +\frac{1}{i\pi} \gamtt{3}{2\tau/3}{1}{\tau}\otimes[{d\tau}]\\
&\,+\frac{2}{3}\,1\otimes\left[d\tau\,g^{(2)}\left(\frac{2\tau}{3},\tau\right)\right] +\frac{1}{i\pi}\,1\otimes\left[d\tau\,g^{(3)}\left(\frac{2\tau}{3},\tau\right)\right]\,.
\esp\eeq
Before we continue to analyze the coaction, let us comment on the eMPLs of length one in the right-hand side of eq.~\eqref{eq:coaction_example_rational_1}. We find
\beq\bsp
\Delta\left(\gamtt{2}{2\tau/3}{1}{\tau}\right) &\,= \gamtt{2}{2\tau/3}{1}{\tau}\otimes 1\,,\\
 \Delta\left(\gamtt{3}{2\tau/3}{1}{\tau}\right) &\,= \gamtt{3}{2\tau/3}{1}{\tau}\otimes 1\,.
\esp\eeq
Hence, these functions must be constant, and equal to their value at the cusp at infinity. This value can easily be obtained using the $q$-expansions of the coefficients of the Eisenstein-Kronecker series~\cite{Broedel:2014vla} (see Appendix~\ref{app:q_exp}). For example, we find
\beq
\lim_{\tau\to i\infty}\gamtt{2}{2\tau/3}{1}{\tau} = \lim_{\tau\to i\infty}\int_0^1dz\,g^{(2)}\left(z-\frac{2\tau}{3},\tau\right) = -2\zeta_2\,.
\eeq
Similarly, we find 
\beq
\lim_{\tau\to i\infty}\gamtt{3}{2\tau/3}{1}{\tau} = 0\,.
\eeq
Equation~\eqref{eq:coaction_example_rational_1} can then be cast into the simpler form
\beq\bsp\label{eq:coaction_example_rational_2}
\Delta(f(\tau)) &\,= f(\tau)\otimes 1 +\frac{2\pi^2}{9}\,1\otimes[d\tau] +\frac{2}{3}\,1\otimes\left[d\tau\,g^{(2)}\left(\frac{2\tau}{3},\tau\right)\right] \\
&\,+\frac{1}{i\pi}\,1\otimes\left[d\tau\,g^{(3)}\left(\frac{2\tau}{3},\tau\right)\right]\,.
\esp\eeq
We can now express all the $g^{(n)}$ functions in terms of $f^{(n)}$ functions using eq.~\eqref{eq:g_to_f}. The latter can themselves be expressed in terms of the Eisenstein series $h^{(n)}_{N,r,s}$, where $N=3$ in our case. We find,
\beq\bsp\label{eq:coaction_example_rational_3}
\Delta(f(\tau)) &\,= f(\tau)\otimes 1 +\frac{2\pi^2}{81}\,1\otimes[d\tau] -\frac{2}{3}\,1\otimes\left[d\tau\,h^{(2)}_{3,0,1}(\tau)\right] -\frac{1}{i\pi}\,1\otimes\left[d\tau\,h^{(3)}_{3,0,1}(\tau)\right]\,,
\esp\eeq
where we have written all Eisenstein series in terms of the basis in eq.~\eqref{eq:Eisenstein_Basis_2}. From eq.~\eqref{eq:coaction_example_rational_3} we immediately see that we can write $f$ as a linear combination of iterated integrals of Eisenstein series for $\Gamma(3)$,
\beq
f(\tau) = c + \frac{2\pi^2}{81}\,I\left(\begin{smallmatrix} 0& 0\\ 0& 0\end{smallmatrix};\tau\right) - \frac{2}{3}\,I\left(\begin{smallmatrix} 2& 3\\ 0& 1\end{smallmatrix};\tau\right)- \frac{1}{i\pi}\,I\left(\begin{smallmatrix} 3& 3\\ 0& 1\end{smallmatrix};\tau\right)\,,
\eeq
where $c$ is a constant that satisfies $\Delta(c)=c\otimes 1$. The value of $c$ can easily be obtained by computing the value at the cusp at infinity of $f$,
\beq
c = \lim_{\tau\to i\infty}\gamtt{0&2}{0&2\tau/3}{1}{\tau} = -\frac{\pi^2}{6}\,.
\eeq
We then find
\beq\bsp\label{eq:coaction_example_rational_4}
f(\tau) &\,= -\frac{\pi^2}{6} + \frac{2\pi^2}{81}\,I\left(\begin{smallmatrix} 0& 0\\ 0& 0\end{smallmatrix};\tau\right) - \frac{2}{3}\,I\left(\begin{smallmatrix} 2& 3\\ 0& 1\end{smallmatrix};\tau\right)- \frac{1}{i\pi}\,I\left(\begin{smallmatrix} 3& 3\\ 0& 1\end{smallmatrix};\tau\right)\\
&\,= -\frac{\pi^2}{6}  - \frac{2}{3}\,I_F\left(\begin{smallmatrix} 2& 3\\ 0& 1\end{smallmatrix};\tau\right)- \frac{1}{i\pi}\,I_F\left(\begin{smallmatrix} 3& 3\\ 0& 1\end{smallmatrix};\tau\right)\,,
\esp\eeq
where in the last line we have made explicit the fact that $f$ lies in the Fourier subspace.
We have checked that eq.~\eqref{eq:coaction_example_rational_4} is correct by comparing the 35 first terms in the $q$-expansion on both sides.

% !TEX root = elliptic_symbols.tex

\section{An elliptic class of hypergeometric functions}
\label{sec:hyper}

As a non-trivial application of our formalism, we consider the class of hypergeometric functions 
studied in ref.~\cite{Broedel:2017kkb}. They are defined by the following integral representation
\begin{align}
\begin{split}
T(n_1,n_3,n_3; \lambda) &= \int_0^1dx\,x^{-1/2+n_1+\alpha_1\eps}\,(1-x)^{-1/2+n_2+\alpha_2\eps}\,
(1-\lambda x)^{-1/2+n_3+\alpha_3\eps} \\
&=\frac{1}{\sqrt{\lambda}}\, \int_0^1 \frac{dx}{y}\,x^{n_1+\alpha_1\eps}\,(1-x)^{n_2+\alpha_2\eps}
\,(1-\lambda x)^{n_3+\alpha_3\eps}\,,
\end{split}
\label{eq:T_def}
\end{align}
where $n_i$ and $\alpha_i$ are integers and we take $0<\lambda<1$. 
Clearly, $y^2=x(x-1)(x-L)$ describes an elliptic curve with $L = 1/\lambda$.\footnote{Note that we use a different notation for the variables here with respect to
ref.~\cite{Broedel:2017kkb} in order to avoid confusion later in the text.} 
As always, we call the two periods of the elliptic curve curve 
$\omega_1$ and $\omega_2$ and the corresponding quasi-periods
$\eta_1$ and $\eta_2$, with
\begin{align}
\omega_1 = 2 {\rm K}(\lambda)\,, \qquad \omega_2 = 2 i\, {\rm K}(1-\lambda)\,.
\end{align}
As usual, the ratio of the two periods is indicated by $\tau$,
\begin{align}
\tau = \frac{\omega_2}{\omega_1}= i\frac{\, {\rm K}(1-\lambda)}{{\rm K}(\lambda)} \,.
\end{align}
By a standard use of integration-by-parts identities, 
all integrals in eq.~\eqref{eq:T_def} can be expressed in terms of two
master integrals, which we choose as
\begin{align}
T_1(\lambda) &\,= T(0,0,0;\lambda)\,, \qquad T_2(\lambda) \,= \frac{1+\lambda}{3 \lambda}\,T(0,0,0;\lambda)-T(1,0,0;\lambda) \,.
\end{align}

In ref.~\cite{Broedel:2017kkb}, it was shown that the coefficients of the expansion in $\eps$ of the 
two master integrals can be expressed in terms of a particular variant of elliptic polylogarithms,
defined as iterated integrals over transcendental and algebraic kernels and denoted $\text{E}_3$ in ref.~\cite{Broedel:2017kkb}.
The goal of this section is to show that the two master integrals $T_1(\lambda)$ and $T_2(\lambda)$ can easily be rewritten in terms of iterated integrals of
Eisenstein series, and that our symbol calculus can be used to render
this derivation particularly transparent.

One of the main results of ref.~\cite{Broedel:2017kkb} was to show that the $\text{E}_3$
functions can always be expressed in terms of the eMPLs $\widetilde{\Gamma}$.
The starting point is a variant of Abel's map, which allows one to associate to each point on the 
elliptic curve defined by the cubic equation $y^2=x(x-1)(x-L)$ a point on the torus defined by the periods $\omega_1$ and $\omega_2$,
\begin{align}
z(a) = \frac{\sqrt{L}}{2}\, \int_{\infty}^a \frac{dx}{y}\,. \label{eq:Abel2F1}
\end{align}
From ref.~\cite{Broedel:2017kkb} we see that the integrands of the $\text{E}_3$ functions have poles at most 
at the points $x\in \{0,1,L\}$, which under the map in eq.~\eqref{eq:Abel2F1} are sent to the half periods,
\begin{align}
z(0) =  \frac{\omega_2}{2}\,, \quad
z(1) =  \frac{\omega_3}{2} = \frac{\omega_1+\omega_2}{2}\,, \quad
z(L)  =  \frac{\omega_1}{2}\,.
\end{align}

Let us see explicitly how this works for the first non-trivial order in the $\eps$ expansion
of the first master integral $T_1(\lambda)$.
\noindent Upon transforming the results obtained in ref.~\cite{Broedel:2017kkb} 
from $\text{E}_3$ to $\widetilde{\Gamma}$, we can write the first master integral as follows
\beq
T_1(\lambda) = \omega_1\, \left( 1 + \sum_{j=1}^\infty \; \eps^j\, \mathcal{T}^{(j)}_1(\lambda)   \right)\,.
\eeq
with
\begin{alignat}{2}
\mathcal{T}_1^{(1)}(\lambda) = 
%&-i \pi  \left(\alpha _1-\alpha _2\right) 
% \Big[ \tau (\tau+1) -4 \gamtt{ 0 & 0 }{ 0 & 0 }{ z_{0,1} }{ \tau  }
%-4 \gamtt{ 0 & 0 }{ 0 & 0 }{ z_{1,1} }{ \tau  }
% \Big] \nonumber \\ & 
&\frac{i \pi}{2}  \left(\alpha _1-\alpha _2\right)  \nonumber 
%+\left(\alpha _2+\alpha _3\right) \Big[2 (\tau +1) \gamtt{ 1 }{ 0 }{ z_{0,1} }{ \tau  }
%-4 \left(\gamtt{ 0 & 1 }{ 0 & 0 }{ z_{1,1} }{ \tau  }
%+\gamtt{ 1 & 0 }{ 0 & 0 }{ z_{0,1} }{ \tau  }\right)\Big]  \nonumber \\ & 
+\left(\alpha _2+\alpha _3\right) \Big[2 \gamtt{ 1 }{ 0 }{ z_{0,1} }{ \tau  }
-4 \gamtt{ 0 & 1 }{ 0 & 0 }{ z_{1,1} }{ \tau  }
+4\gamtt{ 0 & 1 }{ 0 & 0 }{ z_{0,1} }{ \tau  }\Big]  \nonumber \\ & 
%%%%
+\alpha _1 \Big[4 \gamtt{ 0 & 1 }{ 0 & 0 }{ z_{0,1} }{ \tau  }
- 4\gamtt{ 0 & 1 }{ 0 & z_{0,1} }{ z_{0,1} }{ \tau  }
+4\gamtt{ 1 & 0 }{ 0 & 0 }{ z_{1,1} }{ \tau  }
- 4\gamtt{ 1 & 0 }{ z_{0,1} & 0 }{ z_{1,1} }{ \tau  }  \nonumber \\ &
\qquad + 2 \tau \left(   \gamtt{ 1 }{ z_{0,1} }{ z_{1,1} }{ \tau  } - \gamtt{ 1 }{ 0 }{ z_{1,1} }{ \tau  } \right)
\Big]  \nonumber \\ & 
%%%%
+\alpha _2 \Big[4 \gamtt{ 0 & 1 }{ 0 & z_{1,1} }{ z_{1,1} }{ \tau  }
-4\gamtt{ 0 & 1  }{0 & z_{1,1} }{ z_{0,1} }{ \tau  }
-2  \gamtt{ 1 }{ z_{1,1} }{ z_{0,1} }{ \tau  }\Big]  \nonumber \\ & 
%%%%
+\alpha _3 \Big[4 \gamtt{ 0 & 1 }{ 0 & z_{1,0} }{ z_{1,1} }{ \tau  }
-4\gamtt{ 1 & 0 }{ z_{1,0} & 0 }{ z_{0,1} }{ \tau  }
-2 \gamtt{ 1 }{ z_{1,0} }{ z_{0,1} }{ \tau  }\Big] \,,
%%%%
 \label{eq:FDT1eps1}
\end{alignat}
where we introduced the shorthand notation
\beq
z_{r,s} = \frac{r}{2} + \frac{s}{2} \tau\,.
\eeq

We see that eq.~\eqref{eq:FDT1eps1} only involves eMPLs evaluated at rational points, and thus it can be expressed in terms of iterated integrals 
of Eisenstein series of level $N=2$ (see Section~\ref{sec:eMPLs_to_modular}).
We can compute the coaction of eq.~\eqref{eq:FDT1eps1}, and we find
\beq
\Delta \left( \mathcal{T}_1^{(1)}(\lambda) \right)  = 
\mathcal{T}_1^{(1)}(\lambda) \otimes 1  
+ A \otimes \big[ d \tau\big] + B \otimes \big[ d\tau \, h^{(2)}_{2,1,0}(\tau) \big]
+ C \otimes \big[  d \tau \, h^{(2)}_{2,1,1}(\tau)\big]
\eeq
with 
\begin{alignat}{2}
A &=
2 \alpha _1 
 \Big[   \gamtt{ 1 }{ 0 }{ z_{0,1} }{ \tau  }
- \gamtt{ 1 }{ 0 }{ z_{1,1} }{ \tau  } \nonumber \\&
 + \frac{1}{ i \pi } \Big( \gamtt{ 2 }{ 0 }{ z_{0,1} }{ \tau  }
- \gamtt{ 2 }{ 0 }{ z_{1,1} }{ \tau  }   
-\gamtt{ 2 }{ z_{0,1} }{ z_{0,1} }{ \tau  }
+\gamtt{ 2 }{ z_{0,1} }{ z_{1,1} }{ \tau  }
+\frac{\pi ^2}{4} \Big)\Big]\nonumber \\&
%%%%
+\frac{2 \alpha _2}{i \pi } \left[ -\gamtt{ 2 }{ z_{1,1} }{ z_{0,1} }{ \tau  }
+\gamtt{ 2 }{ z_{1,1} }{ z_{1,1} }{ \tau  }+ \frac{\pi ^2}{4} \right] \nonumber \\&
%%%%
+2 \alpha _3 \left[ - \gamtt{ 1 }{ z_{1,0} }{ z_{0,1} }{ \tau  }
+ \gamtt{ 1 }{ z_{1,0} }{ z_{1,1} }{ \tau  }
- \frac{1}{i \pi } \left(\gamtt{ 2 }{ z_{1,0} }{ z_{0,1} }{ \tau  }
-\gamtt{ 2 }{ z_{1,0} }{ z_{1,1} }{ \tau  }\right) \right] \nonumber \\&
%%%%
+ 2 \left(\alpha _2+\alpha _3\right)
\left[ \gamtt{ 1 }{ 0 }{ z_{0,1} }{ \tau  }
-  \gamtt{ 1 }{ 0 }{ z_{1,1} }{ \tau  }
+  \frac{1 }{i \pi } \left(\gamtt{ 2 }{ 0 }{ z_{0,1} }{ \tau  }
- \gamtt{ 2 }{ 0 }{ z_{1,1} }{ \tau  }\right) \right]\,,
\label{eq:coac2F1ab}\\
%%%%%%%
B &= - \frac{\left(\alpha _1+2 \alpha _2+\alpha _3\right)}{i \pi }\,, \\ 
%%%%%%%%
C &= \frac{\left(\alpha _1-\alpha _2-2 \alpha _3\right)}{i \pi }\,. 
\label{eq:coac2F1cd}
\end{alignat}
In order to simplify these expressions further, we can rewrite all eMPLs of length one in terms of iterated integrals of modular forms.
This can be achieved by iterating the procedure above, namely computing their coaction
and using it to re-express them in terms of iterated integrals of modular forms.
We do not show this here explicitly but, 
upon doing this, the expression for $A$ simplifies and we are left with 
\begin{align}
\Delta \left( \mathcal{T}_1^{(1)}(\lambda) \right)  &= \mathcal{T}_1^{(1)}(\lambda)\otimes1
 - \frac{ (\alpha_1+2 \alpha_2+\alpha_3)}{i\pi} \; 1\otimes \big[ d \tau \,h^{(2)}_{2,1,0}(\tau)\big] \nonumber \\ 
&+\frac{ (\alpha_1-\alpha_2-2 \alpha_3)}{i \pi} \; 1\otimes \big[ d \tau \,h^{(2)}_{2,1,1}(\tau)\big] +\frac{i \pi}{2 } (\alpha_1+\alpha_2) \;1\otimes \big[ d \tau \big] \,.
\end{align}
Using eq.~\eqref{eq:ftoh} and the definition in eq.~\eqref{eq:itmod}, 
it is easy to recognize in the previous equation the symbol of the function
\begin{align}
\begin{split}
F(\lambda) &=  
 -\frac{\; (\alpha_1 +2  \alpha_2 + \alpha_3)}{i \pi } \; \IMF{2\; 2}{1\; 0}{\tau}  
 + \frac{ \; (\alpha_1 - \alpha_2 - 2 \alpha_3)}{i \pi} \; \IMF{2\; 2}{1\; 1}{\tau}  \\
&
+ \frac{i \pi}{2} \left( \alpha_1 + \alpha_2 \right) \; \IMF{0\; 0}{0\; 0}{\tau} 
\,.
\end{split}
\end{align}
It is easy to check that 
\beq
\Delta\left( \mathcal{T}_1^{(1)}(\lambda) -F(\lambda) \right) = \left( \mathcal{T}_1^{(1)}(\lambda) -F(\lambda) \right)\otimes 1\,.
\eeq
We cannot yet conclude that $\mathcal{T}_1^{(1)}(\lambda)$ is equal to $F(\lambda)$, because they may differ by terms on which the coaction acts trivially.
By computing the difference $\mathcal{T}_1^{(1)}(\lambda) -F(\lambda) $ 
at the cusp at infinity $\tau = i\, \infty$ (which corresponds to $\lambda=0$, where the integrals in eq.~\eqref{eq:T_def} can easily be evaluated in terms of the Euler's Beta function), we find
\begin{align}
\begin{split}\label{eq:T11_IEI}
 \mathcal{T}_1^{(1)}(\lambda)  &= 
  -\frac{\; (\alpha_1 +2  \alpha_2 + \alpha_3)}{i \pi} \; \IMF{2\; 2}{1\; 0}{\tau}  
 + \frac{\; (\alpha_1 - \alpha_2 - 2 \alpha_3)}{i \pi} \; \IMF{2\; 2}{1\; 1}{\tau}  \\
&
+ \frac{i \pi}{2} \left( \alpha_1 + \alpha_2 \right) \; \IMF{0\; 0}{0\; 0}{\tau} 
- 2  \, (\alpha_1 + \alpha_2)  \log 2
\,.
\end{split}
\end{align}
We have checked that the $q$-expansion of the previous equation agrees
with the corresponding $q$-expansion of the original expression eq.~\eqref{eq:FDT1eps1}.

Before we discuss higher orders in $\epsilon$ and the second master integral, let us make a comment about the branch cut structure of eq.~\eqref{eq:T11_IEI}. 
It is well known that hypergeometric functions have a branch cut starting at $\lambda=1$, but not at $\lambda=0$. The cusp at infinity corresponds to the point $\lambda=0$, and so eq.~\eqref{eq:T11_IEI} has no branch point for $\tau \to i \infty$, i.e., eq.~\eqref{eq:T11_IEI} actually lies in the Fourier subspace,
\begin{align}\bsp
\mathcal{T}_1^{(1)}(\lambda) &=  -\frac{\; (\alpha_1 +2  \alpha_2 + \alpha_3)}{i \pi} \; \IMFF{2\; 2}{1\; 0}{\tau}  
 + \frac{ \; (\alpha_1 - \alpha_2 - 2 \alpha_3)}{i \pi} \; \IMFF{2\; 2}{1\; 1}{\tau}  \\
&
- 2  \, (\alpha_1 + \alpha_2)  \log 2\,,
\esp
\end{align}
where the functions $I_F$ were defined is \eqref{eq:IFourier}.

We can repeat the same exercise for higher orders in $\eps$ and for the second master integral.
All the steps are identical, and here we only present the results for the two master integrals in terms of iterated integrals
of Eisenstein series of level $N=2$ up to $\ord(\eps^2)$.
For the first master integral, we find

\begin{align}
\mathcal{T}_1^{(2)}(\lambda) &=
-\frac{\left(\alpha _1+2 \alpha _2+\alpha _3\right){}^2 }{\pi ^2}
\IEIF 
{2}{2}{1}{0}
{2}{2}{1}{0} \stopIEIF{\tau}
-\frac{1}{2} \left(\alpha _1+\alpha _2\right) \left(\alpha _1+2 \alpha _2+\alpha _3\right) 
\IEIF 
{0}{0}{0}{0}
{2}{2}{1}{0} \stopIEIF{\tau} \nonumber \\ &
+\frac{\left(\alpha _1-\alpha _2-2 \alpha _3\right) \left(\alpha _1+2 \alpha _2+\alpha _3\right) }{\pi ^2} \Big[
\IEIF 
{2}{2}{1}{0}
{2}{2}{1}{1} \stopIEIF{\tau}    
+ \IEIF
{2}{2}{1}{1}
{2}{2}{1}{0} \stopIEIF{\tau}  \Big] \nonumber \\ &
+\frac{6 \left(\alpha _1-\alpha _3\right) \left(\alpha _1+2 \alpha _2+\alpha _3\right) }{\pi ^2} 
\IEIF 
{0}{0}{0}{0}
{4}{2}{1}{0} \stopIEIF{\tau}
-\frac{\left(-\alpha _1+\alpha _2+2 \alpha _3\right){}^2 }{\pi ^2}
\IEIF 
{2}{2}{1}{1}
{2}{2}{1}{1} \stopIEIF{\tau}  \nonumber \\ &
+\frac{1}{2} \left(\alpha _1+\alpha _2\right) \left(\alpha _1-\alpha _2-2 \alpha _3\right) 
\IEIF 
{0}{0}{0}{0}
{2}{2}{1}{1} \stopIEIF{\tau}  \nonumber \\ &
+\frac{6 \left(\alpha _1-\alpha _2\right) \left(\alpha _1+\alpha _2+2 \alpha _3\right) }{\pi ^2}
\IEIF
{0}{0}{0}{0}
{4}{2}{1}{1} 
\stopIEIF{\tau}  \nonumber \\ &
+\frac{3 \left(7 \alpha _1^2+7 \left(\alpha _2+\alpha _3\right) \alpha _1+4 \alpha _2^2+4 \alpha _3^2+\alpha _2 \alpha _3\right) }{2 \pi ^2}
\IEIF
{0}{0}{0}{0}
{4}{2}{0}{0} 
\stopIEIF{\tau}  \nonumber \\ &
+\frac{2 \left(\alpha _1+\alpha _2\right) \left(\alpha _1+2 \alpha _2+\alpha _3\right) \log 2 }{i \pi }
\IEIF
{2}{2}{1}{0} 
\stopIEIF{\tau}  \nonumber \\ &
-\frac{2 \left(\alpha _1+\alpha _2\right) \left(\alpha _1-\alpha _2-2 \alpha _3\right) \log 2 }{i \pi }
\IEIF
{2}{2}{1}{1} 
\stopIEIF{\tau}  \nonumber \\ &
%+\frac{1}{6} \left[ \alpha _1^2 \left(\pi ^2+12 \log^2 2 \right)-\alpha _2 \alpha _1 \left(\pi ^2-24 \log^2 2 \right)+\alpha _2^2 \left(\pi ^2+12 \log^2 2 \right)\right] \,.
+\frac{1}{6} \left[ \left( \alpha _1^2 -\alpha _2 \alpha _1 +\alpha _2^2 \right)  \pi ^2  +   \left( \alpha_1 + \alpha_2\right)^2  12 \log^2 2  \right]\,.
\end{align}

\noindent Similarly, we write the second master integral as

\beq\label{eq:T2_to_T2bar}
T_2(z) = \frac{1}{(1+2(\alpha_1+\alpha_2+\alpha_3)\eps)\,z}\,\left[\frac{2\eta_1}{\omega_1}\,T_1(z) + \overline{T}_2(z)\right]\,,
\eeq
with
\beq
\overline{T}_2(z) = \sum_{j=0}^{\infty} \epsilon^j \, \overline{\mathcal{T}}_2^{(j)}(\lambda)\,.
\eeq
The first three coefficients are given by
\begin{align}
\overline{\mathcal{T}}_2^{(0)}(\lambda) &= 0 \,, \\
\overline{\mathcal{T}}_2^{(1)}(\lambda) &= 
\frac{\omega _1}{3}  \left[ \alpha _1 (-(\lambda +1))+\alpha _2 (2 \lambda -1)-\alpha _3 (\lambda -2)\right]
+\frac{\pi ^2 \left(\alpha _1+\alpha _2\right)}{\omega _1}\,,\\
\overline{\mathcal{T}}_2^{(2)}(\lambda) &=
 \frac{\omega_1}{3 i \pi}
\left(\alpha _1+2 \alpha _2+\alpha _3\right)  \left(\alpha _1 (\lambda +1)+\alpha _2 (1-2 \lambda )+\alpha _3 (\lambda -2)\right) 
 \IEIF{2}{2}{1}{0}\stopIEIF{\tau} \nonumber \\ &
- \frac{\omega_1}{3 i \pi} \left(\alpha _1-\alpha _2-2 \alpha _3\right) \left(\alpha _1 (\lambda +1)+\alpha _2 (1-2 \lambda )+\alpha _3 (\lambda -2)\right)
 \IEIF{2}{2}{1}{1}\stopIEIF{\tau}  \nonumber \\ &
 +\frac{i \pi }{\omega _1} \left(\alpha _1+\alpha _2\right)  \left[ \left(\alpha _1+2 \alpha _2+\alpha _3\right) 
 \IEIF{2}{2}{1}{0}\stopIEIF{\tau} 
 -  \left(\alpha _1-\alpha _2-2 \alpha _3\right) 
 \IEIF{2}{2}{1}{1}\stopIEIF{\tau} 
 \right] \nonumber \\ &
 +\frac{6 \alpha _1 \alpha _2 }{\pi ^2 \omega _1} \Big[
 \IEIF{1}{2}{1}{1}{4}{2}{0}{0}\stopIEIF{\tau} - 8  \IEIF{1}{2}{1}{1}{4}{2}{1}{0}\stopIEIF{\tau} - 8 \IEIF{1}{2}{1}{1}{4}{2}{1}{1}\stopIEIF{\tau} \Big] \nonumber \\ &
 +\frac{4 \alpha _2 \left(\alpha _1+\alpha _2+\alpha _3\right) }{\omega _1} \Big[
 \IEIF{1}{2}{1}{1}{2}{2}{0}{0}\stopIEIF{\tau} - 2 \IEIF{1}{2}{1}{1}{2}{2}{1}{0}\stopIEIF{\tau} - 2  \IEIF{1}{2}{1}{1}{2}{2}{1}{1}\stopIEIF{\tau} \Big] \nonumber \\&
 + \frac{12}{i \pi \omega_1} \left[ \left(\alpha _1-\alpha _3\right)
  \left(\alpha _1+2 \alpha _2+\alpha _3\right) 
 \IEIF{4}{2}{1}{0}\stopIEIF{\tau}  
 +\left(\alpha _1-\alpha _2\right) \left(\alpha _1+\alpha _2+2 \alpha _3\right) 
 \IEIF{4}{2}{1}{1}\stopIEIF{\tau} \right]
 \nonumber \\ &
  +\frac{3 \left(7 \alpha _1^2+7 \left(\alpha _2+\alpha _3\right) \alpha _1+4 \alpha _2^2+4 \alpha _3^2+\alpha _2 \alpha _3\right) }{i \pi  \omega _1}
 \IEIF{4}{2}{0}{0}\stopIEIF{\tau} \nonumber \\ &
 +\frac{2}{3} \left(\alpha _1+\alpha _2\right) \omega _1 \log 2 \left(\alpha _1 (\lambda +1)+\alpha _2 (1-2 \lambda )
 +\alpha _3 (\lambda -2)\right)  \nonumber \\ &
   -\frac{2 \pi ^2 \alpha _2 \left(\alpha _2+\alpha _3\right) }{\omega _1}
 \IEIF{1}{2}{1}{1}\stopIEIF{\tau}
 -\frac{2 \pi ^2 \left(\alpha _1+\alpha _2\right){}^2 \log 2}{\omega _1}\,.
\end{align}

% !TEX root = elliptic_symbols.tex
\section{The sunrise integral in two dimensions}
\label{sec:sunrise}

In this section we apply our formalism to the two-loop sunrise integral with three equal masses. This integral can be written as
\beq
S(p^2,m^2) = \int\frac{\mathfrak{D}^dk_1\,\mathfrak{D}^dk_2}{(k_1^2-m^2)(k_2^2-m^2)((k_1+k_2-p)^2-m^2)}\,,
\eeq
where the integration measure in $d=2-2\eps$ dimensions is given by
\beq
\mathfrak{D}^dk \equiv \frac{e^{\gamma_E\eps}}{i\pi^{d/2}}\,d^dk\,,
\eeq
with $\gamma_E=-\Gamma'(1)$ the Euler-Mascheroni constant. This integral has been considered many times before in the literature, and various different representations for it are known~\cite{Laporta:2004rb,Kniehl:2005bc,Bloch:2013tra,Adams:2013nia,Remiddi:2013joa,Adams:2014vja,Adams:2015gva,Adams:2015ydq,Remiddi:2016gno,Adams:2016xah,Adams:2017ejb,Remiddi:2017har,Hidding:2017jkk,Broedel:2017siw}. In particular, in ref.~\cite{Broedel:2017siw} it was shown that the sunrise integral can be expressed in terms of a variant of elliptic polylogarithms, which can themselves be written in terms of the eMPLs reviewed in Section~\ref{sec:empls}~\cite{Broedel:2017kkb}. Moreover, in ref.~\cite{Adams:2017ejb} the same integral was expressed in terms of iterated integrals of Eisenstein series for $\Gamma_1(12)$. So far, however, it has remained rather mysterious why the sunrise integral admits representations in terms of two seemingly very different classes of special functions, and in particular why only Eisenstein series, and no cusp forms, appear. The purpose of this section is to show how we can use our results, in particular the coaction on eMPLs, to rewrite the expression for the sunrise integral of ref.~\cite{Broedel:2017siw} in terms of iterated integrals of Eisenstein series. The result of this section is not only the first application of our `elliptic symbol calculus' to an actual Feynman integral, but it elucidates at the same time the role of iterated integrals of Eisenstein series in the context of the sunrise integral. For simplicity, we only discuss the case of the sunrise integral in strictly $d=2$ dimensions. Higher orders in $\eps$ (as well as the case of the second master integral) are conceptually similar and do not introduce anything new into the discussion.

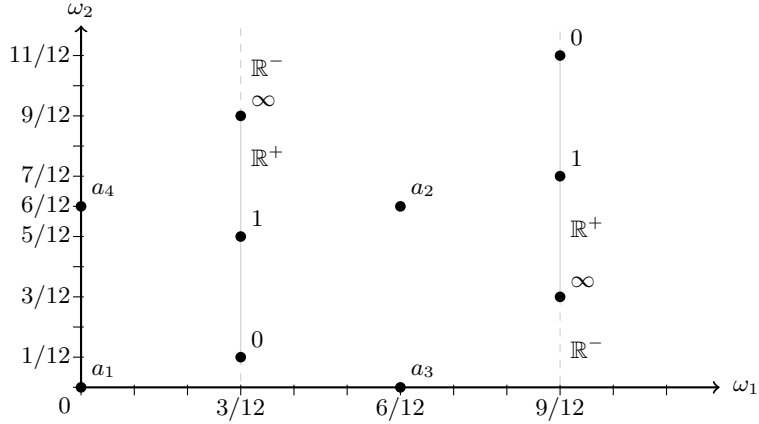
\begin{figure}[t]
    \center
    \begin{tikzpicture}[y=.4cm, x=.7cm, font=\footnotesize]
        % axes
        \draw [->,thick]   (0,0) -- coordinate (x axis mid) (12,0);
        \draw [->, thick]  (0,0) -- coordinate (y axis mid) (0,12);
        % axes labels
        \node at (12.5,0) {$\omega_1$};
        \node at (0,12.5) {$\omega_2$};
        % ticks
        \foreach \x in {0,...,11}
            \draw (\x,1pt) -- (\x,-3pt);
        \foreach \y in {0,...,11}
            \draw (1pt,\y) -- (-3pt,\y);
        % some tick marks
        \node [anchor=north east] at (0,0) {$0$};
        \node [anchor=north] at (3,0) {$3/12$};
        \node [anchor=north] at (6,0) {$6/12$};
        \node [anchor=north] at (9,0) {$9/12$};

        \node [anchor=east]  at (0,1) {$1/12$};
        \node [anchor=east]  at (0,5) {$5/12$};
        \node [anchor=east]  at (0,6) {$6/12$};
        \node [anchor=east]  at (0,9) {$9/12$};
        \node [anchor=east]  at (0,3) {$3/12$};
        \node [anchor=east]  at (0,7) {$7/12$};
        \node [anchor=east]  at (0,11){$11/12$};

        % half periods
        \fill[black] (0,0) circle (2pt);
        \node [anchor=south west] at (0,0) {$a_1$};
        \fill[black] (0,6) circle (2pt);
        \node [anchor=south west] at (0,6) {$a_4$};
        \fill[black] (6,6) circle (2pt);
        \node [anchor=south west] at (6,6) {$a_2$};
        \fill[black] (6,0) circle (2pt);
        \node [anchor=south west] at (6,0) {$a_3$};

        % other punctures
        \draw[gray!40] (3,1) -- (3,9);
        \draw[dashed, gray!40] (3,1) -- (3,0);
        \draw[dashed, gray!40] (3,9) -- (3,12);
        \node [anchor=south west] at (3, 7) {$\mathbb{R}^+$};
        \node [anchor=south west] at (3, 10) {$\mathbb{R}^-$};
        \fill[black] (3,1) circle (2pt);
        \node [anchor=south west] at (3,1) {$0$};
        \fill[black] (3,5) circle (2pt);
        \node [anchor=south west] at (3,5) {$1$};
        \fill[black] (3,9) circle (2pt);
        \node [anchor=south west] at (3,9) {$\infty$};

        % other punctures
        \draw[gray!40] (9,11) -- (9,3);
        \draw[dashed, gray!40] (9,11) -- (9,12);
        \draw[dashed, gray!40] (9,3) -- (9,0);
        \node [anchor=north west] at (9, 6) {$\mathbb{R}^+$};
        \node [anchor=north west] at (9, 2) {$\mathbb{R}^-$};
        \fill[black] (9,11) circle (2pt);
        \node [anchor=south west] at (9,11) {$0$};
        \fill[black] (9,7) circle (2pt);
        \node [anchor=south west] at (9,7) {$1$};
        \fill[black] (9,3) circle (2pt);
        \node [anchor=south west] at (9,3) {$\infty$};
    \end{tikzpicture}
    \caption{Images under Abel's map of the branch points and punctures in the fundamental domain of the torus. Note that every point on the elliptic curve has two images on the torus; we have moved both images of the real line into the fundamental domain using periodicity.}
    \label{fig:fundom}
\end{figure}

Our starting point is the result of the sunrise integral in terms of elliptic polylogarithms $\text{E}_4$ of ref.~\cite{Broedel:2017siw}. In ref.~\cite{Broedel:2017kkb} it was shown that every $\text{E}_4$ function can be written in terms of eMPLs $\widetilde{\Gamma}$. We can write
\beq\bsp
S(p^2,m^2) = \frac{2\omega_1}{(s+m^2)\,\sqrt{a_{12}a_{43}}}\,J(\tau) + \ord(\eps)\,,
\esp\eeq
with $s= -p^2$, $a_{ij} = a_i-a_j$, and
\begin{equation}\begin{split}
  \label{eqn:roots}
a_1 &\,= \frac{1}{2}\left(1-\sqrt{1+\rho}\right)\,,\\
a_2 &\,= \frac{1}{2}\left(1+\sqrt{1+\rho}\right)\,,\\
 a_3&\,= \frac{1}{2}\left(1-\sqrt{1+\overline{\rho}}\right)\,,\\
  a_4&\,=\frac{1}{2}\left(1+\sqrt{1+\overline{\rho}}\right)\,,
\end{split}\end{equation}
where the auxiliary variables $\rho$ and $\bar{\rho}$ are defined by
\begin{equation}
    \rho = -\frac{4m^2}{(m+\sqrt{-s})^2} {\rm~~and~~} \overline{\rho} = -\frac{4m^2}{(m-\sqrt{-s})^2}\,. 
\end{equation}
The two periods of the elliptic curve associated to the sunrise integral are
\beq
\omega_1 = 2\,\textrm{K}(\lambda) {\rm~~and~~}\omega_2 = 2i\,\textrm{K}(1-\lambda)\,,
\eeq
where $\lambda$ denotes the cross-ratio formed out of the four branch points $a_i$,
\beq
\lambda = \frac{a_{13}a_{24}}{a_{12}a_{34}}\,.
\eeq
The function $J(\tau)$ is a linear combination of eMPLs,
\beq\bsp\label{eq:sunrise_J}
J(\tau) &\,= \gamtt{ 0 & 1 }{ 0 & z_{3,1} }{ z_{3,1} }{ \tau  }-\gamtt{ 0 & 1 }{ 0 & z_{3,1} }{ z_{3,5} }{ \tau  }+\gamtt{ 0 & 1 }{ 0 & z_{3,5} }{ z_{3,1} }{ \tau  }-\gamtt{ 0 & 1 }{ 0 & z_{3,5} }{ z_{3,5} }{ \tau  }\\
&\,-2 \gamtt{ 0 & 1 }{ 0 & z_{3,9} }{ z_{3,1} }{ \tau  }+2 \gamtt{ 0 & 1 }{ 0 & z_{3,9} }{ z_{3,5} }{ \tau  }+2 \gamtt{ 0 & 1 }{ 0 & z_{9,3} }{ z_{3,1} }{ \tau  }-2 \gamtt{ 0 & 1 }{ 0 & z_{9,3} }{ z_{3,5} }{ \tau  }\\
&\,-\gamtt{ 0 & 1 }{ 0 & z_{9,7} }{ z_{3,1} }{ \tau  }+\gamtt{ 0 & 1 }{ 0 & z_{9,7} }{ z_{3,5} }{ \tau  }-\gamtt{ 0 & 1 }{ 0 & z_{9,11} }{ z_{3,1} }{ \tau  }+\gamtt{ 0 & 1 }{ 0 & z_{9,11} }{ z_{3,5} }{ \tau  }\\
&\,+\frac{ \tau}{3}  \Big[\gamtt{ 1 }{ z_{3,1} }{ z_{3,5} }{ \tau  }+\gamtt{ 1 }{ z_{3,5} }{ z_{3,1} }{ \tau  }-\gamtt{ 1 }{ z_{3,9} }{ z_{3,1} }{ \tau  }-\gamtt{ 1 }{ z_{3,9} }{ z_{3,5} }{ \tau  }\\
&\,+3 \gamtt{ 1 }{ z_{9,3} }{ z_{3,1} }{ \tau  }-\gamtt{ 1 }{ z_{9,3} }{ z_{3,5} }{ \tau  }-\gamtt{ 1 }{ z_{9,7} }{ z_{3,1} }{ \tau  }-2 \gamtt{ 1 }{ z_{9,11} }{ z_{3,1} }{ \tau  }\\
&\,+\gamtt{ 1 }{ z_{9,11} }{ z_{3,5} }{ \tau  }\Big]-\frac{2\pi i}{9}  \tau ^2\,.
\esp\eeq
The variable $\tau$ is simply the ratio of the two periods,
\beq
\tau = \frac{\omega_2}{\omega_1}=i\frac{\textrm{K}(1-\lambda)}{\textrm{K}(\lambda)}\,,
\eeq
and we have defined
\beq\label{eq:sunrise_z}
z_{r,s} = \frac{r}{12}+\tau\frac{s}{12}\,.
\eeq
Note that, if we assign to $\tau = \gamtt{0}{0}{\tau}{\tau}$ weight zero and length one, then $J(\tau)$ is of uniform length two and weight one.
The points $z_{r,s}$ are the images of the branch points $a_i$ and the points $\{0, 1, \infty\}$ under Abel's map for the elliptic curve $y^2=(x-a_1)\ldots(x-a_4)$ (See fig.~\ref{fig:fundom}).

%%%
%%%
From eqs.~\eqref{eq:sunrise_J} and~\eqref{eq:sunrise_z} we see that the function $J(\tau)$ can be expressed in terms of eMPLs evaluated at rational points. Hence, we expect that $J(\tau)$ can equally-well be expressed in terms of iterated integrals of Eisenstein series for $\Gamma(12)$ (as well as weakly holomorphic modular forms of weight one and level twelve). We can use the same method as for the hypergeometric $_2F_1$ function in the previous section to write $J(\tau)$ in terms of Eisenstein series. We start by computing the coaction of $J(\tau)$. We find
\beq
\Delta(J(\tau)) = J(\tau)\otimes 1 + 1\otimes X_2 + X_{11}\otimes [d\tau]\,,
\eeq
with 
\beq\bsp
X_2 &\,= \frac{4}{\pi^2} \left(\left[ d\tau\, h^{(3)}_{12,0,1}\Big|d\tau\right] - \left[ d\tau\, h^{(3)}_{12,0,5}\Big|d\tau\right] + \left[ d\tau\, h^{(3)}_{12,6,1}\Big|d\tau\right] - \left[ d\tau\, h^{(3)}_{12,6,5}\Big|d\tau\right]\right)\\
&\, +\frac{4}{\pi^2} \left(\left[ d\tau\, h^{(3)}_{12,3,4}\Big|d\tau\right] - \left[ d\tau\, h^{(3)}_{12,3,2}\Big|d\tau\right] + \left[ d\tau\, h^{(3)}_{12,9,4}\Big|d\tau\right] - \left[ d\tau\, h^{(3)}_{12,9,2}\Big|d\tau\right]\right) \\
&\, +\frac{10}{\pi^2} \left(\left[ d\tau\, h^{(3)}_{12,3,1}\Big|d\tau\right]- \left[ d\tau\, h^{(3)}_{12,9,5}\Big|d\tau\right]- \left[ d\tau\, h^{(3)}_{12,3,5}\Big|d\tau\right]  + \left[ d\tau\, h^{(3)}_{12,9,1}\Big|d\tau\right]\right)\,,
\esp\eeq
and
\begin{align}
\nonumber X_{11} &\, = \frac{1}{2\pi i}\Big(\gamtt{ 2 }{ z_{3,1} }{ z_{3,1} }{ \tau  }-\gamtt{ 2 }{ z_{3,1} }{ z_{3,5} }{ \tau  }+\gamtt{ 2 }{ z_{3,5} }{ z_{3,1} }{ \tau  }-\gamtt{ 2 }{ z_{3,5} }{ z_{3,5} }{ \tau  }\\
\nonumber&\,-2\, \gamtt{ 2 }{ z_{3,9} }{ z_{3,1} }{ \tau  }+2\, \gamtt{ 2 }{ z_{3,9} }{ z_{3,5} }{ \tau  }+2\, \gamtt{ 2 }{ z_{9,3} }{ z_{3,1} }{ \tau  }-2\, \gamtt{ 2 }{ z_{9,3} }{ z_{3,5} }{ \tau  }\\
&\,-\gamtt{ 2 }{ z_{9,7} }{ z_{3,1} }{ \tau  }+\gamtt{ 2 }{ z_{9,7} }{ z_{3,5} }{ \tau  }-\gamtt{ 2 }{ z_{9,11} }{ z_{3,1} }{ \tau  }+\gamtt{ 2 }{ z_{9,11} }{ z_{3,5} }{ \tau  }\Big)\\
\nonumber&\,+\gamtt{ 1 }{ z_{3,9} }{ z_{3,1} }{ \tau  }-\gamtt{ 1 }{ z_{3,9} }{ z_{3,5} }{ \tau  }+\frac{2}{3}\, \gamtt{ 1 }{ z_{9,3} }{ z_{3,1} }{ \tau  }-\frac{2}{3}\, \gamtt{ 1 }{ z_{9,3} }{ z_{3,5} }{ \tau  }\\
\nonumber&\,+\frac{1}{6}\, \gamtt{ 1 }{ z_{9,7} }{ z_{3,1} }{ \tau  }-\frac{1}{6}\, \gamtt{ 1 }{ z_{9,7} }{ z_{3,5} }{ \tau  }+\frac{1}{6}\, \gamtt{ 1 }{ z_{9,11} }{ z_{3,1} }{ \tau  }-\frac{1}{6}\, \gamtt{ 1 }{ z_{9,11} }{ z_{3,5} }{ \tau  }\,.
\end{align}
Note that the function $X_2$ can be identified with the symbol of $J(\tau)$, and we see that the symbol only involves Eisenstein series of weight three for $\Gamma(12)$. Using the reflection and distribution identities in eqs.~\eqref{eq:reflect} and~\eqref{eq:latticesum}, we can write the symbol in the very compact form involving only Eisenstein series for $\Gamma(6)$,
\beq
\cS(J(\tau)) = X_2= \frac{2}{\pi^2}\,\left[d\tau\,h_{6,0,1}^{(3)}\Big|d\tau\right] + \frac{2}{\pi^2}\,\left[d\tau\,h_{6,3,4}^{(3)}\Big|d\tau\right] + \frac{5}{\pi^2}\,\left[d\tau\,h_{6,3,1}^{(3)}\Big|d\tau\right]\,.
\eeq
It is easy to convert the symbol into the symbol of iterated integrals of Eisenstein series,
\beq
\cS(J(\tau)) = \cS\left(\cI(\tau)\right)\,,
\eeq
with
\beq
\cI(\tau) = \frac{2}{\pi^2}\,I\left(\begin{smallmatrix} 0& 0\\ 0& 0\end{smallmatrix}\big|\begin{smallmatrix} 3& 6\\ 0&1\end{smallmatrix};\tau\right) + \frac{2}{\pi^2}\,I\left(\begin{smallmatrix} 0& 0\\ 0& 0\end{smallmatrix}\big|\begin{smallmatrix} 3& 6\\ 3&4\end{smallmatrix};\tau\right) + \frac{5}{\pi^2}\,I\left(\begin{smallmatrix} 0& 0\\ 0& 0\end{smallmatrix}\big|\begin{smallmatrix} 3& 6\\ 3&1\end{smallmatrix};\tau\right)\,.
\eeq
While the two functions have the same symbol, we cannot yet conclude that they are equal. We find
\beq
\Delta(J(\tau)-\cI(\tau)) = (J(\tau)-\cI(\tau))\otimes 1\,,
\eeq
and so the two functions can differ at most by a constant. Evaluating the difference at a single point, we find that $J(\tau)=\cI(\tau)$, and so we obtain the following very compact expression for the sunrise integral with equal masses in $d=2$ dimensions,
\begin{align}
S&(p^2,m^2)\\
\nonumber&\, = \frac{2\omega_1}{\pi^2\,(s+m^2)\,\sqrt{a_{12}a_{43}}}\left[{2}\,I\left(\begin{smallmatrix} 0& 0\\ 0& 0\end{smallmatrix}\big|\begin{smallmatrix} 3& 6\\ 0&1\end{smallmatrix};\tau\right) + {2}\,I\left(\begin{smallmatrix} 0& 0\\ 0& 0\end{smallmatrix}\big|\begin{smallmatrix} 3& 6\\ 3&4\end{smallmatrix};\tau\right) + {5}\,I\left(\begin{smallmatrix} 0& 0\\ 0& 0\end{smallmatrix}\big|\begin{smallmatrix} 3& 6\\ 3&1\end{smallmatrix};\tau\right)\right]+\ord(\eps)\,.
\end{align}
We have checked that the previous expression agrees numerically with the sunrise integral by comparing the $q$-expansions of the iterated integrals with a direct numerical evaluation of the Feynman parametric form of the sunrise integral. 

Let us conclude this section with a comment. We started from the representation of the sunrise integral in terms of eMPLs given in ref.~\cite{Broedel:2017siw}, which was obtained by directly performing the Feynman parameter integral in terms of eMPLs. As a consequence, if we combine the results of this section with ref.~\cite{Broedel:2017siw}, we have shown that it is possible to obtain a representation of the sunrise integral in terms of eMPLs and iterated integrals of modular forms starting from the Feynman parameter integral and simply performing all the integrations, without the need for any additional input.

% !TEX root = elliptic_symbols.tex

\section{Conclusions}
\label{sec:conc}

In this paper we have presented for the first time an explicit construction of
(a variant of) the coaction and the symbol map that are applicable to elliptic multiple polylogarithms.  These findings rely on a number of recent results in the mathematical
literature. In particular, a crucial ingredient for our results is the general
construction of a {coaction} on unipotent periods due to Brown. This construction unifies MPLs and more general iterated integrals over some classes of differential forms into one single mathematical framework.

Following this path, we have been able to extend many of the well-known
techniques applied to ordinary MPLs to their elliptic analogues. In addition, we have shown how they
can be used to systematically derive non-trivial functional relations between
eMPLs, similarly to what can be done for MPLs.  Moreover, in our investigation
of eMPLs, we have been naturally lead to consider iterated integrals of
quasi-modular forms, which arise when evaluating eMPLs at some special rational
points of the corresponding elliptic curve.  Our construction for the coaction
is general enough to be applied equally well also to this class of functions.
In particular, it has provided us with a very efficient computational tool to
transform linear combinations of eMPLs evaluated at rational points into linear
combinations of iterated integrals of Eisenstein series.  As a first
non-trivial application of these ideas, we have shown how a special class of
elliptic hypergeometric functions, which can naturally be expressed in terms of
eMPLs, can indeed be represented in terms of iterated integrals of modular
forms.  As a second and physically relevant application, we have considered the, by now famous, massive two-loop sunrise graph.  While both eMPLs and iterated
integrals of modular forms have been shown to appear in different forms in the
evaluation of this Feynman integral, the connection between the two
representations had largely remained unclear until today.  With the use of the
coaction defined in this paper, we have been able  to make this connection
manifest, thus paving the way for a new approach to the
study of Feynman integrals that cannot be expressed in terms of standard MPLs.

Our results suggest a number of interesting directions for future research.
First, in our construction of the coaction we have seen that one fundamental
ingredient was the definition of symbols for certain classes of pairs of
differential forms called de Rham periods.  In particular, this required
dealing exclusively with functions that fulfill {unipotent} differential
equations, i.e. differential equations with trivial homogeneous part.  This
should be contrasted with the fact that in the physics literature the symbol
map is closely connected to {pure functions}, i.e., combinations of MPLs
without any rational functions. This concept of purity is also at the very core
of the definition of a {canonical basis} for MPL-like Feynman integrals.  Since
the requirement of unipotency seems to be more general than that of purity, it
will be interesting to investigate the consequences of this in the definition
of a generalized version of a canonical basis also for non MPL-like Feynman
integrals (see also ref.~\cite{Adams:2018yfj}).  A second direction worth
exploring is the connection with the recent proposal for the construction of a
{diagrammatic coaction} directly at the level of the loop Feynman
integrals~\cite{Abreu:2017enx}.  Until now, such a construction was based on
the coaction defined for MPLs and could therefore be made concrete only for
one-loop Feynman integrals~\cite{Abreu:2017ptx,Abreu:2017mtm}.  With the
results of our paper, a generalization of these ideas to encompass multi-loop
Feynman integrals could become possible, thereby potentially endowing some of the
concepts introduced in this paper with an intrinsic physical meaning.

\section*{Acknowledgments}
We would like to thank Herbert Gangl and Don Zagier for useful discussions
about modular forms, and in particular Don Zagier for providing us with proofs
for some of the finding in Section~\ref{sec:eMPLs_to_modular}. JB, CD and FD
would like to thank the Bethe Center for Theoretical Physics and the Hausdorff
Institute for Mathematics in Bonn for hospitality during part of this project.
This research was supported by the the ERC grant 637019 ``MathAm'', and the U.S.
Department of Energy (DOE) under contract DE-AC02-76SF00515.

\appendix 

% !TEX root = elliptic_symbols.tex

\section{Hopf algebras and comodules}
\label{app:algebras}

In this Appendix we collect definitions of some of the algebraic structures encountered in this paper, in particular Hopf algebras and comodules.
For simplicity, we only discuss vector spaces instead of modules, and we consider all vector spaces to be defined over $\mathbb{Q}$.

A \emph{coalgebra} is a vector space $C$ together with a coproduct $\Delta:C\to C\otimes C$ that is coassociative,
\beq
(\Delta\otimes\textrm{id})\Delta = (\textrm{id}\otimes\Delta)\Delta\,,
\eeq
and is equipped with a counit, i.e., a map $\varepsilon:C\to \mathbb{Q}$ such that
\beq
(\textrm{id}\otimes \varepsilon)\Delta = (\varepsilon\otimes\textrm{id})\Delta = \textrm{id}\,.
\eeq

A \emph{Hopf algebra} is an algebra that is at the same time a coalgebra such that the product and coproduct are compatible,
\beq
\Delta(a\cdot b) = \Delta(a)\cdot \Delta(b)\,,
\eeq 
and it is equipped with an antipode $S:H\to H$ such that 
\beq
S(a\cdot b) = S(b)\cdot S(a)\,,
\eeq
and
\beq
m(S\otimes \textrm{id})\Delta = m(\textrm{id}\otimes S)\Delta = \varepsilon u\,,
\eeq
where $m$ denotes the multiplication in $H$ and $u:\mathbb{Q}\to H$ is the unit (i.e., $u(1)$ is the unit in $H$). The algebra $\cP^{\mathfrak{dr}}$ of all de Rham periods is a Hopf algebra with the coproduct $\Delta^{\mathfrak{dr}}:\cP^{\mathfrak{dr}}\to \cP^{\mathfrak{dr}}\otimes \cP^{\mathfrak{dr}}$ given by
\beq
\Delta^{\mathfrak{dr}}([\omega,\omega']) = \sum_i[\omega,\omega_i]\otimes [\omega_i,\omega']\,.
\eeq

A (right-)comodule over a coalgebra $C$ is a vector space $M$ together with a map $\rho:M\to M\otimes C$ such that
\beq
(\rho\otimes\textrm{id})\rho = (\textrm{id}\otimes\Delta)\rho {\rm~~and~~}(\textrm{id}\otimes \varepsilon)\rho = \textrm{id}\,.
\eeq
The algebra $\cP^{\mathfrak{m}}$ of motivic periods is a comodule over $\cP^{\mathfrak{dr}}$, with the coaction $\rho = \Delta^{\mathfrak{m}}$ given by eq.~\eqref{eq:motivic_coaction}.

% !TEX root = elliptic_symbols.tex

\section{Homotopy-invariant iterated integrals}
\label{app:iterated_integrals}

In this appendix we review some standard material on homotopy-invariant integrals.
If $\gamma$ is a path and $\omega_i$, $1\le i\le n$, are differential one-forms, then we consider iterated integrals defined in the following way. Let $t\in[0,1]$ be a local coordinate parametrizing the path $\gamma$, and we write $\omega_i=dt\,f_i(t)$. We define
\beq
\int_\gamma\omega_1\ldots \omega_n \equiv \int_{0\le t_1\le\ldots\le t_n\le 1}dt_1\,f_1(t_1)\ldots dt_n\,f_n(t_n)\,.
\eeq
Iterated integrals satisfy the following three basic properties. First, they form a shuffle algebra, 
\beq\label{eq:II_shuffle}
\int_{\gamma}\omega_1\ldots \omega_k\,\int_{\gamma}\omega_{k+1}\ldots \omega_n = \sum_{\sigma\in\Sigma(k,n-k)}\int_{\gamma}\omega_{\sigma(1)}\ldots \omega_{\sigma(k)} \omega_{\sigma(k+1)}\ldots \omega_{\sigma(n)}\,.
\eeq
Second, if the path is a composition of two paths, $\gamma=\gamma_1\gamma_2$, then we can write the integral as a combination of integrals over each path separately,
\beq\label{eq:path_composition}
\int_{\gamma_1\gamma_2}\omega_1\ldots \omega_n = \sum_{k=0}^n\int_{\gamma_1}\omega_1\ldots \omega_k\,
\int_{\gamma_2}\omega_{k+1}\ldots \omega_n\,,
\eeq
where by definition the integral over an empty sequence is 1.
Finally, if $\gamma^{-1}$ denotes the path $\gamma$, but in the reverse direction, then
\beq\label{eq:II_reversal}
\int_{\gamma^{-1}}\omega_1\ldots \omega_n = (-1)^n\,\int_\gamma\omega_n\ldots \omega_1\,.
\eeq

In applications, we are usually interested in \emph{homotopy-invariant} iterated integrals, i.e., iterated integrals that do not depend on the details of the path $\gamma$, but only on its endpoints (more precisely, the integral depends on the homotopy class of the path). We would like to have a criterion -- the \emph{integrability condition} -- for when an iterated integral is homotopy-invariant.

 In the case of a single integral, $n=1$, the solution to this problem is well known: the integral $\int_{\gamma}\omega_1$ is homotopy-invariant if and only if the one-form $\omega_1$ is closed, $d\omega_1=0$. We now review a construction which allows one to extend this simple criterion to sequences of one-forms that define iterated integrals. In particular, we will define a differential on sequences of one-forms, and a sequence of one-forms defines a homotopy-invariant iterated integral if and only if the sequence vanishes under the action of this differential. 
 
 We start by giving a formal definition of the concept of `sequence of differential forms' (the latter are often referred to as `words'). We denote by $B$ the vector space spanned by all words of arbitrary length of differential forms (not just one forms). We denote such a word by $[\omega_1|\ldots|\omega_n]$. More formally, $B$ is simply the tensor algebra on the vector space of differential forms.\footnote{In this context, it is customary to denote tensors by $[a|b]$ instead of $a\otimes b$, in order not to confuse it with the tensor sign in the coproduct in eq.~\eqref{eq:Delta_dec}. This notation also gives the name to $B$: it is usually referred to as the `bar-construction'.} This space is equipped with a natural Hopf algebra structure, where the product is given by the shuffle product of tensors, e.g.,
 \beq\bsp
 [\omega_1|\ldots|\omega_k]\sha  [\omega_{k+1}|\ldots|\omega_n] &\,= \sum_{\sigma\in\Sigma(k,n-k)}[\omega_{\sigma(1)}
 \ldots |\omega_{\sigma(k)} |\omega_{\sigma(k+1)}|\ldots |\omega_{\sigma(n)}]\,,
 \esp\eeq
 and the coproduct is given by the deconcatenation of words,
 \beq\label{eq:Delta_dec}
 \Delta_{\textrm{dec}}([\omega_1|\ldots|\omega_n]) = \sum_{k=0}^n [\omega_1|\ldots|\omega_k]\otimes [\omega_{k+1}|\ldots|\omega_n]\,,
 \eeq
 where empty sequences are $1$ by definition. The antipode is simply given by the reversal of words (up to a sign),
 \beq
 S([\omega_1|\ldots|\omega_n]) = (-1)^n\,[\omega_n|\ldots|\omega_1]\,.
 \eeq
 The shuffle product, the coproduct and the antipode on $B$ are in direct correspondence to the three basic properties of iterated integrals in eq.~\eqref{eq:II_shuffle}, \eqref{eq:path_composition} and~\eqref{eq:II_reversal}.
 
 Besides being a Hopf algebra, $B$ can be equipped with more structure. In particular, we can define two linear maps $D_i:B\to B$, $i=1,2$, such that their sum $D=D_1+D_2$ defines a differential on $B$, i.e., $D$ satisfies a variant of the Leibniz rule with respect to the shuffle product (which is the multiplication on $B$) and it squares to zero, $D^2=0$.
 Since we are mostly interested in sequences of \emph{one}-forms, we content ourselves with giving the definition of the differential only in the case of words of one-forms, where some of the expressions simplify. If $[\omega_1|\ldots|\omega_n]$ is a sequence of one-forms, we have
 \beq\bsp\label{eq:bar_differential}
 D_1([\omega_1|\ldots|\omega_n]) &\,= -\sum_{i=1}^n[\omega_1|\ldots|\omega_{i-1}|d\omega_i|\omega_{i+1}|\ldots|\omega_n]\,,\\
 D_2([\omega_1|\ldots|\omega_n]) &\,= \phantom{-}\sum_{i=1}^{n-1}[\omega_1|\ldots|\omega_{i-1}|\omega_i\wedge\omega_{i+1}|\omega_{i+2}|\ldots|\omega_n]\,.
 \esp\eeq
The differential $D$ allows one to generalize the integrability condition to words of one-forms. 
 A linear combination of words is called \emph{integrable} if it lies in the kernel of $D$. If $\xi$ is a linear combination of words, then a theorem by Chen~\cite{ChenSymbol} states that $\xi$ defines a homotopy-invariant iterated integral precisely when $\xi$ is integrable. In other words, the integrability condition can be cast in the compact form $D\xi=0$. We denote the space of integrable words by $B_0\equiv \textrm{Ker}\,D$.\footnote{Technically speaking, $B$ equipped with the differential $D$ is a differential graded algebra, and the space of integrable words can be identified with the zero-th cohomology group of $B$, i.e., $B_0=H^0B$.}  
Let us discuss some of the properties of the space of integrable words. It is easy to see from the Leibniz rule that $B_0$ is closed under forming shuffle products. It can also be shown that the coproduct closes on $B_0$, and more precisely that $B_0$ is actually a sub-Hopf algebra of $B$.

In the following, we show some special cases of the integrability condition in which they take a simpler form.
\begin{enumerate}
\item In the case of a single integration, $n=1$, $D_2$ does not contribute, and $D_1$ is equivalent to the ordinary differential on differential forms,
\beq
D[\omega_1] = D_1[\omega_1] = -[d\omega_1]\,.
\eeq
We see that we recover the usual prescription for $n=1$.
\item In the case where all one-forms are closed, $d\omega_i=0$, the contribution from $D_1$ vanishes, and the integrability condition takes the form familiar from MPLs, e.g.,
\beq\bsp
D_2([\omega_1|\ldots|\omega_n])
&\,=\sum_{i=1}^{n-1}[\omega_1|\ldots|\omega_{i-1}|\omega_i\wedge\omega_{i+1}|\omega_{i+2}|\ldots|\omega_n] = 0\,.
\esp\eeq
It is easy to check that in the case of `$d\log$'-forms this condition reduces to the integrability condition for symbols of MPLs in eq.~\eqref{eq:MPL_integrability}.
\item Finally, in the case of a one-dimensional problem, all one-forms are necessarily closed and the wedge product of any two one-forms always vanishes, $\omega_i\wedge\omega_{i+1}=0$. Hence, every sequence of one-forms is integrable for a one-dimensional problem. This covers in particular the case of hyperlogarithms, e.g., the case of harmonic polylogarithms where $\omega_i\in\{d\log x, d\log(1-x),d\log(1+x)\}$, and also the case of iterated integrals of modular forms.
\item The matrix $T_A$ in eq.~\eqref{eq:TA_series} contains only integrable words, provided that $dA=A\wedge A$. Indeed, we have
\begin{align}
 DT_A&\, = D_1T_A+D_2T_A \\
\nonumber&\,= -[dA] -[dA|A]^R-[A|dA]^R+[A\wedge A]^R +[A\wedge A|A]^R + [A|A\wedge A]^R +\ldots\\
\nonumber&\, = [A\wedge A-dA]^R +[A\wedge A-dA|A]^R + [A|A\wedge A-dA]^R + \ldots\\
\nonumber&\,=0\,.
\end{align}
\end{enumerate}

% !TEX root = elliptic_symbols.tex

\newcommand{\tG}{\widetilde{\Gamma}}

\section{The total differential of eMPLs}
\label{app:proof}

In this appendix we sketch the proof of the formula for the total differential of eMPLs in eq.~\eqref{eq:gamma_differential}. We start by collecting the terms in eq.~\eqref{eq:gamma_differential} according to the differentials they multiply, and we write
\beq\bsp\
d\widetilde{\Gamma}&\left(A_1\cdots A_k;z,\tau \right) = dz\,\cD_z + d\tau\,\cD_{\tau} + \sum_{p=1}^{k}dz_p\,\cD_{z_p}\,,
\esp\eeq
where the functions $\cD_x$ in the right-hand side are given by (with $2\le p\le k-1$)
\begin{align}
\label{eq:dz}
\cD_z &=g^{(n_1)}(z-z_1,\tau)\,\widetilde{\Gamma}(A_2\cdots A_k;z,\tau)\,,\\
\nonumber
\cD_{z_1} &=- g^{(n_1)}(z-z_1,\tau)\,\tG(A_{2}\cdots A_k;z,\tau) -(-1)^{n_2} g^{(n_{1}+n_2)}(z_2-z_{1})\,\tG(\,^0 _0\, A_{3}\cdots A_k;z,\tau)\\
\label{eq:dz1}&+\sum_{r=0}^{n_2} \binom{n_{1}+r-1}{n_{1}-1}g^{(n_2-r)}(z_{1}-z_2)\,\tG(A_1^{[r]}\, A_{3}\cdots A_k;z,\tau)\\
\nonumber&+ \sum_{r=0}^{n_{1}} \binom{n_2+r-1}{n_2-1}g^{(n_{1}-r)}(z_2-z_{1})\,\tG(A_2^{[r]}\, A_{3}\cdots A_k;z,\tau)\,,\\
\nonumber
\cD_{z_k} &=g^{(n_{k})}(-z_{k},\tau)\, \tG(A_1\cdots A_{k-1},z,\tau) +(-1)^{n_k} g^{(n_{k-1}+n_k)}(z_k-z_{k-1})\, \tG(A_1\cdots A_{k-2}\,^0_0,z)\\
\label{eq:dzk}&-\sum_{r=0}^{n_k} \binom{n_{k-1}+r-1}{n_{k-1}-1}g^{(n_k-r)}(z_{k-1}-z_k)\,\tG(A_1\cdots A_{k-1}^{[r]},z,\tau)\\
\nonumber&- \sum_{r=0}^{n_{k-1}} \binom{n_k+r-1}{n_k-1}g^{(n_{k-1}-r)}(z_k-z_{k-1})\,\tG(A_1\cdots A_{k-2} A_{k}^{[r]},z,\tau)\,,\\
\nonumber\cD_{z_p} &= (-1)^{n_p} g^{(n_{p-1}+n_p)}(z_p-z_{p-1})\, \tG(A_1\cdots A_{p-2}\,{^{0}_{0}}\,A_{p+1}\cdots  A_k;z,\tau)\\
\label{eq:dzi}&-(-1)^{n_{p+1}}g^{(n_{p}+n_{p+1})}(z_{p+1}-z_{p})\,\tG(A_1 \cdots A_{p-1}\,{^{0} _{0}}\,A_{p+2}\cdots A_k ;z,\tau)\\
\nonumber&-\sum_{r=0}^{n_p} \binom{n_{p-1}+r-1}{n_{p-1}-1}g^{(n_p-r)}(z_{p-1}-z_p)\,\tG(A_1 \cdots \, A_{p-1}^{[r]}\hat{A}_p \cdots A_k;z,\tau)\\
\nonumber&-\sum_{r=0}^{n_{p-1}} \binom{n_p+r-1}{n_p-1}g^{(n_{p-1}-r)}(z_p-z_{p-1})\, \tG(A_1\cdots\hat{A}_{p-1} \, A_{p}^{[r]} \cdots A_k;z,\tau)\\
\nonumber&+\sum_{r=0}^{n_{p+1}} \binom{n_{p}+r-1}{n_{p}-1}g^{(n_{p+1}-r)}(z_{p}-z_{p+1})\,\tG(A_1\cdots A_p^{[r]} \hat{A}_{p+1} \cdots A_k;z,\tau)\\
\nonumber&+ \sum_{r=0}^{n_{p}} \binom{n_{p+1}+r-1}{n_{p+1}-1}g^{(n_{p}-r)}(z_{p+1}-z_{p})\,\tG(A_1 \, \cdots \hat{A}_p\, A_{p+1}^{[r]} \cdots A_k;z,\tau)\,,\\
\nonumber\cD_{\tau} &= \sum_{p=1}^{k-1}(-1)^{n_{p+1}}\frac{n_p\!+\!n_{p+1}}{2\pi i}\,g^{(n_p+n_{p+1}+1)}(z_{p+1}-z_p)\widetilde{\Gamma}\!\left(A_1\cdots A_{p-1}\; ^0 _0 \; A_{p+2}\cdots A_k;z,\tau \right)\\
\label{eq:dtau}&\,+\sum_{p=1}^{k}\sum_{r=0}^{n_p+1}\Bigg[\!\binom{n_{p-1}+r-1}{n_{p-1}-1}\,\frac{n_p-r}{2\pi i}\,g^{(n_p-r+1)}(z_{p-1}-z_p)\\
\nonumber&\phantom{+\sum_{p=1}^{k}\sum_{r=0}^{n_p+1}\Bigg[}\times \widetilde{\Gamma}\left(A_1\cdots A_{p-1}^{[r]} \; \hat{A}_{p} \; A_{p+1}\cdots A_k;z,\tau \right)\\
\nonumber&\,\phantom{\sum_{p=1}^{k}\sum_{r=0}^{n_p+1}\Big[}
-\binom{n_{p+1}+r-1}{n_{p+1}-1}\,\frac{n_p-r}{2\pi i}\,g^{(n_p-r+1)}(z_{p+1}-z_p)\\
\nonumber&\phantom{+\sum_{p=1}^{k}\sum_{r=0}^{n_p+1}\Bigg[}\times\widetilde{\Gamma}\left(A_1\cdots A_{p-1} \; \hat{A}_{p} \; A_{p+1}^{[r]}\cdots A_k;z,\tau \right)\Bigg]\,.
\end{align}

 The proof of eq.~\eqref{eq:gamma_differential} then becomes equivalent to showing that
\beq\label{eq:to_be_proved}
\partial_x\widetilde{\Gamma}\left(A_1\cdots A_k;z,\tau \right) = \cD_x\,,\quad \forall x\in\{\tau,z,z_1,\ldots,z_k\}\,.
\eeq
This relation is obviously true for $x=z$. We now show that it is also true for the other cases.
 
 We start by computing the partial derivative of $\widetilde{\Gamma}\left(A_1\cdots A_k;z,\tau \right)$ with respect to $z_p$, $2\le p\le k-1$. Differentiating under the integration sign, and using eq.~\eqref{eq:proof_elements}, we find,
 \begin{align}
{\partial_{z_p}}&\tG(A_1 \cdots A_k;z,\tau)\\
\nonumber&=\,\int_0^z dt_1 g^{(n_1)}(t_1-z_1,\tau)\dots \int_0^{t_{p-1}}dt_p\, {\partial_{z_p}} g^{(n_p)}(t_p-z_p,\tau)\tG(A_{p+1}\cdots A_k;t_p,\tau)\\
\nonumber&=\,-\int_0^z dt_1 g^{(n_1)}(t_1-z_1,\tau)\dots \int_0^{t_{p-1}}dt_p\, {\partial_{t_p}} g^{(n_p)}(t_p-z_p,\tau)\tG(A_{p+1}\cdots A_k;t_p,\tau)\\
\nonumber&=\,-\int_0^z dt_1 g^{(n_1)}(t_1-z_1,\tau)\dots \int_0^{t_{p-1}}dt_p\,  \Bigg\{{\partial_{t_p}}\Big[g^{(n_p)}(t_p-z_p,\tau)\tG(A_{p+1}\cdots A_k;t_p,\tau)\Big]\\
\nonumber&\phantom{=\,}\,\,-g^{(n_p)}(t_p-z_p,\tau){\partial_{t_p}}\tG(A_{p+1}\cdots A_k;t_p,\tau)\Bigg\}\\
\nonumber&=\,-\int_0^z dt_1 g^{(n_1)}(t_1-z_1,\tau)\dots \int_0^{t_{p-2}}dt_{p-1}\, g^{(n_{p-1})}(t_{p-1}-z_{p-1},\tau)g^{(n_p)}(t_{p-1}-z_p,\tau)\\
\nonumber&\,\qquad \qquad\times\tG(A_{p+1}\cdots A_k;t_{p-1},\tau)\\
\nonumber&\phantom{=\,}\,\,+\int_0^z dt_1 g^{(n_1)}(t_1-z_1,\tau)\dots \int_0^{t_{p-1}}dt_p\,  g^{(n_p)}(t_p-z_p,\tau)g^{(n_{p+1})}(t_p-z_{p+1},\tau)\\
\nonumber&\,\qquad \qquad\times\tG(A_{p+2}\cdots A_k;t_{p},\tau)\,.
\end{align}
Note that in the previous equation we have assumed that the variables $\{z_1,\ldots,z_k,z,\tau\}$ are all distinct, so that the derivative only acts in the $p$-th position in the iterated integral. To proceed, we apply the Fay identity~\eqref{eq:Fay} to each of the two terms, and it is easy to check that we can then perform all the integrals in terms of eMPLs, and the resulting expression immediately matches eq.~\eqref{eq:dzi}. The proof for $x\in \{z_1,z_k\}$ is very similar. The only difference lies in the fact that some of the boundary terms do not depend on any integration variable anymore, and so we do not need to apply the Fay identity in that case, resulting in fewer terms proportional to binomial coefficients, cf. eqs.~\eqref{eq:dz1} and~\eqref{eq:dzk}.

Finally let us turn to the proof of eq.~\eqref{eq:to_be_proved} in the case $x=\tau$. While conceptually very similar to the other cases, it involves some subtleties, and we therefore discuss this case in some detail. We start by differentiating under the integral sign and we apply eq.~\eqref{eq:proof_elements}. We find
\begin{align}
2\pi i\,&\partial_{\tau}\widetilde{\Gamma}\left(A_1\cdots A_k;z,\tau \right)\\
\nonumber&=\,2\pi i \sum_{p=1}^k\int_0^z dt_1 g^{(n_1)}(t_1-z_1,\tau)\dots \int_0^{t_{p-1}}dt_p\, {\partial_{\tau}} g^{(n_p)}(t_p-z_p,\tau)\tG(A_{p+1}\cdots A_k;t_p,\tau)\\
\nonumber&=\,- \sum_{p=1}^kn_p\int_0^z dt_1 g^{(n_1)}(t_1-z_1,\tau)\dots \int_0^{t_{p-1}}dt_p\, {\partial_{z_p}} g^{(n_p+1)}(t_p-z_p,\tau)\tG(A_{p+1}\cdots A_k;t_p,\tau)\\
\nonumber&=\,- \sum_{p=1}^kn_p\,{\partial_{z_p}}\tG(A_1 \cdots A_p^{[1]}\cdots A_k;z,\tau)\,.
\end{align}
We see that we can obtain the derivative with respect to $\tau$ from our results for the derivatives with respect to $z_p$ with $n_p\to n_p+1$. This gives the following representation for the derivative with respect to $\tau$,
\beq
2\pi i\,\partial_{\tau}\widetilde{\Gamma}\left(A_1\cdots A_k;z,\tau \right) = T_0 + T_1\,,
\eeq
with
\begin{align}
\nonumber T_0&\,=\sum_{p=1}^{k-1}n_p\,(-1)^{n_{p+1}}\,g^{(n_p+n_{p+1}+1)}(z_{p+1}-z_p,\tau)\,\tG(A_1 \cdots A_{p-1}\,{^0_0}\,A_{p+2}\cdots A_k;z,\tau)\\
&\,+\sum_{p=2}^{k}n_p\,(-1)^{n_{p}}\,g^{(n_{p-1}+n_{p}+1)}(z_{p}-z_{p-1},\tau)\,\tG(A_1 \cdots A_{p-2}\,{^0_0}\,A_{p+1}\cdots A_k;z,\tau)\,,\\
\nonumber&\,=\sum_{p=1}^{k-1}(n_p+n_{p+1})\,(-1)^{n_{p+1}}\,g^{(n_p+n_{p+1}+1)}(z_{p+1}-z_p,\tau)\,\tG(A_1 \cdots A_{p-1}\,{^0_0}\,A_{p+2}\cdots A_k;z,\tau)\,,
\end{align}
and
\begin{align}
\nonumber &T_1  = \sum_{p=1}^k\sum_{r=0}^{n_p+1}\Bigg[n_p\binom{n_{p-1}+r-1}{n_{p-1}-1}g^{(n_p-r+1)}(z_{p-1}-z_p,\tau)\,\tG(A_1\cdots A_{p-1}^{[r]}\,A_{p+1}\cdots A_k;z,\tau)\\
\nonumber&\,\,\,\,\phantom{\sum_{p=1}^kn_p\sum_{r=0}^{n_p+1}}-n_p\binom{n_{p+1}+r-1}{n_{p+1}-1}g^{(n_p-r+1)}(z_{p+1}-z_p,\tau)\,\tG(A_1\cdots A_{p-1}\,A_{p+1}^{[r]}\cdots A_k;z,\tau)\Bigg]\\
\label{eq:T1_eq_in_proof}&\,+\sum_{p=1}^{k-1}\sum_{r=0}^{n_p}n_{p+1} \binom{n_{p+1}+r}{n_{p+1}}g^{(n_p-r)}(z_{p+1}-z_p,\tau)\,\tG(A_1\cdots A_{p-1}\,A_{p+1}^{[r+1]}\cdots A_k;z,\tau)\\
\nonumber &\,-\sum_{p=2}^{k}\sum_{r=0}^{n_p}n_{p-1}\binom{n_{p-1}+r}{n_{p-1}}g^{(n_p-r)}(z_{p-1}-z_p,\tau)\,\tG(A_1\cdots A_{p-1}^{[r+1]}\,A_{p+1}\cdots A_k;z,\tau)\,.
\end{align}
The contribution from $T_0$ immediately matches the contribution in the first line of eq.~\eqref{eq:dtau}.
For the first sum in $T_1$, we have been extending the summation ranges using
the convention $n_0=n_{k+1}=0$ and $(z_0,z_{k+1})=(z,0)$, respectively. The
terms for $p=1,r=0$ and $p=k,r=0$ capture then the contributions $n_1
g^{(n_1+1)}(z-z_1)\,\tG(A_{2}\cdots A_k;z,\tau)$ and $-n_k
g^{(n_k+1)}(-z_k)\tG(A_1\cdots A_{k-1},z,\tau)$ from $-n_1\cD_{z_1}$ and
$-n_k\cD_{z_k}$ after employing $\binom{-1}{-1}=1$.  To proceed, we note that
we can as well extend the sums in the last two lines in
eq.~\eqref{eq:T1_eq_in_proof} to the whole range $[1,k]$.  We shift $r\to r-1$
in the last two lines, and obtain
\begin{align}
\nonumber &T_1  = \sum_{p=1}^k\Bigg\{\sum_{r=0}^{n_p+1}\Bigg[n_p\binom{n_{p-1}+r-1}{n_{p-1}-1}g^{(n_p-r+1)}(z_{p-1}-z_p,\tau)\,\tG(A_1\cdots A_{p-1}^{[r]}\,A_{p+1}\cdots A_k;z,\tau)\\
\nonumber&\phantom{\sum_{p=1}^k\Bigg\{n_p\sum_{r=0}^{n_p+1}}-n_p\binom{n_{p+1}+r-1}{n_{p+1}-1}g^{(n_p-r+1)}(z_{p+1}-z_p,\tau)\,\tG(A_1\cdots A_{p-1}\,A_{p+1}^{[r]}\cdots A_k;z,\tau)\Bigg]\\
&+\!\sum_{r=1}^{n_p+1}\!\Bigg[n_{p+1} \!\binom{n_{p+1}+r-1}{n_{p+1}}g^{(n_p-r+1)}(z_{p+1}-z_p,\tau)\tG(A_1\cdots A_{p-1}\,A_{p+1}^{[r]}\cdots A_k;z,\tau)\\
\nonumber &\!\phantom{\sum_{r=1}^{n_p+1}\Bigg[}-n_{p-1}\binom{n_{p-1}+r-1}{n_{p-1}}g^{(n_p-r+1)}(z_{p-1}-z_p,\tau)\,\tG(A_1\cdots A_{p-1}^{[r]}\,A_{p+1}\cdots A_k;z,\tau)\Bigg]\Bigg\}\,.
\end{align}
At this point we note that
\beq\bsp
n_{\alpha}\, \binom{n_{\alpha}+r-1}{n_{\alpha}} &\,= r\,\binom{n_{\alpha}+r-1}{n_{\alpha}-1}\,
\esp\eeq
which allows to include the terms $r=0$ in the last two lines and combine
everything.  We then obtain
\begin{align}
\nonumber &T_1  =\sum_{p=1}^{k}\sum_{r=0}^{n_p+1}\Bigg[\!\binom{n_{p-1}+r-1}{n_{p-1}-1}\,(n_p-r)\,g^{(n_p-r+1)}(z_{p-1}-z_p) \widetilde{\Gamma}\left(A_1\cdots A_{p-1}^{[r]}\; A_{p+1}\cdots A_k;z,\tau \right)\\
&
\,-\binom{n_{p+1}+r-1}{n_{p+1}-1}\,(n_p-r)\,g^{(n_p-r+1)}(z_{p+1}-z_p)\widetilde{\Gamma}\left(A_1\cdots A_{p-1} \;  A_{p+1}^{[r]}\cdots A_k;z,\tau \right)\Bigg]\,,
\end{align}
in agreement with eq.~\eqref{eq:dtau}. This finishes the proof of eq.~\eqref{eq:gamma_differential}.
 
% \beq\bsp\label{eq:gamma_differential}
%d\widetilde{\Gamma}&\left(A_1\cdots A_k;z,\tau \right) = \sum_{p=1}^{k-1}(-1)^{n_{p+1}}\,\widetilde{\Gamma}\left(A_1\cdots A_{p-1}\; ^0 _0 \; A_{p+2}\cdots A_k;z,\tau \right)\,\omega_{p,p+1}^{(n_p+n_{p+1})}\\
%&\,+\sum_{p=1}^{k}\sum_{r=0}^{n_p+1}\Bigg[\binom{n_{p-1}+r-1}{n_{p-1}-1}\,\widetilde{\Gamma}\left(A_1\cdots A_{p-1}^{[r]} \; \hat{A}_{p} \; A_{p+1}\cdots A_k;z,\tau \right)\,\omega_{p,p-1}^{(n_p-r)}\\
%&\,\phantom{\sum_{p=1}^{k}\sum_{r=0}^{n_p+1}\Big[}
%-\binom{n_{p+1}+r-1}{n_{p+1}-1}\,\widetilde{\Gamma}\left(A_1\cdots A_{p-1} \; \hat{A}_{p} \; A_{p+1}^{[r]}\cdots A_k;z,\tau \right)\,\omega_{p,p+1}^{(n_p-r)}\Bigg]\,,
%\esp\eeq

% !TEX root = elliptic_symbols.tex

\section{The coaction and symbols for ordinary MPLs}
\label{sec:Brown_MPLs}
In this appendix we review how to recover the well-known story of symbols and the coaction on ordinary MPLs from the construction in Section~\ref{sec:de-Rham-symbols}. All the formulas in this section mirror the corresponding formulas for eMPLs in Section~\ref{sec:empls_coaction}.
For simplicity, we only discuss the generic case, i.e., we consider all the indices $a_i$ generic and distinct.

From the differential equation in eq.~\eqref{eq:MPL_tot_diff}, it is clear that the weight of MPLs is strictly lowered by one unit by differentiation, and so MPLs are unipotent. In general, $G(\vec a;z)$ is part of a vector $I$ that satisfies a first order differential equation of the form, $dI = A\,I$, where $A$ is strictly upper-triangular. The vector $I$ is given by
\beq\label{eq:I_vec_def_MPL}
I = \left(I_{\vec b}\right)_{\vec b\subseteq \vec a}^T = \left(\int_0^z\omega_{\vec b}\right)_{\vec b\subseteq \vec a}^T\,.
\eeq
In the previous equation $\omega_{\vec b}$ denotes the sequence of differential forms
\beq
\omega_{\vec b} = \left[\frac{dt}{t-b_k}\big|\ldots\big|\frac{dt}{t-b_1}\right]\,,\quad \vec b = (b_1,\ldots,b_k)\,,
\eeq
and by convention we set
\beq
\int_0^z\omega_{\emptyset} \equiv G(;z) = 1\,.
\eeq
As an example, in the case where $\vec a=(a_1,a_2,a_3)$, we have
\beq
I = (G(a_1,a_2;z),G(a_1;z),G(a_2;z),1)^T\,.
\eeq
This vector satisfies the differential equation $dI=AI$, with
\beq
A = \left(\begin{array}{cccc}
0& 0&d\log\frac{a_1-z}{a_1-a_2} & 0\\
0&0&0&d\log\frac{a_1-z}{a_1}\\
0&0&0&d\log\frac{a_2-z}{a_2}\\
0&0&0&0
\end{array}\right)\,.
\eeq
Using eq.~\eqref{eq:dr_rec}, we see that the symbol of a pair $[\omega_{\vec b},\omega_{\vec a}]$ satisfies the recursion
\beq\label{eq:S_dR_MPL}
\symb([\omega_{\vec b},\omega_{\vec a}]) = \sum_{\vec b\subseteq \vec c\,\subseteq \vec a}\left[\symb([\omega_{\vec b},\omega_{\vec c}])\Big| A_{\vec a\vec c}\right]\,.
\eeq
In the previous equation we have labeled the entries of the matrix $A$ using vectors of indices, in the same way as in the definition of the vector $I$ in eq.~\eqref{eq:I_vec_def_MPL}.
The summation range is restricted to $\vec b\subseteq \vec c\subseteq \vec a$ because $A$ is upper triangular. In the case of the example above, we find
\beq\bsp
\symb([\omega_{\emptyset},\omega_{a_1a_2}]) &=\left[\symb([\omega_{\emptyset},\omega_{\emptyset}])\Big| A_{a_1a_2,\emptyset}\right]+\left[\symb([\omega_{\emptyset},\omega_{a_1}])\Big| A_{a_1a_2,a_1}\right] \\
& + 
\left[\symb([\omega_{\emptyset},\omega_{a_2}])\Big| A_{a_1a_2,a_2}\right] + 
\left[\symb([\omega_{\emptyset},\omega_{a_1a_2}])\Big| A_{a_1a_2,a_1a_2}\right]\\
&=
\left[A_{a_2,\emptyset}\Big| A_{a_1a_2,a_2}\right]\\
&=
\left[d\log\frac{a_2-z}{a_2}\Big|d\log\frac{a_1-z}{a_1-a_2}\right]\,.
\esp\eeq
We can now immediately write down the formula for the coaction in eq.~\eqref{eq:Delta_def} in the case of MPLs,
\beq\bsp\label{eq:Delta_u_MPL}
\Delta(G(\vec a;z)) &\,= \sum_{\vec b\subseteq \vec a} G(\vec b;z)\otimes \symb([\omega_{\vec b},\omega_{\vec a}])
%\\
%&\, = 1\otimes \cS([\omega_{\emptyset}^\vee,\omega_{\vec a}]^{\mathfrak{dr}}) + \sum_{\emptyset\neq\vec b\subseteq \vec a} G(\vec b;z)\otimes \cS([\omega_{\vec b}^\vee,\omega_{\vec a}]^{\mathfrak{dr}})
\,.
\esp\eeq
Working out what the previous equation becomes for our example, we find
\beq\bsp
\Delta(G(a_1,a_2;z)) &\,= 1\otimes\left[d\log\frac{a_2-z}{a_2}\Big|d\log\frac{a_1-z}{a_1-a_2}\right] + G(a_2;z)\otimes\left[d\log\frac{a_1-z}{a_1-a_2}\right] \\
&\, + G(a_1,a_2;z)\otimes 1\,.
\esp\eeq

At this point, our formalism does not seem to be directly related to the coaction on MPLs reviewed in Section~\ref{sec:mpls}, because the formula for the coaction in eq.~\eqref{eq:Delta_u_MPL} looks rather different from the formula for the coaction on MPLs in eq.~\eqref{eq:Delta_MPL}. We now show how the coaction in eq.~\eqref{eq:Delta_u_MPL} is related to the more conventional definition of $\Delta_{\textrm{MPL}}$ in eq.~\eqref{eq:Delta_MPL} and eq.~\eqref{eq:Delta_MPL_decompose}.
Let $\pi_0$ denote the projection onto MPLs of weight zero. Since there are no MPLs of negative weight, $\pi_0$ preserves the multiplication, $\pi_0(xy) = \pi_0(x)\pi_0(y)$, and so does its composition with $\Delta$, which can be identified with the symbol of an MPL,
\beq
\cS(G(\vec a;z)) = (\pi_0\otimes\textrm{id})\Delta(G(\vec a;z)) = 1\otimes \symb([\omega_{\emptyset},\omega_{\vec a}]) \,.
\eeq
For our example above, we find
\beq\label{eq:SG(a1,a2,z)_example}
\cS(G(a_1,a_2;z)) = \symb([\omega_{\emptyset},\omega_{a_1a_2}]) = \left[d\log\frac{a_2-z}{a_2}\Big|d\log\frac{a_1-z}{a_1-a_2}\right]\,.
\eeq
It is easy to check that this is indeed the correct symbol of $G(a_1,a_2;z)$ using more conventional definitions and notations, upon performing the replacement
\beq
\left[d\log\frac{a_2-z}{a_2}\Big|d\log\frac{a_1-z}{a_1-a_2}\right] \to \frac{a_2-z}{a_2}\otimes\frac{a_1-z}{a_1-a_2}\,.
\eeq
We can apply the same reasoning to the functions $G_{\vec b}(\vec a;z)$ that appear in the right-hand side of eq.~\eqref{eq:Delta_MPL}, and we see that
\beq
\cS(G_{\vec b}(\vec a;z)) = (\pi_0\otimes\textrm{id})\Delta(G_{\vec b}(\vec a;z))= 1\otimes \symb([\omega_{\vec b},\omega_{\vec a}])\,.
\eeq
Hence, we can write the coaction in eq.~\eqref{eq:Delta_u_MPL} in the alternative form
\beq\bsp\label{eq:Delta_to_Delta_MPL}
\Delta(G(\vec a;z)) &\,= \sum_{\vec b\subseteq \vec a} G(\vec b;z)\otimes \cS(G_{\vec b}(\vec a;z))
\,.
\esp\eeq
Comparing the previous equation to eq.~\eqref{eq:Delta_MPL}, we see that we can obtain $\Delta$ from $\Delta_{\textrm{MPL}}$ by replacing the second entries by their symbol (cf. eq.~\eqref{eq:Delta_MPL_decompose}),
\beq\label{eq:Delta_to_Delta_MPL_2}
\Delta(G(\vec a;z)) = (\textrm{id}\otimes \cS)\Delta_{\textrm{MPL}}(G(\vec a;z))\,.
\eeq

%%%%%%%%%%%%%%%%%%%%%%%%%%%

% !TEX root = elliptic_symbols.tex

\section{Modular forms}
\label{app:modular_forms}

\subsection{Modular forms}
In this Appendix we give a short review on the mathematical background on modular and quasi-modular forms needed in the context of this paper.
In the following, $\Gamma$ denotes a congruence subgroup of some level $N$ of the modular group $SL(2,\mathbb{Z})$, cf. eq.~\eqref{eq:congruence_subgroups}.
$\Gamma$~acts on the extended upper half-plane $\overline{\mathbb{H}}=\mathbb{H}\cup \mathbb{Q}\cup\{i\infty\}$ via M\"obius transformations, see eq.~\eqref{eq:modular_trafo}. It is easy to see that $\Gamma$ acts separately on $\mathbb{H}$ and $\mathbb{Q}\cup\{i\infty\}$. The equivalence classes of $\mathbb{Q}\cup\{i\infty\}$ under the action of $\Gamma$ are called the \emph{cusps} of $\Gamma$. We call the \emph{cusp at infinity} the equivalence class that contains $i\infty$.

$\Gamma$ also acts on functions defined on the extended upper half-plane. A \emph{modular function} is a function $f:\overline{\mathbb{H}}\to \mathbb{C}$ that is invariant under M\"obius transformations for $\Gamma$,
\beq
f(\gamma\cdot\tau) = f(\tau)\,.
\eeq
One can show that every non-constant modular function must have a pole, and so it is not possible to find non-constant functions that are at the same time holomorphic and invariant under M\"obius transformations for $\Gamma$. 

The lack of holomorphic modular functions prompts one to consider functions with more general modular transformation properties. In particular, for every positive integer $n$ we can define an action of $\Gamma$ on functions on $\overline{\mathbb{H}}$ by
\beq
(f|_n\gamma)(\tau) = (c\tau+d)^{-n}\,f(\gamma\cdot \tau)\,,\quad \gamma=\left(\begin{smallmatrix}a&b\\c&d\end{smallmatrix}\right) \in \Gamma\,.
\eeq
We say that a function is \emph{weakly modular of weight $n$ for $\Gamma$} if it is invariant under this action,
\beq
(f|_n\gamma)(\tau) = f(\tau)\,.
\eeq
This condition is equivalent to the transformation behaviour in eq.~\eqref{eq:weakly_modular}.
A \emph{modular form of weight $n$ for $\Gamma$} is a function $f:\overline{\mathbb{H}}\to \mathbb{C}$ such that
\begin{enumerate}
\item $f$ is weakly modular of weight $n$ for $\Gamma$,
\item $f$ is holomorphic on $\mathbb{H}$,
\item $f$ is holomorphic at the cusps of $\Gamma$, i.e., $(f|_n\gamma)$ is holomorphic at $i\infty$ for all $\gamma\in SL(2,\mathbb{Z})$.
\end{enumerate}
We denote the vector space of modular forms of weight $n$ for $\Gamma$ by $\cM_n(\Gamma)$. It can be shown that $\cM_n(\Gamma)$ is always finite-dimensional. Moreover, the space of all modular forms is a graded algebra
\beq
\cM_{\bullet}(\Gamma) = \bigoplus_{n=0}^\infty\cM_n(\Gamma){\rm~~and~~} \cM_m(\Gamma)\cdot \cM_n(\Gamma)\subseteq \cM_{m+n}(\Gamma)\,.
\eeq

%%%%%%%%%%%%%%%%%%%%%%%%%%%%%%%%%%%%%%%%%%%%%%%%%

\subsection{Quasi-modular forms}

In Section~\ref{sec:modular_forms} we have only discussed iterated integrals of modular forms, and in Section~\ref{sec:eMPLs_to_modular} we have argued that eMPLs evaluated at rational points naturally lead to iterated integrals of Eisenstein series for $\Gamma(N)$ for some value of $N$. However, not every Eisenstein series is a modular form. In particular, the Eisenstein series $G_2$ of weight two does not transform as a weakly modular function, cf. eq.~\eqref{eq:G2_trafo}. This prompts us to consider functions with a slightly more general transformation property than eq.~\eqref{eq:weakly_modular}.

A \emph{quasi-modular form of weight $n$ and depth $p$ for $\Gamma$} is a function $f:\overline{\mathbb{H}}\to \mathbb{C}$ that is holomorphic on $\mathbb{H}$ and at the cusps such that
\beq\label{eq:quasi_modular}
(f|_n\gamma)(\tau) = f(\tau)  +  \sum_{r=1}^p f_r(\tau)\,\left(\frac{c}{c\tau+d}\right)^r\,,
\eeq
where $f_1,\ldots,f_p$ are holomorphic functions. Let us discuss some examples of quasi-modular forms. It is easy to see that every modular form is also a quasi-modular form of depth zero. Comparing eq.~\eqref{eq:quasi_modular} to the transformation of the Eisenstein series of weight two in eq.~\eqref{eq:G2_trafo}, we see that $G_2(\tau)$ is a quasi-modular form of weight two and depth one. Finally, derivatives of modular forms are in general not modular, but they are quasi-modular forms of depth one. More generally, unlike the space of modular forms, the space of quasi-modular forms is closed under differentiation.
One can show (cf. Proposition 20 of ref.~\cite{ZagierModular}) that these are essentially the only quasi-modular forms. More precisely, every quasi-modular form can be written as polynomial in $G_2(\tau)$ with coefficients that are modular forms. Alternatively, and more interesting in the context of iterated integrals, one can show that the space of quasi-modular forms is generated as a vector space by all modular forms and $G_2(\tau)$, as well as their derivatives. Since in the context of iterated integrals we can always integrate away all derivatives, we see that we only need to add $G_2(\tau)$ as a new integration kernel in order to describe \emph{all} iterated integrals of quasi-modular forms (cf. ref.~\cite{Matthes:QuasiModular} for a study of iterated integrals of quasi-modular forms of level $N=1$). Hence, all the results from Section~\ref{sec:modular_forms} immediately generalize to quasi-modular forms if we include $G_2(\tau)$ into our set of integration kernels.

%%%%%%%%%%%%%%%%%%%%%%

\subsection{A basis for Eisenstein series for $\Gamma_1(N)$}
In eqs.~\eqref{eq:Eisenstein_Basis_2} and~\eqref{eq:Eisenstein_Basis} we presented a (conjectural) basis for $\cE_n(\Gamma(N))$. As $\cE_n(\Gamma(N))$ contains the Eisenstein series for $\Gamma_1(N)$ as a special case, we can write every Eisenstein series for $\Gamma_1(N)$ in terms of the basis in eq.~\eqref{eq:Eisenstein_Basis}. Here we present an explicit basis for $\cE_n(\Gamma_1(N))$ in terms of the basis for $\Gamma(N)$ in eq.~\eqref{eq:Eisenstein_Basis_2}.

We start by analyzing how $\Gamma_1(N)$ acts on the functions $h_{N,r,s}^{(n)}$. From eq.~\eqref{eq:trafo} we see that the matrix $\left(\begin{smallmatrix}a& b\\c&d\end{smallmatrix}\right)=\left(\begin{smallmatrix}1& b\\0&1\end{smallmatrix}\right)\!\!\mod N$ acts on $h_{N,r,s}^{(n)}$ in a very simple way,
\beq\bsp
 h^{(n)}_{N,r,s}\left(\frac{a\tau + b}{c\tau+d}\right) &\,=(c\tau+d)^n\,h^{(n)}_{N,r+sb,s}(\tau)\,,\quad\left(\begin{smallmatrix}a& b\\c&d\end{smallmatrix}\right)\in\Gamma_1(N)\,.
\esp\eeq
Next, let us consider an element $h\in \cE_n(\Gamma_1(N))$. We can write $h$ in the basis of eq.~\eqref{eq:Eisenstein_Basis_2},
\beq
h(\tau) = \sum_{(r,s)\in B_N}c_{r,s}\,h_{N,r,s}^{(n)}(\tau)\,.
\eeq
The behaviour under $\Gamma_1(N)$ transformations implies the following constraint on the linear combination,
\beq\label{eq:weak_constraint}
\sum_{(r,s)\in B_N}c_{r,s}\,h_{N,r,s}^{(n)}(\tau) = \sum_{(r,s)\in B_N}c_{r,s}\,h_{N,r+bs,s}^{(n)}(\tau)\,.
\eeq
This relation must hold for every $\left(\begin{smallmatrix}a& b\\c&d\end{smallmatrix}\right)\in\Gamma_1(N)$, and in particular, for 
$\left(\begin{smallmatrix}1& 1\\0&1\end{smallmatrix}\right)\in\Gamma_1(N)$ (note that $\left(\begin{smallmatrix}1& 1\\0&1\end{smallmatrix}\right)\notin\Gamma(N)$), and so we obtain a constraint that is independent of $b$,
\beq\label{eq:strong_constraint}
\sum_{(r,s)\in B_N}c_{r,s}\,h_{N,r,s}^{(n)}(\tau) = \sum_{(r,s)\in B_N}c_{r,s}\,h_{N,r+s,s}^{(n)}(\tau)\,.
\eeq
Although eq.~\eqref{eq:strong_constraint} is a special case of eq.~\eqref{eq:weak_constraint}, eq.~\eqref{eq:strong_constraint} implies eq.~\eqref{eq:weak_constraint}, because invariance under shifts by one unit implies invariance under shifts by $b$ units.
We then see that elements of $\cE_n(\Gamma_1(N))$ correspond to `cyclic sums' when expressed in terms of the basis of $\Gamma(N)$ in eq.~\eqref{eq:Eisenstein_Basis}.

In the following we present a basis for $\cE_n(\Gamma_1(N))$ in terms of `cyclic sums' of basis elements of $\cE(\Gamma_1(N))$. We find that
\beq\label{eq:Eisenstein_G1_Basis_2}
\cE_n(\Gamma_1(N)) = \Big\langle \sigma_{N,r,s}^{(n)}: (r,s) \in B^1_N\Big\rangle_{\mathbb{C}}\,,
\eeq
with $B^1_N = B^1_{N,1}\cup B^1_{N,2}\cup B^1_{N,3}$ and
\begin{align}\label{eq:Eisenstein_G1_Basis}
B^1_{N,1}&\, = \{(r,s): 0\le r<\textrm{gcd}(N,s)\textrm{ and } 0<s\le\lceil N/2-1\rceil\textrm{ and } \textrm{gcd}(N,r,s) = 1\}\,, \\
\nonumber B^1_{N,2}&\, = \{(r,0): 1\le r\le N/2\textrm{ and } \textrm{gcd}(N,r) = 1\}\,, \\
\nonumber B^1_{N,3}&\, = \left\{\begin{array}{ll}\displaystyle \{(r,N/2): 1\le r\le N/4\textrm{ and } \textrm{gcd}(N,r,N/2) = 1\}\,,& \textrm{ $N$ even}\,,\\
\emptyset\,, &\textrm{ $N$ odd}\,.
\end{array}\right.
\end{align}
The functions $\sigma_{N,r,s}^{(n)}$ are given by
\beq
\sigma_{N,r,s}^{(n)}(\tau) = \sum_{k=0}^{n_{N,r,s}}h_{N,r+k\, g_{N,s},s}^{(n)}(\tau)\,,
\eeq
with 
\beq
n_{N,r,s} = \left\lfloor\frac{N-r-1}{g_{N,s}}\right\rfloor \textrm{ and } g_{N,s} = \textrm{gcd}(N,s)\,.
\eeq
Just like in the case of $\Gamma(N)$, the formula does not hold for small weights and level, because some basis elements are absent. It is easy to see that the functions $\sigma_{N,r,s}^{(n)}$ satisfy the constraint in eq.~\eqref{eq:strong_constraint}, and so they lie in the Eisenstein subspace for $\Gamma_1(N)$. Moreover, they are all linearly independent in $\cE_n(\Gamma(N))$, and thus also in $\cE_n(\Gamma_1(N))$. Finally, we have checked that the number of elements $\sigma_{N,r,s}^{(n)}$ in eq.~\eqref{eq:Eisenstein_G1_Basis_2} is equal to the dimension of $\cE_n(\Gamma_1(N))$, at least up to $N=100$. For $N>4$, the dimension of $\cE_n(\Gamma_1(N))$ is given by the closed formula
\beq
\textrm{dim}\,\cE_n(\Gamma_1(N)) = \frac{1}{2}\sum_{d|N}\phi(d)\phi(N/d)\,,
\eeq
where $\phi$ is Euler's totient function, i.e., $\phi(d)$ is the number of integers less than $d$ that are coprime to $d$. Based on our analysis, we conclude that the $\sigma_{N,r,s}^{(n)}$ form a basis of $\cE_n(\Gamma_1(N))$ at least through level $N=100$, and we conjecture that this holds in general.

% !TEX root = elliptic_symbols.tex

\section{$q$-expansion for the functions $g^{(n)}$}
\label{app:q_exp}

The integration kernels $g^{(n)}(z,\tau)$ employed in definition
\eqref{eq:gamt_def} can be expanded in a $q$-series. Writing out the
$q$-expansions for the Jacobi $\theta$ functions the right-hand side of
eq.~\eqref{eq:Eisenstein-Kronecker} and sorting the powers of $\alpha$ leads to
the following expansions 
\begin{align}
g^{(1)}(z,\tau) &= \pi  \cot(\pi z) + 4\pi \sum_{m=1}^{\infty}  \sin(2\pi m z) \sum_{n=1}^{\infty} q^{mn}
\label{kron7.6}\\
g^{(k)}(z,\tau) \Big|_{k=2,4,\ldots} &= - 2 \Big[ \zeta_k + \frac{ (2\pi i)^k }{(k-1)!} \sum_{m=1}^{\infty}  \cos(2\pi m z) \sum_{n=1}^{\infty}n^{k-1} q^{mn} \Big]
\label{kron7.10} \\
g^{(k)}(z,\tau) \Big|_{k=3,5,\ldots}&= - 2i \frac{ (2\pi i)^k }{(k-1)!} \sum_{m=1}^{\infty}  \sin(2\pi m z) \sum_{n=1}^{\infty}n^{k-1} q^{mn}\,,
\label{kron7.11} 
\end{align}
where the simple pole in $g^{(1)}$ is represented by $\pi \cot(\pi z)=\frac{1}{z}+ {\cal O}(z)$.

\bibliography{bib}

\end{document}